\documentclass[usenatbib,useAMS]{mn2e}

\usepackage{amssymb}
\usepackage{graphicx}

\title[Runaway Stars]{Ejection velocities of high Galactic latitude runaway stars}
\author[M. D. V. Silva and R. Napiwotzki]{M.~D.~V.~Silva$^1$\thanks{E-mail: m.d.d.v.silva@herts.ac.uk} and R.~Napiwotzki$^1$\\
$^1$Centre for Astrophysics Research, STRI, University of Hertfordshire, College Lane, Hatfield AL10 9AB}

\date{2010 October}

\pagerange{\pageref{firstpage}--\pageref{lastpage}} \pubyear{2010}

\begin{document}

\maketitle

\begin{abstract}
We estimate the distribution of ejection velocities for the known population of high galactic latitude runaway stars. The initial sample is a collection of 174 early-type stars selected from the literature. The stars are first classified according to their evolutionary status in order to obtain a homogeneous sample of 96 genuine main sequence stars. Their present velocities and flight times are then estimated using proper motion data from various astrometric catalogues (including Tycho-2, UCAC2, and USNO-B) and the  ejection velocities are computed by tracing their orbits back in time, based on a galactic potential. The potential used is constructed from a mass density model chosen to fit the most recent observational constraints.

We find evidence for two different populations of runaway stars: a ``high'' velocity population, with a maximum ejection velocity of about 400 -- 500 $\mathrm{km}\, \mathrm{s}^{-1}$, and a ``low'' velocity population, with a maximum ejection velocity of about $300\ \mathrm{km}\, \mathrm{s}^{-1}$. We argue that the observed limit of $500\ \mathrm{km}\, \mathrm{s}^{-1}$ and the bimodality of the observed ejection velocity distribution are natural consequences of the so-called Binary Ejection Mechanism. We discuss the connection between the ``high'' velocity population and the so-called hypervelocity stars, showing how previously studied hypervelocity stars are consistent with the results obtained.

We also find that some stars that were once thought to be best explained as being formed in the Halo are compatible with a runaway hypothesis once proper motions are included in the analysis. However, three stars in the selected sample appear to be inconsistent with ejection from the galactic disc. Possible scenarios are discussed, including a possible formation in the galactic halo. 
\end{abstract}


\section{Introduction}
Star formation in our galaxy is believed to be confined to the star forming regions of the disc. ``Runaway'' stars are young, blue early-type stars that have been observed outside these star forming regions (OB associations, open clusters) and have usually kinematics different from typical early-type, main sequence stars. A survey of blue stars performed by \citet{greenstein} revealed the presence of these main sequence stars in the Halo. In a later work, \citet{tobin1} argued that a low-resolution spectrum alone does not permit us to distinguish normal luminosity main sequence stars from the later hot evolutionary stages of low mass stars. Other explanations for the presence of runaway stars in the Halo were proposed: formation \textit{in situ} in the Halo, and ejection from the disc by a powerful ejection mechanism.
Two ejection mechanisms have been proposed:
\begin{enumerate}
\item {\bf the binary ejection mechanism (BEM)}: this was first proposed by \citet{blaauw} to explain the ejection of runaway O and B stars out of the galactic plane. In this scenario the secondary star of a close binary receives its ejection velocity when the primary explodes as a supernova. Because one of the stars explodes, this model predicts that runaway stars should never be found in binaries (composed of two main sequence stars). Calculations by \citet{zwart} predict a negative mass-ejection velocity correlation (secondary stars with lower masses receive the greatest kicks) and a maximum ejection velocity of $\lesssim 300\ \mathrm{km}\, \mathrm{s}^{-1}$. Similar results were found by \citet{leonard3}. More recently, \citet{przybilla} estimated that velocities up to $\simeq 400\ \mathrm{km}\, \mathrm{s}^{-1}$ are possible in binaries containing an early B and a Wolf-Rayet star.
\item {\bf the dynamical ejection mechanism (DEM)}: this was first proposed by \citet*{poveda} as an alternative to produce runaway stars. Dynamical interactions between stars inside young, open clusters can give large kicks to one or both stars involved in a collision, i.e. a close encounter. The large ejection velocities are achieved most efficiently by collisions of two close binaries (since they have larger cross-sections for the collision). Calculations by \citet{leonard2} predict a binary fraction among runaway stars of about 10 per cent, a negative mass-ejection velocity correlation, and a maximum ejection velocity of $\lesssim 200\ \mathrm{km}\, \mathrm{s}^{-1}$. Later simulations by \citet{leonard5}, where a much greater number of experiments were conducted, revised the value of the maximum ejection velocity to $\lesssim 400\ \mathrm{km}\, \mathrm{s}^{-1}$, although very rare events producing velocities up to $\lesssim 1000\ \mathrm{km}\, \mathrm{s}^{-1}$ are possible (by setting the parameters \textit{a posteriori} to maximize the velocity). More recently, \citet*{gvaramadze1} have shown that collisions between binaries and very massive stars ($M \gtrsim 50\ \mathrm{M}_{\odot}$) can also eject stars with velocities up to $300$ -- $400\ \mathrm{km}\, \mathrm{s}^{-1}$.
\end{enumerate}

Both mechanisms were found to operate in nature in a study by \citet*{hoogerwerf} by tracing back in time the orbits of runaway stars to their parent clusters. Two main sequence runaway stars were traced back to the same region of the Orion association Ori OB1, which is evidence for the DEM. In contrast, a runaway star and a pulsar were traced back to the same region of the Sco OB2 association, which is evidence for the BEM. This study used proper motion (high precision astrometry was made available by Hipparcos) and radial velocity data to trace back in time the orbits of runaway stars to the parent cluster. This method makes it possible to estimate the moment in time when the star left the cluster and also the velocity at that instant (the ejection velocity). Since the orbits are computed in a realistic galactic potential, actual estimates of these quantities were obtained instead of lower limits as was done in previous studies which used only radial velocity information and/or ignored the galactic potential. In this work we use a similar method but we apply it to distant stars.

The predicted ejection velocity distribution is similar in both models but there are some differences, in particular the DEM predicts an enhanced high velocity tail. Studies in the past have not been able to properly constrain the ejection velocity distribution for two reasons: small, biased samples, and lack of proper motion measurements for the more distant stars. Since the more distant stars correspond to the higher velocities, this second aspect is of particular relevance. This is made clear if we remember that the more systematic studies to date were based on Hipparcos data \citep{martin2,allen2}, with a limiting magnitude of $V\simeq 12.4$ and complete up to $V\simeq 9$. On the other hand, studies based on the Palomar-Green \citep{green1} and Edinburgh-Cape \citep{stobie1} surveys reach fainter magnitudes but many stars in these studies have unreliable ejection velocity estimates due to the lack of proper motion measurements.

Evidence in recent studies for a link between runaway stars and hypervelocity stars has been mounting up. Hypervelocity stars are a different class of stars moving with even higher velocities than runaway stars, in unbound orbits. They are generally believed to be ejected by the Supermassive Black Hole in the centre of the Milky Way \citep*{brown3}. A systematic search for hypervelocity stars undertaken by \citet*{brown2} resulted in the finding of seven stars still bound to the galaxy. Although these stars were still interpreted as hypervelocity stars (based on their high spatial velocities and large distances) and consequently assumed to have been ejected from the centre of the galaxy, it is possible that at least some of them were ejected far from the galactic centre as well, since no dynamical analysis was performed to verify their places of origin. The hypervelocity stars HD~271791 and HIP~60350 were determined to have been ejected from regions far from the galactic centre and with velocities compatible with present models (\citealt{heber2} and \citealt{przybilla}) and for HIP~60350 \citep{irrgang}.

The possibility that some young stars that are observed in the halo are not consistent with an ejection from the galactic plane scenario is still under debate. This is usually interpreted to mean that they were formed \emph{in situ} in the halo. This possibility has been investigated in many studies (cf. \citealt{hambly3} and references therein), however usually the conclusion is that, once the full kinematical information is taken into account, it is not possible to rule out the runaway hypothesis. A good example is the star PHL~346 (cf. \citealt{keenan4}) which was shown to be consistent with ejection from the galactic plane by \citet*{ramspeck}, however new candidates were presented in the same paper, in particular the stars PHL~159, PG~1511+467, SB~357, and HS~1914+7139. 

The purpose of our study is to perform a systematic and homogeneous analysis of the sample of known runaway stars at high galactic latitudes. This will overcome the major limitations of the previous studies: lack of proper motion data in many cases, no error analysis, and limited range in apparent magnitudes/height above the galactic plane. In particular, the \citet{martin2} and \citet{allen2} studies excluded almost all stars in the high velocity tail of ejection velocity distribution due to the restriction to bright stars, whereas the several studies on the Palomar-Green and Edimburgh-Cape surveys \citep{stobie1} lacked proper motion data for many stars, although they cover a greater range in brightness (and distance). Our aim was to produce more solid constraints to the theoretical models by estimating the ejection velocity distribution, shed light on the relationship between runaway and hypervelocity stars, and investigate the possible existence of candidates for star formation in the halo.

This article is organised as follows: in Section~\ref{s:sample} we describe the selected sample of runaway stars candidates, in Section~\ref{s:evol} we describe the \emph{criteria} and method used to classify the evolutionary status of the stars in the initial sample, including an extensive discussion of the influence of rotation, ending with a short description of the final sample of main sequence stars. In Section~\ref{s:orb} we describe the method used to compute the orbits of the sample of main sequence stars, using the position and full space velocities as input, in Section~\ref{s:results} we present our results regarding the ejection velocities and flight times, where we try to make a case for the existence of an upper limit to the ejection velocity and a link between runaway stars and low-velocity ``hypervelocity'' stars, finally we summarize all results in the final Section~\ref{s:conclusion}.

\section{Sample}
\label{s:sample}
Candidate runaway stars were compiled from previous studies. We only selected stars for which a spectroscopic analysis is published, so we could be sure of their main sequence status and know their atmospheric parameters. The 
quality of the analysis for the different samples is not homogeneous so we found it convenient to separate the list of 
candidate stars in two groups. We have included in the first group (Group A, Table~\ref{table1}) the samples selected from the Palomar-Green (PG) survey \citep{saffer}, the Edinburgh-Cape (EC) survey, and the sample from \citet{ramspeck}. In the second group (Group B, Table~\ref{table2}) we have included the samples from the papers by \citet{conlon2}, \citet{martin1}, and \citet{behr}.

The ``complete'' sample of 28 stars selected by \citet{saffer} from the PG survey and the sub-samples, including a ``complete''
sample of 13 stars, selected by \citet{rolleston1}, \citet{magee2}, and \citet{lynn2} from the EC survey, constitute our 
main sources of candidates since they provide good coverage (in a statistical sense) of both hemispheres.
The PG sample was studied in a series of papers: \citet{hambly1}, \citet{rolleston2}, and \citet{lynn1}. These studies performed high resolution spectroscopic analyses, including the determination of radial velocities and atmospheric parameters, and detailed abundance analyses.
The EC sample was similarly studied in a series of papers by \citet{rolleston1}, \citet{magee2} and \citet{lynn2}.
We have also included candidates found in four other sources. The sample from \citet{ramspeck} which is of special interest because it includes many candidates at high distances (up to $7\, \mathrm{kpc}$) from the galactic plane (if they are indeed on the main sequence).
Another important source is the study by \citet{conlon3} which contains seven candidates, many of which appear to be at very high distances from the plane (more than $4\, \mathrm{kpc}$).

The sample of \citet{conlon2} provides 32 candidates which are also part of the Hipparcos catalogue. These 32 stars were 
studied in a series of papers (\citealt*{keenan1}; \citealt*{keenan2}; \citealt*{keenan3}; \citealt{conlon1}). Finally we have included a few candidates found in \citet{martin1} and \citet{behr}. \citet{martin1} performed an extensive analysis of its candidates. It should be noted that the study by \citet{behr} focuses on the distribution of the projected rotational velocity of Blue Horizontal Branch stars, hence its atmospheric parameters estimates are not appropriate for the parameter range of runaway stars.
It is important to note that there are overlaps between these different samples.
The total number of initial candidate stars was then 174.

\begin{table*}
\centering
\begin{minipage}{150mm}
\centering
\caption{Group A stars, including stars from the Palomar-Green (PG) survey \citep{saffer}, the Edinburgh-Cape survey, and the sample from \citet{ramspeck}. The ``?'' symbol indicates no data was available, whereas a ``(?)'' is used to distinguish cases where a MS status is less well established. One $\checkmark$ corresponds to $40\ \mathrm{km}\, \mathrm{s}^{-1}\le v\, \sin i < 70\ \mathrm{km}\, \mathrm{s}^{-1}$, two $\checkmark$ correspond to $70\ \mathrm{km}\, \mathrm{s}^{-1}\le v\, \sin i < 150\ \mathrm{km}\, \mathrm{s}^{-1}$, and three $\checkmark$ to $v\, \sin i > 150\ \mathrm{km}\, \mathrm{s}^{-1}$.}

\label{table1}
\tiny
\begin{tabular}{ll|cccc|lc}
\hline
Name & Alternative & Parallax & $v\, \sin i$ & Consistent & Inconsistent & Verdict & References \\
 & name & $<2\,\sigma$ & & abundances & with PAGB nature & & \\
\hline
PG 0122+214 & & ? & $ \checkmark \checkmark $ & Yes & Yes & MS & (6),(7) \\
PG 1511+367 & & ? & $ \checkmark \checkmark $ & Yes & Yes & MS & (6) \\
PG 1533+467 & & ? & $ \checkmark \checkmark \checkmark $ & Yes & Yes & MS & (6) \\
PG 1610+239 & & ? & $ \checkmark \checkmark $ & Yes & Yes & MS & (6) \\
PHL 159 & & ? & - & ? & Yes & MS (?) & (6) \\
PHL 346 & & ? & $ \checkmark $ & Yes & No & MS & (6) \\
SB 357 & & ? & $ \checkmark \checkmark \checkmark $	& Yes & Yes & MS & (2),(6) \\
BD --15 115 & & Yes & - & Yes & Yes & MS & (2),(5),(6),(8) \\
HS 1914+7139 & & ? & $ \checkmark \checkmark \checkmark $ & Yes & Yes & MS & (6) \\
PG 0009+036 & & ? & $ \checkmark \checkmark \checkmark $ & ? & Yes & MS & (3) \\
PG 0855+294 & & ? & $ \checkmark \checkmark $ & Yes & Yes & MS & (3),(7) \\
PG 0914+001 & & ? & $ \checkmark \checkmark \checkmark $ & ? & Yes & MS & (3) \\
PG 0934+145 & & ? & - & No & Yes & Non-MS & (3) \\
PG 0936+109 & & ? & - & No & Yes & Non-MS & (3) \\
PG 0954+049 & & ? &	? & No & Yes & Non-MS & (3) \\
PG 0955+291 & & ? & $ \checkmark \checkmark \checkmark $ & Yes & Yes & MS & (3) \\
PG 1011+293 & & ? & - & No & Yes & Non-MS & (3) \\
PG 1205+228 & & Yes & $ \checkmark \checkmark \checkmark $ & Yes & Yes & MS & (1),(3),(7) \\
PG 1209+263 & & ? & $ \checkmark \checkmark $ & No & Yes & MS & (3) \\
PG 1212+369 & & ? & ? & No & ? & Non-MS & (3) \\
PG 1213+456 & & ? & ? & No & Yes & Non-MS & (3) \\
PG 1243+275 & & ? & ? & No & No & Non-MS & (3) \\
PG 1310+316 & PB 3408 & ? & ? & No & Yes & Non-MS & (3) \\
PG 1332+137 & Feige 84 & Yes & $ \checkmark \checkmark $ & Yes & Yes & MS & (3),(7),(8) \\
PG 1351+393 & PB 890 & ? & ? & No & Yes & Non-MS & (3) \\
PG 2111+023 & & ? & $ \checkmark \checkmark $ & Yes & Yes & MS & (3) \\
PG 2120+062 & & ? & - & Yes & No & Non-MS & (3) \\
PG 2128+146 & & ? & - & No & No & Non-MS & (3) \\
PG 2134+049 & & ? & - & No & No &	Non-MS & (3) \\
PG 2146+087 & & ? & - & No & Yes & Non-MS & (3) \\
PG 2159+051 & & ? & - & No & Yes & Non-MS & (3) \\
PG 2214+184 & & ? & - & No & Yes & Non-MS & (3) \\
PG 2219+094 & & ? & $ \checkmark \checkmark \checkmark $ & Yes & Yes & MS & (3),(6),(7) \\
PG 2229+099 & & ? & - & No & Yes & MS (?) & (3) \\
PG 2237+178 & & ? & ? & ? & ? & ? & (3) \\
PG 2345+241 & & ? & - & Yes & Yes & MS & (3),(7) \\
PG 2351+198 & & ? & - & No & Yes & Non-MS & (3) \\
PG 2356+167 & & ? & - & No & Yes & Non-MS & (3) \\
EC 04420--1908 & & ? & $ \checkmark \checkmark \checkmark $ & Yes & Yes & MS & (5) \\
EC 01483--6804 & CPD --68 91 & ? & - & No & No & Non-MS & (5) \\
EC 05515--6231 & HD 40031 & ? & $ \checkmark $ & No & Yes & MS (?) & (5) \\
EC 06012--7810 & & ? & - & No & Yes & Non-MS & (5) \\
EC 09470--1433 & & ? & - & Yes & Yes & Non-MS & (5) \\
EC 19071--7643 & HIP 94513 & Yes & - & Yes & Yes & MS & (5) \\
EC 19337--6743 & HD 184308 & No & $ \checkmark \checkmark \checkmark $ & Yes & Yes & MS & (5) \\
EC 19476--4109 & HD 187311 & Yes & $ \checkmark \checkmark $ & Yes & Yes & MS & (5) \\
EC 19489--5641 & BPS CS 22896 -0165 & ? & ? & ? & ? & ? & (5) \\
EC 19490--7708 & & ? & - & No & Yes & Non-MS & (5) \\
EC 19579--4259 & & ? & - & No & Yes & Non-MS & (5) \\
EC 19586--3823 & CPD --38 7924 & ? & $ \checkmark \checkmark \checkmark $ & Yes & Yes & MS & (5) \\
EC 19596--5356 & & ? & $ \checkmark \checkmark \checkmark $ & Yes & Yes & MS & (9) \\
EC 20011--5005 & & ? & - & Yes & Yes & MS & (5) \\
EC 20089--5659 & HD 191466 & ? & $ \checkmark \checkmark $ & Yes & Yes & MS & (5) \\
EC 20104--2944 & & ? & $ \checkmark $ & Yes & Yes & MS & (5) \\
EC 20252--3137 & & ? & $ \checkmark $ & No & Yes & MS (?) & (5) \\
EC 20485--2420 & & ? & - & No & No & Non-MS & (5) \\
EC 03240--6229 & & ? & $ \checkmark \checkmark \checkmark $ & Yes & Yes & MS & (4) \\
EC 03462--5813 & & Yes & $ \checkmark \checkmark \checkmark $ & Yes & Yes & MS & (4) \\
EC 05229--6058 & & ? & - & No & No & Non-MS & (4) \\					
EC 05438--4741 & & ? & - & Yes & Yes & MS & (4) \\					
EC 05490--4510 & & ? & - & No & Yes & MS (?) & (4) \\
EC 05515--6107 & & ? & $ \checkmark \checkmark \checkmark $ & Yes & Yes & MS & (4) \\
EC 05582--5816 & & ? & $ \checkmark \checkmark \checkmark $ & Yes & Yes & MS & (4) \\
EC 06387--8045 & & ? & $ \checkmark \checkmark \checkmark $ & Yes & Yes & MS & (4) \\
EC 09414--1325 & & ? & $ \checkmark \checkmark \checkmark $ & Yes & Yes & MS & (4) \\
EC 09452--1403 & & ? & $ \checkmark \checkmark $ & ? & Yes & MS (?) & (4) \\		
EC 10087--1411 & & ? & $ \checkmark \checkmark \checkmark $ & Yes & Yes & MS & (4) \\
EC 10500--1358 & & ? & $ \checkmark \checkmark $ & Yes & Yes & MS & (4) \\
EC 10549--2953 & & ? & $ \checkmark \checkmark \checkmark $ & Yes & Yes & MS & (4) \\
EC 11074--2912 & & ? & - & No & Yes & Non-MS & (4) \\
EC 13139--1851 & & ? & $ \checkmark $ & Yes & Yes & MS & (4) \\
EC 20140--6935 & & Yes & $ \checkmark $ & Yes & Yes & MS & (4) \\
EC 20153--6731 & & ? & $ \checkmark \checkmark $ & Yes & Yes & MS & (4) \\
EC 20292--2414 & & Yes & $ \checkmark \checkmark \checkmark $ & Yes & Yes & MS & (4) \\
EC 20411--2704 & & ? & - & Yes & Yes & Non-MS & (4) \\
EC 23169--2235 & & ? & $ \checkmark \checkmark $ & Yes & Yes & MS & (4) \\
\hline
\end{tabular}
References: (1) \citet{conlon2}; (2) \citet{conlon3}; (3) \citet{saffer}; (4) \citet{rolleston1}; (5) \citet{magee2}; (6) \citet{ramspeck}; (7) \citet{behr}; (8) \citet{martin1}; (9) \citet{lynn2}.
\end{minipage}
\end{table*}

\begin{table*}
\centering
\begin{minipage}{150mm}
\centering
\caption{Group B stars, including the samples from the papers of \citet{conlon2}, \citet{martin1}, and \citet{behr}. The symbols have the same meaning as in Table \ref{table1}.}
\centering
\label{table2}
\tiny
\begin{tabular}{ll|cccc|lc}
\hline
Name & Alternative & Parallax & $v\, \sin i$ & Consistent & Inconsistent & Verdict & References \\
 & name & $<2\,\sigma$ & & abundances & with PAGB nature & & \\
\hline
PB 5418 & & ? & $ \checkmark $ & Yes & Yes & MS & (2) \\
Ton S 195 & SB 463 & ? & - & No & Yes & Non-MS & (2) \\
Ton S 308 & & ? & $ \checkmark \checkmark $ & Yes & Yes & MS & (2) \\
PHL 2018 & & ? & $ \checkmark \checkmark \checkmark $ & Yes & Yes & MS & (2) \\
BD --2 3766 & HD 121968 & Yes & $ \checkmark \checkmark \checkmark $ & Yes & Yes & MS & (2),(8) \\
BD +00 0145 & & ? & - & No & & Non-MS & (7) \\
BD +36 2242 & & Yes & $ \checkmark \checkmark $ & ? & Yes & MS (?) & (7) \\
HD 7374 & & No & - & ? & Yes & Non-MS & (7) \\
HD 27295 & & No & - & ? & Yes & Non-MS & (7) \\
HD 128801 & & No & - & No & Yes & Non-MS & (7) \\
HD 135485 & & No & - & No & Yes & Non-MS & (7) \\
PG 1530+212 & & ? & $ \checkmark \checkmark $ & ? & Yes & MS (?) & (7) \\
HIP 1241 & HD 1112 & Yes & $ \checkmark \checkmark $ & No & Yes & MS (?) & (7),(8) \\
HIP 1511 & & Yes & - & No & Yes & Non-MS & (8) \\
HIP 1904 & HD 1999 & Yes & ? & Yes & Yes & MS & (1) \\
HIP 2702 & HD 3175 & Yes & - & Yes & Yes & MS & (1),(5) \\
HIP 3812 & JL 212 & Yes & $ \checkmark \checkmark \checkmark $ & Yes & Yes & MS & (1),(5) \\
HIP 6419 & HD 8323 & Yes & ? & ? & Yes & ? & (1),(8) \\
HIP 11809 & Feige 23 & Yes & $ \checkmark \checkmark \checkmark $ & Yes & Yes & MS & (1),(8) \\
HIP 11844 & HD 15910 & Yes & $ \checkmark \checkmark $ & Yes & Yes & MS & (8) \\
HIP 12320 & Feige 25 & Yes & $ \checkmark \checkmark \checkmark $ & Yes & Yes & MS & (1),(8) \\
HIP 13800 & Feige 29 & Yes & $ - $ & Yes & Yes & MS & (2),(7) \\
HIP 15967 & HD 21305 & Yes & - & No & Yes & Non-MS & (8) \\
HIP 16130 & HD 21532 & Yes & $ \checkmark $ & Yes & Yes & MS & (1),(8) \\
HIP 16466 & HD 21996 & No & - & Yes & Yes & Non-MS & (1) \\
HIP 16758 & HD 22586 & Yes & $ \checkmark \checkmark $ & No & No & MS (?) & (1),(5) \\
HIP 28132 & HD 40267 & Yes & $ \checkmark \checkmark $ & Yes & Yes & MS & (8) \\
HIP 37903 & BD +61 996 & Yes & $ \checkmark \checkmark \checkmark $ & Yes & No & MS & (8) \\
HIP 41979 & & Yes & - & No & Yes & Non-MS & (8) \\
HIP 45904 & HD 233622 & Yes & $ \checkmark \checkmark \checkmark $ & No & Yes & MS (?) & (7),(8) \\
HIP 48394 & HD 237844 & Yes & $ \checkmark \checkmark \checkmark $ & No & Yes & MS (?) & (8) \\
HIP 50750 & BD +16 2114 & Yes & ? & ? & Yes & ? & (8) \\
HIP 51624 & HD 91316 & Yes & ? & ? & No & ? & (1) \\
HIP 52906 & BD +38 2182 & Yes & $ \checkmark \checkmark \checkmark $ & Yes & Yes & MS & (1),(8) \\
HIP 55051 & HD 97991 & Yes & $ \checkmark \checkmark \checkmark $ & ? & No & MS (?) & (1) \\
HIP 55461 & Feige 40 & Yes & $ \checkmark \checkmark $ & Yes & Yes & MS & (1),(7),(8) \\
HIP 56322 & HD 100340 & Yes & $ \checkmark \checkmark \checkmark $ & Yes & Yes & MS & (7),(8) \\
HIP 58046 & HD 103376 & Yes & $ \checkmark \checkmark \checkmark $ & Yes & Yes & MS & (1),(7),(8) \\
HIP 59067 & HD 105183 & Yes & $ \checkmark \checkmark $ & Yes & Yes & MS & (1),(7),(8) \\
HIP 59955 & HD 106929 & Yes & $ \checkmark \checkmark \checkmark $ & Yes & Yes & MS & (1),(8) \\
HIP 60578 & BD +49 2137 & Yes & - & ? & Yes & ? & (7),(8) \\
HIP 60615 & BD +36 2268 & Yes & $ \checkmark $ & No & Yes & Non-MS & (1),(7),(8) \\
HIP 61800 & HD 110166 & Yes & $ \checkmark \checkmark \checkmark $ & Yes & Yes & MS & (8) \\
HIP 65388 & PB 166 & Yes & - & No & Yes & Non-MS & (8) \\
HIP 69247 & HD 123884 & Yes & - & No & Yes & Non-MS & (8) \\
HIP 70275 & HD 125924 & Yes & $ \checkmark $ & ? & Yes & MS (?) & (1),(7),(8) \\
HIP 71667 & BD +20 3004 & Yes & $ \checkmark \checkmark $ & ? & Yes & MS (?) & (7),(8) \\
HIP 75577 & HD 137569 & Yes & - & Yes & Yes & Non-MS & (7),(8) \\
HIP 76161 & HD 138503 & Yes & ? & Yes & No & MS (?) & (8) \\
HIP 77131 & HD 140543 & Yes & $ \checkmark \checkmark \checkmark $ & No & No & MS (?) & (8) \\
HIP 77716 & BD +33 2642 & Yes & - & No & No & Non-MS & (7),(8) \\
HIP 79649 & HD 146813 & No & $ \checkmark \checkmark $ & Yes & Yes & MS (?) & (1),(8) \\
HIP 81153 & HD 149363 & Yes & $ \checkmark \checkmark $ & No & No & MS (?) & (8) \\
HIP 82236 & BD +13 3224& Yes & ? & ? & Yes & ? & (8) \\
HIP 96130 & HD 183899 & No & $ \checkmark $ & Yes & Yes & MS (?) & (8) \\
HIP 98136 & HD 188618 & Yes & $ \checkmark \checkmark $ & Yes & No & MS (?) & (8) \\
HIP 104931 & & Yes & ? & ? & & ? & (8) \\
HIP 105912 & HD 204076 & Yes & $ \checkmark \checkmark $ & Yes & No & MS (?) & (1) \\
HIP 107027 & HD 206144 & Yes & $ \checkmark \checkmark \checkmark $ & Yes & No & MS (?) & (1),(8) \\
HIP 108215 & HD 208213 & Yes & $ \checkmark \checkmark \checkmark $ & Yes & Yes & MS & (1) \\
HIP 109051 & HD 209684 & Yes & $ \checkmark \checkmark $ & Yes & Yes & MS & (8) \\
HIP 111396 & HD 213781 & Yes & - & Yes & Yes & MS & (1),(7),(8) \\
HIP 111563 & HD 214080 & Yes & $ \checkmark \checkmark $ & Yes & No & MS (?) & (1) \\
HIP 112790 & HD 216135 & Yes & $ \checkmark \checkmark $ & Yes & Yes & MS & (1),(8) \\
HIP 113735 & HD 217505 & Yes & ? & Yes & Yes & MS & (1) \\
HIP 114569 & HD 218970 & Yes & $ \checkmark \checkmark $ & No & Yes & MS (?) & (1),(8) \\
HIP 114690 & HD 219188 & Yes & $ \checkmark \checkmark \checkmark $ & Yes & Yes & MS & (1) \\
HIP 115347 & HD 220172 & Yes & - & Yes & No & MS (?) & (1) \\
HIP 115729 & HD 220787 & Yes & - & Yes & Yes & MS & (1),(7),(8) \\
\hline
\end{tabular}
References:~(1)~\citet{conlon2}; (2) \citet{conlon3}; (3) \citet{saffer}; (4) \citet{rolleston1}; (5) \citet{magee2}; (6) \citet{ramspeck}; (7) \citet{behr}; (8) \citet{martin1}; (9) \citet{lynn2}.
\end{minipage}
\end{table*}

\begin{figure*}

\includegraphics[width=124mm]{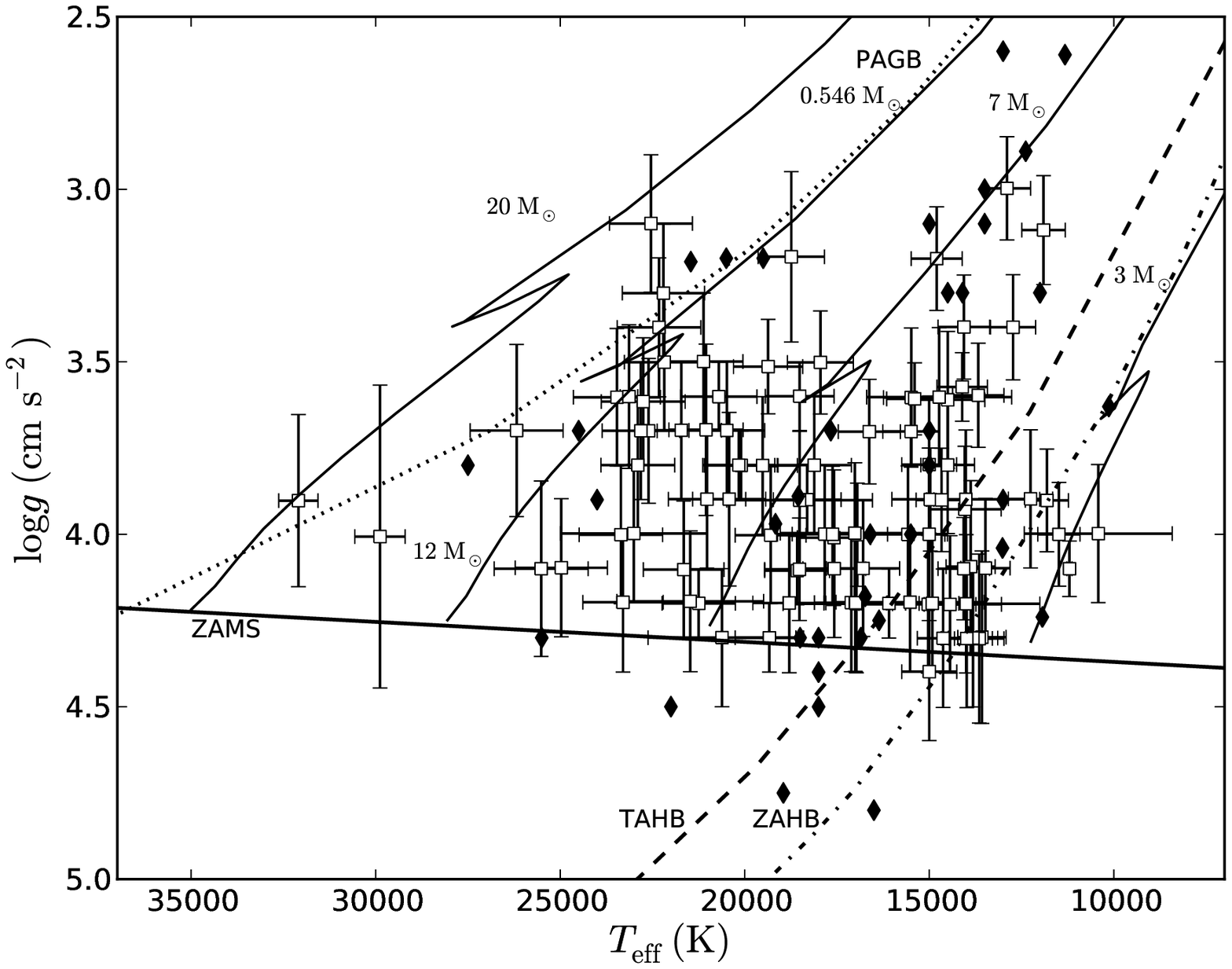}
\caption{$T_{\mathrm{eff}}-\log g$ diagram. The white squares with error bars represent stars classified as main sequence. The black diamonds represent stars classified as evolved stars, mostly blue horizontal branch. Also shown are theoretical tracks for masses in the range 3 -- 20 $\mathrm{M}_{\odot}$ \citep{schaller} as solid lines. The zero age horizontal branch (ZAHB) is shown as a dashed-dotted line and the terminal age horizontal branch (TAHB) is shown as a dashed line, both for a Helium mass fraction of $Y=0.247$ and $\mathrm{[Fe/H]}=-1.48$ \citep{dorman}. Note that the tracks for the runaway stars are for solar metallicity, in contrast with the (low) metallicity assumed for the HB stars, which is appropriate for halo metal-poor stars. The theoretical track of a PAGB star of $0.546\ \mathrm{M}_{\odot}$ \citep{schoenberner} is shown as a dotted line.}
\label{fig:1}
\end{figure*}

\section{Evolutionary status}
\label{s:evol}
In principle, it should be possible to identify main sequence stars from the position they occupy on a $T_{\mathrm{eff}}-\log g$  diagram, however the region of the diagram which corresponds to O and B stars is crossed by low mass stars in post main sequence evolutionary stages \citep{tobin1}. Low mass stars (initial mass $M\lesssim 2\ M_{\odot}$) will evolve to the horizontal branch after the Helium flash. Stars in the horizontal branch phase burn Helium in the core and Hydrogen in a shell. The effective temperature of a star when it enters the horizontal branch will be directly linked to the mass lost during the red giant phase. The Horizontal Branch (hereafter HB) stars which occupy the region of interest ($10000\ \mathrm{K} < T_{\mathrm{eff}} < 30000\ \mathrm{K}$) correspond to the hotter end of this mass sequence and are called blue horizontal branch stars (hereafter BHB stars). The post-HB evolution of these stars depends on the mass of the envelope which determines the strength of the Hydrogen burning shell \citep*{dorman}: when the envelope mass is greater than a given critical mass, the star will evolve to the AGB, and after a period of rapid mass loss, it will enter the Post-AGB (hereafter PAGB) phase; on the other hand, when the envelope mass is less than the critical value, the star will either not reach the tip of the AGB, or not even enter the AGB, staying hot until it enters the white dwarf cooling sequence. Since BHB stars have envelopes with small masses they correspond to this second group. In Fig.~\ref{fig:1}, we have plotted a $T_{\mathrm{eff}}-\log g$ diagram where theoretical tracks for main sequence (hereafter MS) \citep{schaller} and low mass PAGB stars \citep*{schoenberner} and the zero age horizontal branch \citep{dorman} are shown, giving an indication of the regions occupied with stars in different evolutionary stages.

Stars in our sample were classified as main sequence stars or old evolved stars based on their position on the $T_{\mathrm{eff}}-\log g$  diagram, abundance pattern, projected rotation velocity, and parallax. In Fig.~\ref{fig:1}, we show a $T_{\mathrm{eff}}-\log g$  diagram with the stars that were selected from the initial sample. These are the stars we believe are on, or near, the main sequence and whose orbits were computed, as described in Section~\ref{s:orb}.

\subsection{Atmospheric parameters}
Temperatures and gravities were computed from Str\"omgren $uvby\beta$ photometry (\citealt*{moon1} calibration; recalibrated and implemented by \citealt*{napiwotzki}) whenever it was available. The main sources of photometry were \citet{hauck} and \citet{mooney}, the latter concentrating only on stars from the PG survey. The temperature and gravity estimates obtained in spectroscopic studies were also considered, whenever available. These estimates always assume LTE model atmospheres (with the exception of \citealt{lynn2}). There are two groups of estimates: those based on medium-resolution spectroscopy and estimates based on high-resolution spectroscopy. Those based on high resolution spectroscopy were preferred, followed by the estimates obtained from the Str\"omgren $uvby\beta$ photometry. The assumed errors on $\log g$ (using the $\beta$ filter) and $T_{\mathrm{eff}}$ photometric estimates were of $0.2\ \mathrm{dex}$ and 10 per cent, respectively. The errors on $\log g$ and $T_{\mathrm{eff}}$ estimates taken from the literature range between $\simeq 0.1 - 0.25$ and $\simeq 10 - 20$~per~cent, respectively. \citet{martin1} estimated $\log g$ by searching the closest match in a model grid. This method is of low precision since it can easily produce large errors, potentially twice the size of the separation between grid points which was $0.25\ \mathrm{dex}$. Results from other investigations were preferred if available. The values for the stars classified as main sequence are shown in Appendix \ref{ap:2}.

The position of a star in the $T_{\mathrm{eff}}-\log g$ diagram can often be used to rule out a PAGB nature. Indeed, as shown in Fig.~\ref{fig:1}, $T_{\mathrm{eff}}$ and $\log g$ of most runaway stars are only consistent with either a MS or BHB evolutionary status. In Fig.~\ref{fig:1} we can see that the BHB stars populate only a narrow strip in the $T_{\mathrm{eff}}-\log g$ diagram, however this phase is relatively long lived (when compared with the lifetime of OB main sequence stars), lasting about $100\ \mathrm{Myr}$, which means this is the region of the $T_{\mathrm{eff}}-\log g$ diagram where we expect most contamination by low mass evolved stars. However, most stars classified as MS are in the region between the PAGB tracks and the terminal BHB. Given that BHB stars spend only $\sim 10\ \mathrm{Myr}$ in the post-BHB phase (which is sometimes called ``AGB-\emph{manque}'', as these stars do not ascend the AGB), the contamination in this region is expected to be minimal. The number of selected stars in the region where the PAGB tracks intersect the main sequence is even smaller. As an aside, we note that the lack of stars in the PAGB region is due to the small number of O stars in our sample. In summary, BHB and post-BHB stars occupy the region of the $T_{\mathrm{eff}}-\log g$ diagram between the ZAHB and the PAGB tracks, however they spend 90 per cent of their lifes between the ZAHB and the TAHB, and only 10 per cent between the TAHB and the PAGB track. Hence the contamination after the TAHB should be minimal.

\subsection{Abundances}
An abundance analysis was, in general, part of all the spectroscopic studies. This analysis uses the respective atmospheric 
parameters estimates and assumes an LTE atmosphere. The abundance pattern permitted us to distinguish MS stars from BHB stars, as the atmospheres of the latter are dominated by diffusion and show strong deviations from the approximately solar mix seen in MS stars. In particular, there is usually an He depletion and depletion or enhancement of metals, depending on the evolutionary history, and balance between the effects of diffusion and levitation of the heavier metals due to radiation pressure \citep{behr3}. In the case of PAGB stars, some elements can be modified by dredge ups, but the absence of strong modification of the abundance pattern is usually not conclusive. The abundances found in the literature were compared with the normal abundances for B stars, found in \citet{kilian1,kilian2}. A star with abundances that were not consistent with normal abundances, within errors, was marked as such.

\subsection{Projected rotation velocity}
\label{s:prv}
A high projected rotation velocity, or $v\sin i$, where $i$ is the angle between the line of sight and the rotation axis, is an excellent indicator of a young age as old evolved stars do not rotate as fast as young objects. The maximum rotation velocity for a blue HB star appears to be around $30 - 40\ \mathrm{km}\, \mathrm{s}^{-1}$ \citep{behr}. As such, we have considered a star with a projected rotation velocity in excess of $40\ \mathrm{km}\, \mathrm{s}^{-1}$ to be likely a young object, a star with a projected rotation velocity in excess of $70\ \mathrm{km}\, \mathrm{s}^{-1}$ to be very likely a 
young object, and a star with a projected rotation velocity in excess of $150\ \mathrm{km}\, \mathrm{s}^{-1}$ to be extremely likely a young object. This is noted in Tables~\ref{table1}~and~\ref{table2} by 1, 2 or 3 ticks respectively.

It is known that a high rotation velocity will make a star appear cooler and more luminous (lower surface gravity implying a larger radius) if observed at high inclination angles \citep{fremat,wenske}. This effect is sometimes called \emph{gravitational darkening} and is caused by the increasing effective gravitational acceleration as a function of the latitude, with the equatorial regions becoming cooler than the polar regions of the star. The selected sample contains three Be stars: SB~357 \citep{ramspeck}, PG~0914+001 \citep{rolleston2} and HIP~3812 \citep{magee2}. Typical rotation velocities of a Be star are very high (on average 88 per cent of the break-up velocity according to \citealt{fremat}), we would expect significant effects due to gravitational darkening. Indeed, if the measured surface gravity for PG~0914+001 is taken at face value then the estimated critical velocity ($v_{crit}=273\ \mathrm{km}\, \mathrm{s}^{-1}$) will be lower than the measured projected rotation velocity ($v\sin i=325\ \mathrm{km}\, \mathrm{s}^{-1}$), which strongly suggests that the surface gravity has been underestimated implying a luminosity overestimate. This would also explain the very large distance derived for the star PG~0914+001 of $\simeq 35$ kpc, based on a low measured surface gravity. Although the gravitational darkening effect is more important in Be stars, because of their high rotation velocity, it should not be ignored in normal B stars as pointed out by \citet{wenske}. According to \citet{abt1} B stars rotate with velocities of 40 -- 50 per cent of the critical velocity on average (depending on the exact spectral type). On the other hand, the effect only becomes significant for velocities greater than $\simeq 60$ per cent of the critical velocity \citep{fremat}.

We have corrected the Be stars observed temperatures and surface gravities using the theoretical models by \citet{fremat} assuming a rotation velocity of 99 per cent of the break-up velocity. Although there is some indication that many Be stars are actually rotating slower than this (88 per cent is the most probable value according to \citealt{fremat}, and 70 per cent the lower limit according to \citealt{ekstrom1}), we know that at least in the case of PG~0914+001 the rotation must be very close to critical because of the high $v\sin i$. The case for rotation close to the critical velocity is weaker for the two other Be stars, but the same correction is applied due to homogeneity considerations. Nevertheless, the difference between a correction assuming a rotation of 99 and 80 per cent of critical velocity is minimal for these two other Be stars ($\sim 0.1\, \mathrm{dex}$ for $\log g$ and $1000\, \mathrm{K}$ for $T_{\mathrm{eff}}$).

In the case of the normal B stars, the gravitational darkening effect only becomes significant for velocities greater than 60 per cent, as was mentioned previously. Nevertheless, the model for 80 per cent of break-up velocity was preferred because some stars have a projected rotation velocity value incompatible with a rotation velocity lower than 60 per cent of the critical velocity. Both approaches produce similar results (the corrections for $\log g$ and $T_{\mathrm{eff}}$ are, on average, $0.1\, \mathrm{dex}$ and $400\, \mathrm{K}$ for the 60 per cent case, and $0.1\, \mathrm{dex}$ and $485\, \mathrm{K}$ for the 80 per cent case).

In principle we would like to correct the gravitational darkening effect for all the stars in the sample, but, since the true rotation is unknown for any given star, the correction is done in a statistical sense only. The chosen model implies a very large true rotation velocity (80 per cent of the critical rotation velocity), hence we would overestimate the correction for most stars with small $v\sin i$ values. To account for this problem we have done the correction only for stars with $v\sin i$ larger than 35 per cent of the break-up velocity. Moreover, the correction was not applied in cases where rotation velocity measurements were not available (HIP~1904, HIP~113735 and HD~138503), and when the effective temperature was much higher than the maximum temperature ($27000\, \mathrm{K}$) available in the theoretical model grid (HD~140543 and HD~149363).

\subsection{Selection results}
By applying the selection \textit{criteria}, we have classified 96 stars of the initial sample of 174 as being likely or very likely on the main sequence. The remaining 78 stars are most likely halo population, evolved stars, mostly stars on the horizontal branch judged from their position on the $T_{\mathrm{eff}}-\log g$ diagram. The 96 selected stars have been further classified according to how strong their case for being on the main sequence is. In the more convincing cases there is good evidence for a normal abundance pattern, and/or for high rotation velocity, whereas the less convincing cases either present weaker evidence for normal abundance coupled with low or non-measured rotation velocity. These two cases correspond to the verdicts of MS and MS (?) in Tables~\ref{table1}~and~\ref{table2}, respectively. All relevant data for this sample of main sequence stars is shown in Appendix~\ref{ap:2}.

The selected sample covers a range in magnitudes of $6.5<V<14.5$, which corresponds to a range in heights above the galactic plane of $0.3 - 30.5\, \mathrm{kpc}$. It is this sample of main sequence stars that is the focus of interest throughout the remainder of this paper.

\section{Orbital analysis}
\label{s:orb}
The method used to estimate the ejection velocities consists in tracing the orbits of the stars backwards in time until the first intersection with the galactic plane. The present space velocity is obtained from the radial velocity and proper motion (given the distance), which are available for every star in our sample. Given the velocity and the position of the star, we integrated the orbit in the galactic gravitational potential after inverting the velocity direction (to go back in time). This is essentially the method used to good effect by \citet{hoogerwerf}, and \citet{ramspeck}, among others. The final output is then the instant of intersection with the galactic plane -- which is equal to the time the star spent on the orbit after ejection (or flight time) -- and the velocity at that instant (the ejection velocity).
The orbits were computed using the program \textsc{orbit6} developed by \citet*{odenkirchen}. This program integrates the orbits of test particles in a modified version of the \citet*{allen1} potential, where the disc has a scale-length of $3\, \mathrm{kpc}$. The changes made to the original galactic potential are detailed in Appendix~\ref{ap:1}. The program takes as input the spatial coordinates and full space velocity (in a right-handed galactocentric frame of reference where the $X$ axis passes through the position of the Sun, pointing in the opposite direction, and $Z$ points to the north galactic pole). The steps involved in computing these values, in chronological order, were:
\begin{itemize}
\item the mass was obtained from the atmospheric parameters, by interpolating the theoretical tracks shown in Fig.~\ref{fig:1};
\item the distance was computed from the visual magnitude and the absolute magnitude, which in turn was obtained from the mass and atmospheric parameters;
\item the coordinates and velocities in the galactocentric Cartesian frame of reference were obtained from the 
measured equatorial coordinates, radial velocities and proper motions, and distances.
\end{itemize}
Each of these steps was implemented as a separate subroutine in a \textsc{fortran} program. Errors were propagated using a Monte Carlo scheme by assuming a Gaussian distribution and choosing a random point from the input parameter space on each iteration. All steps will be expanded on in later subsections.

The orbit for each star was integrated 10000 times (in the Monte Carlo scheme previously described) using a timestep of $0.2\, \mathrm{Myr}$, and the program stopped, on any given iteration, when the orbit reached the galactic plane (which corresponds to $Z=0$ in the galactocentric Cartesian reference frame) or when the flight time exceeded $250\, \mathrm{Myr}$ (to avoid the program running forever in those situations where the simulated star did not reach the plane), whichever happened first.
The velocity of the star when it reached the galactic plane, after being corrected for the rotation of the galactic disc, was taken as the ejection velocity. The time since the beginning of the integration until the crossing of the plane was taken as flight time. Instances with the star not reaching the galactic plane within $250\, \mathrm{Myr}$ were not included in the calculation of velocities and flight times.

\subsection{Masses, ages and distances}
\label{s:mass}
The masses and ages were obtained by interpolating between the theoretical evolutionary tracks of 
\citet{schaller} (Schaller tracks). We note that for this reason, the age of a star is more properly termed \textit{evolutionary age}. The tracks were converted to the $\mathrm{T}_{\mathrm{eff}}-\log g$ plane. A metallicity corresponding to $Z=0.02$ (close to solar) was assumed. As a test of the uncertainty caused by different input physics, the evolutionary ages were compared to different estimates obtained using the theoretical tracks by \citet*{bressan1} (Padova tracks), which are the same as the tracks by \citet*{girardi1} during the main sequence phase, and to the theoretical tracks by \citet*{pietrinferni1} (BaSTI tracks). This latter tracks are only available for stars with less than $10\, \mathrm{M}_{\odot}$ (note that this covers 80 per cent of the sample). In the case of the Padova tracks, the relative difference between the two determinations is typically less than 10 per cent for stars older than $50\ \mathrm{Myr}$ and less than 20 per cent for stars younger than $50\ \mathrm{Myr}$. Note that the increased discrepancy for younger stars can be partly attributed to the fact that the two models start at different evolutionary stages. In the case of the BaSTI tracks, the relative differences in evolutionary age are typically less than 10 per cent for all stars (masses lower than $10\, \mathrm{M}_{\odot}$). In both comparisons, the differences can at least partly be explained by the different treatment of the overshoot into the convective layers. Note that these differences between sets of theoretical tracks are in most cases much smaller than the errors in the determination of the evolutionary age, resultant from the uncertainty in the effective temperature and surface gravity determinations.

The distance was obtained from the distance modulus. Absolute magnitudes were computed for given $\log g$, $T_{\mathrm{eff}}$ and mass using $V$ band fluxes from \textsc{ATLAS} model atmospheres. We note here that the apparent magnitudes were corrected for interstellar reddening using either Str\"omgren $uvby\beta$ photometry (\citealt{napiwotzki} re-calibration for B stars of the \citealt{moon1} original calibration) when available, or the reddening maps of \citet*{schlegel1}. These reddening maps include the total galactic reddening in any given direction. However we are dealing with stars in the Halo only and most reddening occurs in the disc. This was verified by comparing the Str\"omgren $uvby\beta$ photometry reddening estimates with the Schlegel values, which were in good agreement.

The spectroscopic distance determination can be tested, if accurate trigonometric parallaxes exist. However, even the accuracy achievable with Hipparcos is not sufficient for the sample of runaway stars. Only seven stars have parallax errors below 50 per cent and the best case corresponds to an error $> 30$~per~cent. However, we could carry out a test of the spectroscopic method using the sample of early type stars from the Table~4 of \citet{napiwotzki}. Parameters were derived in a fashion similar to many investigations of runaway stars: $T_{\mathrm{eff}}$ was determined from Str\"omgren photometry and $\log g$ from Balmer line fitting. All these stars have accurate Hipparcos parallaxes. We found that the ratio of distances determined using our method and the trigonometric parallax determination is on average 1.03, with a standard deviation of 0.11. This dispersion is of the same order of the expected error in distance corresponding to the best determinations of $\log g$, which is $\sim 10$~per~cent for an error of 0.1~dex.

\subsection{Position and velocity}
Current positions in equatorial coordinates were taken from the Hipparcos and the UCAC 2 catalogues, using the equinox 2000.0 transformation as given by SIMBAD, and then converted to three-dimensional galactocentric Cartesian coordinates, $(X,Y,Z)$.
Velocities were computed from measured proper motions and radial velocity (relative to the local standard of rest -- LSR) and transformed to galactocentric Cartesian coordinates by accounting for the motion of the Sun and of the LSR. The adopted Sun -- Galactic Centre distance was $8.0\ \mathrm{kpc}$ and the Sun velocity in galactocentric Cartesian coordinates was (11.0,5.3,7.0) $\mathrm{km}\, \mathrm{s}^{-1}$ \citep{dehnen}.

The sources used for the proper motions were: the UCAC2 catalogue \citep{zacharias}, the USNO-B catalogue \citep{monet}, 
the Tycho-2 \citep{Hog} and Hipparcos \citep{leeuwen1} catalogues, the SuperCOSMOS science archive, the SPM catalogue \citep{girard}, and the NPM2 catalogue \citep{hanson1}.
The different proper motion estimates were combined by a weighted average following the prescription in Section~2 of \citet{pauli1}. The values obtained are shown in Appendix \ref{ap:2}.

Many proper motion measurements in the USNO-B catalogue lack estimates of the associated error, or in other cases an error of zero is given. To account for these two situations we have estimated the typical proper motion error for the stars in the sample. This was accomplished by comparing the USNO-B proper motions with the other catalogues, in particular with Tycho-2 and UCAC2, and computing the USNO-B error which minimized the weighted (with the errors) average difference between the proper motions from USNO-B and the other catalogues.

It was determined that the quality of the SuperCOSMOS proper motion measurements tended to decrease with increasing brightness. For this reason, SuperCOSMOS proper motions were considered only for stars fainter than $V=14$. This limit was determined from comparisons with all other proper motion sources.

\subsection{Error propagation}
The orbit obtained at the end is a very complicated function of the various input parameters, but still it is  
important to have an estimate of the errors in the ejection velocities and flight times. An error analysis was ignored in most previous studies with the notable exception of the studies by \citet{martin2} and \citet{ramspeck}. Since it is impossible 
to estimate these errors using the standard error propagation formula we implemented a Monte Carlo simulation where each 
input parameter is a random point taken from a Gaussian distribution which reflects the error distribution of 
the measurement. A further advantage of this procedure is that it permits a study of the correlations between 
the different variables.

\section{Results and discussion}
\label{s:results}
The final results of the analysis described in Section~\ref{s:orb}, performed on the 96 main sequence stars selected from the initial sample of 174, are the flight time (i.e. the time spent in the orbit since ejection until the present), the ejection velocity (i.e. the orbital velocity for $Z=0$, relative to the star's LSR at the moment of ejection) and spatial $(X_0,Y_0)$ coordinates of the ejection point, and their associated errors. The distributions of these output variables are highly skewed, hence the median is used as a measure of central tendency value instead of the average, and the percentiles (confidence interval bounds) corresponding to 15.9 and 84.1 per cent of the distribution (corresponding to $\pm 1\sigma$ for a Gaussian distribution) were used instead of the standard deviation as a measure of the error range.

\begin{figure}
\includegraphics[width=84mm]{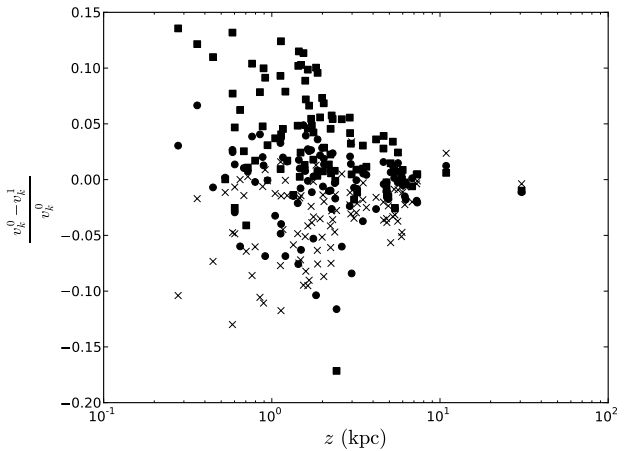}
\caption{Relative difference of the ejection velocities between the tested galactic potential models. The circles represent the short scalelength disc model, the crosses represent the long scalelength disc model, and the squares represent the heavier halo model.}
\label{f:2}
\end{figure}

\begin{figure}
\includegraphics[width=84mm]{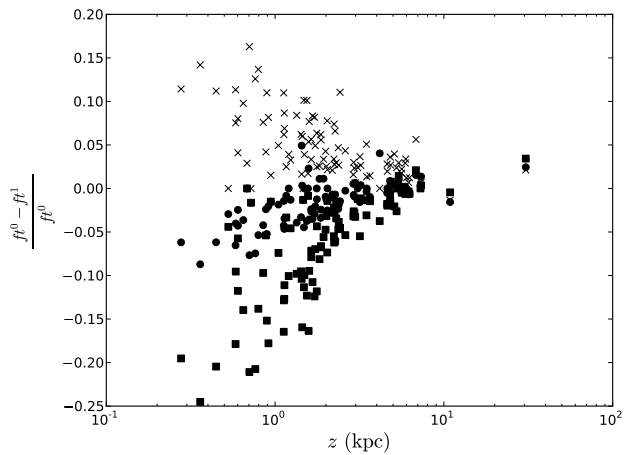}
\caption{Relative difference of the flight times between the tested galactic potential models. The symbols have the same meaning as in the ejection velocities plot.}
\label{f:3}
\end{figure}

The model of the galactic potential is an obvious source of uncertainty in the final results. A detailed analysis of the galactic potential parameter space is outside the scope of our study. However we have tested a number of alternative models as an attempt to quantify the effect on the flight times and ejection velocities, by choosing different, reasonable, parameters, starting from the base model (cf.~Appendix~\ref{ap:1}). We have tested three modified models: one with a disc scale-length shorter by 50 per cent, one with a disc scale-length longer by 50 per cent, and one with a total mass higher by 30 per cent. In all cases the disc component mass, and the halo component mass and power law index were changed to fit the galaxy's circular rotation curve to the observed value of $220\, \mathrm{km}\, \mathrm{s}^{-1}$ at the position of the Sun. The plot of the relative difference of each model to the default model, as a function of height above the galactic plane, is shown on Fig.~\ref{f:2}, for the flight times, and Fig.~\ref{f:3}, for the ejection velocities. It is noticeable that the difference between the models drops with distance to the disc. It is not surprising that the influence of the disc component becomes less significant for large distances. Note in particular that the model with a heavier halo only affects the outcome because the choice of a heavier halo forces a much lighter disc in order to keep the local circular velocity unchanged. However, in the outer regions, where the disc component loses its influence, the results are almost unchanged which suggests that the mass of the halo is not crucial for the obtained results, as long as the local orbital velocity is kept at $\simeq 220\, \mathrm{km}\, \mathrm{s}^{-1}$. The fact that the results for the stars which are further away are mostly insensitive to the choice of potential is particularly important, since these are the stars that have the highest ejection velocities. The difference is more dramatic for the flight times which reach a difference of 25 per cent. Nevertheless, most cases in both the ejection velocities and flight times are contained in the interval 10--15 per cent in relative difference.

\subsection{Flight times}
\label{s:ft}
\begin{figure}
\includegraphics[width=84mm]{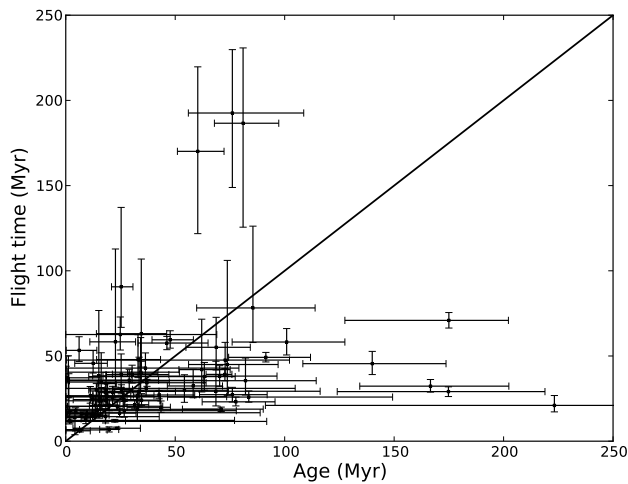}
\caption{Flight time \textit{versus} evolutionary ages including error bars.}
\label{f:4}
\end{figure}

If the disc ejection hypothesis is correct it follows that the computed flight times must be less than, or equal, to the corresponding evolutionary ages (cf. Section~\ref{s:mass}). The flight times of all stars in our sample are plotted in Fig.~\ref{f:4}. We consider them to be consistent within the errors if the flight time estimate minus its error, is equal to or lower than the evolutionary age estimate plus its error. We find that the flight times are consistent with the computed evolutionary ages to within the errors for 83 per cent of the sample and to within twice the errors for 97 per cent of the sample (corresponding to only three discrepant objects). Moreover, we note that for most stars $T_\mathrm{f}\simeq T_\mathrm{l}$, as it is expected from theory for both the BES and DES mechanisms. \citet{zwart} estimates, for the BES, that a late-type B star (mass in the interval 3--5 $\mathrm{M}_\odot$) will spend more than 75 per cent of its lifetime as a runaway star, and \citet{leonard2} estimate, for the DES, a time for ejection from a cluster of the order of 10 million years, corresponding to more than 75--80 per cent of the lifetime spent as runaway, in the case of late-type B star. The fact that we find this agreement with theory suggests that the models are adequate or conversely, if we assume the models are correct, that the method used to compute the flight times does not have a fatal flaw. In particular, we are led to believe that the assumed galactic potential (cf. Section~\ref{s:orb} and Appendix~\ref{ap:1}) is realistic enough for our purposes. This conclusion is supported by the discussion in the beginning of the section on the sensivity of the orbits to changes in the galactic potential.
\begin{figure}
\includegraphics[width=84mm]{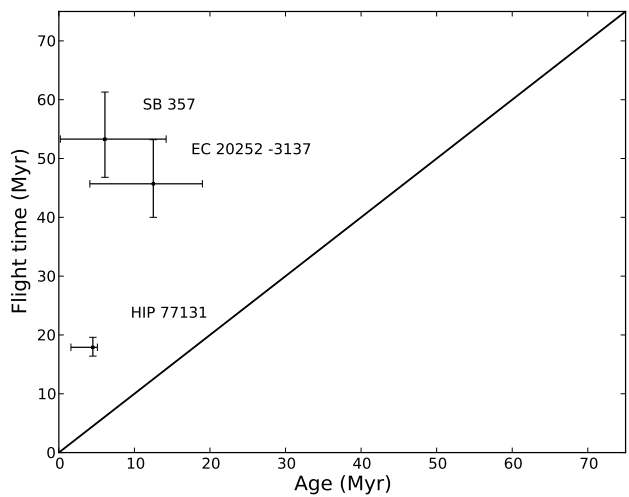}
\caption{Flight time \textit{versus} evolutionary ages including error bars. Only stars deviating more than twice the errors in both directions are plotted.}
\label{f:5}
\end{figure}
In spite of this overall good agreement of observations and theoretical expectation, we find three stars in our sample for which the computed flight time is in serious disagreement (more than twice the errors in both directions) with our estimate of their evolutionary age (see Fig.~\ref{f:5}). Taken at face value this could indicate that these stars were born in the galactic halo, a scenario proposed by \citet*{hambly3} for the star PHL~346. If confirmed, this could have very important consequences for our understanding of star formation. However, first we have to consider other possible mechanisms and effects capable of extending the lifetime of main sequence B stars. These are:

\begin{enumerate}
\item rotation induced mixing: early-type stars have generally high rotation velocities, in particular in the case of Be stars which rotate with velocities near the critical velocity. The centrifugal forces created by the fast rotation create extra mixing acting as an extra \textit{overshoot} diffusing the elements produced in the core. This effect would increase the lifetime of a B star at most by 20--25 per cent \citep{maeder1}.
\item lower than solar metallicity: as described in Section~\ref{s:mass}, ages were computed assuming a metallicity close to solar. A significant scatter of metallicities is observed for young stars in the disc. The metallicity distribution, as derived by \citet{fuhrmann1} from a sample of main sequence stars from B to G spectral types within 25 pc of the Sun, has a standard deviation of 1.4 times the solar metallicity, which is comparable to the radial scatter in the thin disc \citep{cescutti1}. This effect is relevant because stars with lower metallicities stay longer on the main sequence, and the correspondent ZAMS (zero age main sequence) is bluer, implying longer lifetimes for given effective temperatures and surface gravities, when compared with stars with higher metallicity.
\item blue straggler: the DEM may result in the ejection of a binary which could then merge after the ejection, creating a rejuvenated star analogous to the blue stragglers observed in clusters. This situation would be effectively the same as a star with a mass equal to the sum of the two stars in the progenitor binary being ejected from a position higher above the galactic plane. Moreover, since the binary stars would have a spectral type later than B (to be able to produce a B star after the merger) this would increase dramatically the lifetime of the system. However, the ejection of intact binaries is predicted to be a rare event and the ejection velocities lower than about $100\ \mathrm{km}\, \mathrm{s}^{-1}$ \citep{leonard2}. More recently \citet*{perets} has demonstrated that this mechanism could explain, in principle, the extreme youth of the these distant stars, but the problem with the small numbers and velocities still remains.
\end{enumerate}
The alternatives to these mechanisms would be a formation in the Halo scenario, or errors in the analysis of the stars' atmospheres (which would propagate to the estimated distance) and/or errors in the measurement of proper motions. The three stars under consideration, that have flight times higher than the evolutionary ages within two times the errors, are: SB~357, EC~20252--3137 and HIP~77131. Although a few 2-sigma outliers are expected (about five in a sample of 100 objects), these stars have small errors, which makes it difficult to explain the discrepancies as statistical fluke. The estimated ages for SB~357, EC~20252--3137 and HIP~77131 are $6.1_{-5.9}^{+8.1}\ \mathrm{Myr}$, $12.5_{-8.4}^{+6.5}\ \mathrm{Myr}$ and $4.5_{-2.9}^{+0.6}\ \mathrm{Myr}$, and the flight times $53.3_{-6.5}^{+8.0}\ \mathrm{Myr}$, $45.7_{-5.7}^{+7.8}\ \mathrm{Myr}$ and $17.9_{-1.5}^{+1.7}\ \mathrm{Myr}$ respectively. We have computed the ages the stars would have if we take into account the maximum effects of metallicity and rotation. The results are summarized in Table~\ref{tab:6}. A more detailed discussion of each star follows:

\begin{table}
\caption{The evolutionary ages of SB~357, EC~20252--3137 and HIP~77131 if we assume a mass fraction of metals of $Z=0.008$, and we take into account the effect of rotation. The first column gives the ages adjusted for rotation, the second column gives the ages for low metallicity, and the third column gives the age for metallicity together with effect of rotation. All ages are given in Myr. The evolutionary ages for low metallicites were estimated from evolutionary tracks by \citet{fagotto1}. We have assumed rotation increases the lifetimes by 25 per cent (cf. \citealt{maeder1}).}
\label{tab:6}
\begin{tabular}{@{}lccc}
\hline
 & & & metallicity and\\
Star & rotation & metallicity & rotation\\
\hline
SB 357 & $7.6_{-7.3}^{+10.2}$ & $18_{-12}^{+10}$ & $22.5_{-15}^{+12.5}$\\
&&&\\
EC 20252--3137 & $15.6_{-10.5}^{+8.2}$ & $22_{-11}^{+4}$ & $27.5_ {-13.8}^{+5}$\\
&&&\\
HIP 77131 & $5.6_{-3.6}^{+0.8}$ & $6_{-1.5}^{+3}$ & $7.5_{-1.9}^{+3.8}$\\
\hline
\end{tabular}
\end{table}

\paragraph*{SB 357}
This is a Be star, so its MS status appears to be well supported. However a high value for the projected rotation velocity also means that no metal abundances were determined for this star, although the Helium abundance appears to be normal \citep{ramspeck}. The ages given in Table~\ref{tab:6} are justified by the fact that this star is a fast rotator coupled with the fact that there is no evidence for Solar metallicity. However, even when both effects are combined, the difference between ages and flight times is still significant within (the equivalent to) $1.5\ \sigma$. A different possibility is that the gravitational darkening may have been overestimated in this particular case. If the correction is ignored the computed flight time is $65_{-9}^{+11.9}\ \mathrm{Myr}$, whereas the age is at most (assuming a low metallicity and adding the effect of rotation) $46_{-2.3}^{+7.8}\ \mathrm{Myr}$, implying that even in this situation the difference is still significant within the error bounds. We note, however, that since SB~357 is a Be star this is a situation where the correction is the most important. As a final test, we compared the age estimate with the one obtained using the theoretical tracks by \citet{bressan1}, but this increases the discrepancy even more since this new determination is a factor of 2 smaller than the original one.

\paragraph*{EC 20252--3137}
The classification of this star as being on the main sequence appears to be solid. It is rotating with a (projected) velocity of $60\ \mathrm{km}\, \mathrm{s}^{-1}$ and has a normal abundance pattern, although there is some evidence for slightly supersolar metallicities \citep{magee2}. Given this evidence for high metallicity the assumption of low metallicty is completely arbitrary. Moreover, the lack of evidence for a very high rotation velocity may mean that the increase in lifetime due to rotation may be much smaller than 25 per cent in this case, implying small effects from extra mixing in the stellar interior. Even if we ignore these considerations, the corrections just barely make the ages and flight times consistent within (the equivalent to) $2\, \sigma$. Since this star was not corrected for the gravitational darkening effect, given the small rotation velocity, it is possible that the difference between age and flight time has been overestimated. If the correction is applied then the flight time increases to $46.2_{-5.6}^{+8.2}\ \mathrm{Myr}$ and the age at most to $22.5_{-10}^{+10}\ \mathrm{Myr}$ (assuming a low metallicity and adding the effect of rotation). Once again, we also compared the age determination with the one obtained form the tracks by \citet{bressan1} but this determination ($11.9\ \mathrm{Myr}$) is very similar to the original one. We note again that the difference in this situation is still significant within the errors.

\paragraph*{HIP 77131}
This star was classified as being on the main sequence based on the very high (projected) velocity of $250\ \mathrm{km}\, \mathrm{s}^{-1}$. The abundances of the elements N, C, and Si were measured by \citet{martin1} and, although the abundances of nitrogen and carbon are consistent with a main sequence status, silicon appears to be overabundant. As in the case of EC~20252--3137, the assumption of low metallicity is not justified in this instance, but even if it were the flight time is still inconsistent with the evolutionary age as can be seen in Table~\ref{tab:6}. Since this star was not corrected for the gravitational darkening effect, as explained in Section~\ref{s:prv}, once again the difference between age and flight time may have been overestimated. The correction for HIP~77131 was done assuming, for the purpose of this analysis, $0.1\ \mathrm{dex}$ for the gravity and $3000\ \mathrm{K}$ for the effective temperature, based on a rough extrapolation (assuming a constant function) from the values corresponding to the highest temperature on the theoretical model \citep{fremat}. The flight time obtained after the correction was $17.9_{-1.5}^{+1.7}\ \mathrm{Myr}$, which is still inconsistent with the corrected age of $6.3_{-2}^{+4}\ \mathrm{Myr}$ (computed assuming low metallicity and adjusted for increased lifetime due to high rotation, after correction for the gravitational darkening effect), even within twice the error bounds. Again in this case we compared the original age estimate with the one obtained using the theoretical tracks by \citet{bressan1} but this determination ($4.2\ \mathrm{Myr}$) is very similar to the original one.

In summary, all three stars appear to be inconsistent with ejection from the disc even if we consider effects which would increase their lifetimes. In all cases new observations to confirm the atmospheric parameters are highly warranted, in particular in the case of HIP~77131 since it would be very important to have spectroscopic estimates. In that respect, the most intriguing case is probably the star SB~357, since its atmospheric parameters were obtained from high resolution spectra and so should be particularly reliable. However, since the composition of the star is unknown, it is possible that its metallicty is well below solar, in contrast to the other two stars (EC~20252--3137 and HIP~77131). For this reason SB~s357 is probably the strongest case yet for a star formed in the halo. Note that the blue straggler scenario is not discussed because Mass/Luminosity ratios, rotation velocities, and chemical abundances of B type blue stragglers in open clusters are indistinguishable from ``normal'' stars (\citealt*{schonberner2}, \citealt*{andrievsky1}, \citealt*{schonberner3}). However, as was mentioned before in this section, the ejection of an intact binary with a high ejection velocity should be an extremely rare occurrence. The estimated fraction of binaries among runaway stars is of $\sim 1$ per cent (\citealt{perets} and references therein), which is a factor of 3 lower than the observed fraction (3 in 96 stars).

\subsection{Ejection velocities}
\label{s:ejection}
Simulations for both ejection mechanisms (\citealt{leonard2} and \citealt{zwart}) predict a two dimensional mass-energy distribution. In particular, this distribution should show an inverse relation between mass and ejection velocity. It is known that runaway stars will usually be observed near the apex of their orbits, where the velocity is the lowest, (see \citealt{martin2}) -- explaining why the distribution of measured (heliocentric) radial velocities in our sample has an average of only $41\ \mathrm{km}\, \mathrm{s}^{-1}$. Hence, stars with high ejection velocities will typically be further away than stars with lower ejection velocities. This fact introduces a bias in the detection of the highest ejection velocity stars, i.e. the stars which constitute the tail of the ejection velocity distribution, where the two ejection mechanisms are expected to differ. 

\begin{figure}
\includegraphics[width=84mm]{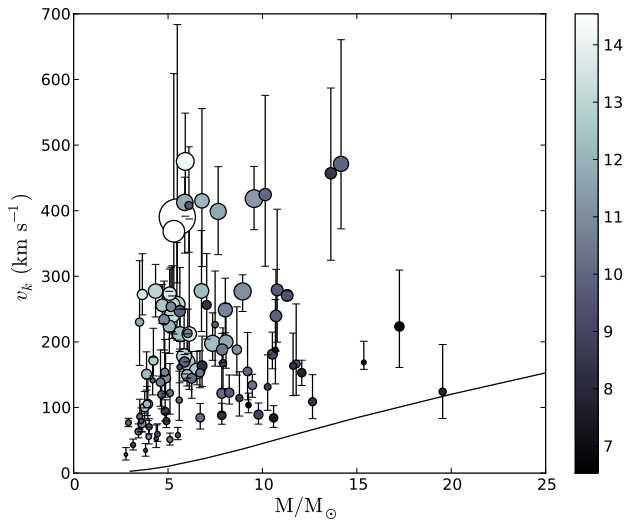}
\caption{Ejection velocity -- Mass distribution. The apparent visual magnitude is given by the grayscale, from black to white, black being the brightest and white the faintest. The size of the circles is proportional to the height above the galactic plane. The line indicates the minimum velocity needed to reach a height of $1\ \mathrm{kpc}$.}
\label{f:6}
\end{figure}

\begin{figure}
\includegraphics[width=84mm]{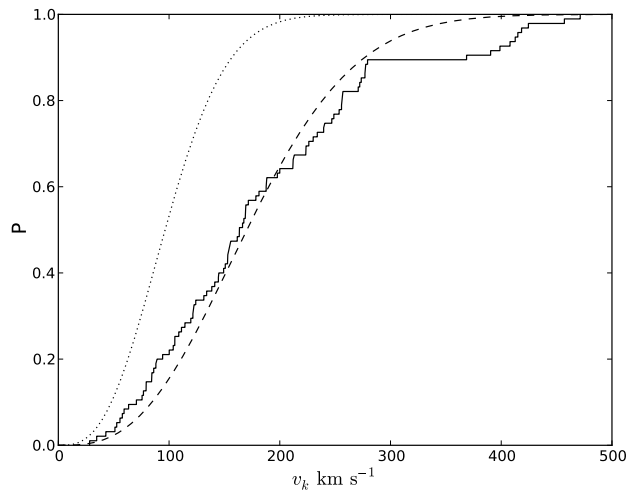}
\caption{Ejection velocity distribution. The filled line is the empirical cumulative distribution function, the dashed line is the best fit Maxwellian distribution (peaking at $156\ \mathrm{km}\, \mathrm{s}^{-1}$), and the dotted line is the the predicted Maxwellian distribution \citep{leonard4} (peaking at $100\ \mathrm{km}\, \mathrm{s}^{-1}$).}
\label{f:7}
\end{figure}

\begin{figure}
\includegraphics[width=84mm]{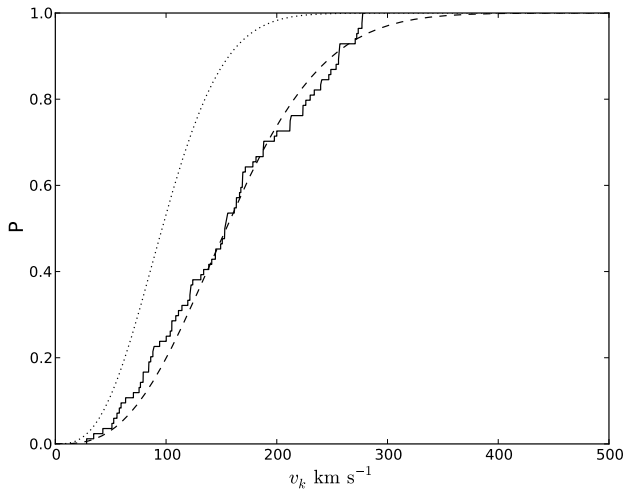}
\caption{Ejection velocity distribution for the sub-sample of stars with ejection velocities lower than $350\ \mathrm{km}\, \mathrm{s}^{-1}$. The filled line is the empirical cumulative distribution function, the dashed line is the best fit Maxwellian distribution (peaking at $141\ \mathrm{km}\, \mathrm{s}^{-1}$), and the dotted line is the the predicted Maxwellian distribution \citep{leonard4} (peaking at $100\ \mathrm{km}\, \mathrm{s}^{-1}$).}
\label{f:8}
\end{figure}

The mass-ejection velocity distribution for our sample is shown on Fig.~\ref{f:6}, with the size of the symbols being proportional to the (vertical) distance to the disc (larger symbols meaning higher distances), and the colour indicating the apparent visual magnitude. The minimum velocity needed for a star to reach a height of 1 kpc above the disc is shown as a solid line, explaining why no stars are found below said line. This is a result of the selection criteria which excluded nearby runaway stars. Nevertheless, the distribution appears to show a trend of decreasing ejection velocity for higher masses as predicted by theory (the trend is very weak and depends on the three more massive stars, including one of the problematic stars mentioned in Section \ref{s:ft}, the most massive one). This is the reason why the sample is dominated by late type B stars, which is potentially problematic because it means that the stars that we need to observe are also the fainter. We note that the surveys used as a source go as deep as $V=15$ (considering only runaway star candidates), however the faintest runaway observed has a magnitude of $V\sim 14.5$. We also note that the trend of increasing distance to the disc (and corresponding decrease in brightness) for higher ejection velocities appears to be true only up to about 300 $\mathrm{km}\, \mathrm{s}^{-1}$, with the maximum ejection velocity being about 400 -- 500 $\mathrm{km}\, \mathrm{s}^{-1}$. The fact that we find a mix of bright and relatively faint stars, in a magnitude range of $12<V<14.5$, clustered around 400 -- 500 $\mathrm{km}\, \mathrm{s}^{-1}$ reflects a real drop in ejection velocity distribution. In fact, this is the highest ejection velocity predicted by \citet{leonard2}, \citet{gvaramadze1}, and \citet{przybilla}. In Fig.~\ref{f:7} we have plotted the cumulative distribution function (CDF) of the distribution function and the best Maxwellian fit which peaks at $156\ \mathrm{km}\, \mathrm{s}^{-1}$. It can be seen that the fit is good for velocities up to $\simeq 300 \mathrm{km}\, \mathrm{s}^{-1}$, however there is a suggestion of bimodality which could indicate the existence of two different populations. If we assume that this group of high velocity runaway stars (velocities higher than $350\ \mathrm{km}\, \mathrm{s}^{-1}$, corresponding to 11 stars above this threshold and 85 below) corresponds to a different population and remove them from the distribution, then the fit to a Maxwellian distribution is much better, peaking at $141\ \mathrm{km}\, \mathrm{s}^{-1}$, as can be seen on Fig.~\ref{f:8}. Hence, the ``slow'' group appears to be consistent with the standard ejection scenarios. Moreover, since \citet{leonard4} predicts a Maxwellian distribution peaked at 50 -- 100 $\mathrm{km}\, \mathrm{s}^{-1}$ for the ejection velocity, the observed distribution seems to indicate that we could not be missing many objects in the tail of the distribution, if we remember that we are missing many low velocity objects because we have selected only stars high above the galactic plane. Note that the selection of high galactic latitudes may induce an overestimate of the number of high velocity stars, since the observed cone volume increases with distance. Although this effect may explain the strong high velocity tail observed, it does not explain the apparent bimodality. Possible selection effects and completeness of the sample will be discussed in the next section. The significance and possible identity of the apparent high velocity population will be discussed later.

\subsection{Number estimate and completeness}
An estimate of the proportion of runaways in the local population of early type stars is important because it provides a quantitative measure of the accuracy of the ejection velocity result, and an important constraint on the theoretical models. The number of early type stars in the solar neighbourhood was estimated from the Hipparcos Catalogue \citep{leeuwen1}, by applying a colour cut of $B-V<0$ (corresponding to a temperature larger than $9500\ \mathrm{K}$, see \citealt{napiwotzki}), and a parallax $>5\ \mathrm{mas}$, corresponding to a distance of 200 pc. The derived surface number density of early type stars in the solar neighbourhood was $1.1\times 10^4\, \mathrm{kpc}^{-2}$. The estimate of the local number density of runaway stars requires an estimate of the completeness of our sample.

There are a few different effects influencing the completeness of the sample: the height above the galactic plane and the lifetime on the main sequence are linked to the ejection velocity, and to the brightness of the stars. It is impossible to get a firm grasp on the influence of these selection effects without doing extensive modeling. Nevertheless a rough estimate of completeness is useful for the reasons described. The brightness is directly related with the distance, which can be decomposed in a radial component and a perpendicular component (height above the galactic plane). Hence, for a given radius we will start losing stars when we go above a certain height. If we assume that the ejection of runaway stars is an isotropic process, then the distribution of the radial component (projection on the galactic plane) of the runaway stars contained inside a cylinder with a given radius should be the same as the distribution of bright young stars in the disc, inside a circle with the same radius. For this reason, we have estimated completeness by computing the distribution of the radial component within a circle of radius (distance on the galactic plane) $r$, and comparing it with a realistic simulated isotropic distribution of stars in the disc, scaled to have the same frequency as the sample at radius $r$. We then varied $r$ until we found a good fit between both distributions on the interval 0 -- $r$. Using this method we estimated the sample to be complete up to a radius of $\simeq 1\ \mathrm{kpc}$. This means that up to a radial component of 1 kpc we are complete in the height component of the distance. Note that this is a volume limited sample, since we have selected only stars above $\simeq 280\, \mathrm{pc}$. This lower limit in height coupled with the lower limit of $20\degr$ (absolute value) in galactic latitude explains why we are complete up to a radius of 1 kpc, which translates into a height of $\simeq 360\, \mathrm{pc}$ for this galactic latitude. This is the effect mentioned in Section~\ref{s:ejection}, which increases the relative number of stars at greater distances from the galactic plane or, equivalently the number of stars with high ejection velocities.

From the volume limited complete sample of runaway stars we have estimated the local surface number density to be $8.6\, \mathrm{kpc}^{-2}$. This corresponds to 27 stars inside a cylinder with radius of 1 kpc. \citet{mdzi1} compiled a list of 61 runaway star candidates inside a cylinder of 2 kpc around the Sun using Hipparcos proper motion data. Since Hipparcos data is complete only up to $V\simeq 9$ we know this sample is highly incomplete, and will consist mostly of nearby stars, however this means that it is a good complement to our sample. If we combine the two samples we find that the local number density is $13.4\, \mathrm{kpc}^{-2}$. Note that the true value is probably smaller since the stars from the \citet{mdzi1} sample have not been confirmed as real runaway stars, nevertheless this number is a good estimate of the upper limit. It is interesting to note that our estimate is not far from the one by \citet{leonard2}, who estimates a value of $\simeq 10\, \mathrm{kpc}^{-2}$.

We have then extrapolated the local densities of runaway stars and early type main sequence stars to their total number in the galaxy, using the simulated distribution of young stars to compute the proportion of the local samples relative to the whole galaxy. We estimate the population of runaway stars to be $\simeq 0.1\ \textrm{per cent}$ of the total number of young, early type main sequence stars, which appears to be a reasonable number, if we assume that the number density decreases for later types. This is consistent with estimates of space frequencies of runaway stars: 30 per cent for O type and 4 per cent for the range B0 to B5 \citep*{stone1}.

\subsection{Comparison with other studies}
In this section we divide the literature overlapping with our work in four groups, noting that we are interested in comparing results regarding the kinematical analysis:
\begin{enumerate}
\item the papers dealing with the Palomar-Green sample, here abbreviated to PG papers \citep{rolleston2,lynn1};
\item the papers dealing with the Edinburgh-Cape sample, here abbreviated to EC papers \citep{rolleston1,magee2,lynn2};
\item the papers dealing with the Hipparcos sample \citep{allen2,conlon2,martin1,martin2};
\item the remaining papers \citep{conlon3,ramspeck,behr}.
\end{enumerate}
The studies focused on the Hipparcos sample are by themselves unsuitable to study the ejection velocity distribution, given that we are interested in the tail of the aforementioned distribution, where the difference between the BEM and DEM should be more evident. The Hipparcos brightness limit (see Table~\ref{tab:7}) is crippling if we want to have an estimate of the ejection velocity for the more distant, therefore potentially faster, objects. This means that although proper motions are used in the kinematical analysis performed by \citet{allen2} and \cite{martin2}, these studies do not give a full picture regarding the nature of the ejection velocity distribution.

\begin{table}
\caption{Limiting magnitudes and completeness estimates. The completeness estimates have a dependency on galactic latitude and spectral type.}
\label{tab:7}
\begin{tabular}{@{}lcc}
\hline
 Survey & limiting & completeness \\
 & magnitudes & range \\
\hline
Hipparcos & $V \sim 12.4$ & $7.3\lesssim V\lesssim 9$\\
&&\\
Edinburgh-Cape (EC) & $V=15$ & $11\lesssim V\lesssim 15\, ^*$\\
&&\\
Palomar-Green (PG) & $11\lesssim V\lesssim 14.5$ & $13\leq B_{\mathrm{PG}}\leq 14.6\, ^{**}$\\
\hline
\end{tabular}
*~--~for the spectral type interval B0~--~B5 \citep{lynn2}
**~--~Photographic magnitudes $B_{\mathrm{PG}}$ given for PG survey
\end{table}

On the other hand, the studies dealing with the EC and PG surveys suffer from the fact that they explore only a narrow range of brightnesses (see Table~\ref{tab:7}) and proper motion information was not used for the kinematical analysis of the studied objects. In the EC series of papers, the kinematical analysis is incomplete in the sense that proper motion data was not available for the stars EC~20011--5005 and EC~20105--2944. Similarly in the PG series of papers, the kinematical analysis lacked proper motion data for five stars (40 per cent). The other studies are not systematic, consisting of observations of only a small number of objects. The stars from the studies by \citet{behr} and \citet{conlon3} cover a range in magnitude that is barely higher than the limiting magnitude from the Hipparcos sample and they do not include proper motion data in the kinematical analysis. On the other hand, the sample studied by \citet{ramspeck} covers the range $10\lesssim V \lesssim 14$ in brightness and does include proper motion data in the kinematical analysis performed for all but four stars: PHL~159, PG~1511+467, SB~357, and HS~1914+7139. The star SB~357 is also analysed by \citet{conlon3}.

By combining these samples we were able to better constrain the ejection velocity distribution, because we cover the near objects (Hipparcos sample) and distant objects (the other samples), corresponding to a range in distance of 0.57 -- 35 kpc. The fact that our kinematical analysis includes proper motion data for all stars also constitutes an improvement over previous analysis, because it has permitted us to compute the orbits with all information necessary (position and velocity). In relation to the Hipparcos sample nothing new has emerged from our study, which is not surprising given the fact that the data from Hipparcos was already available and it has good quality. On the other hand, we have new results on stars from the other samples. All the stars mentioned earlier in this section lacking proper motion data on previous studies have now more precise estimates of the flight times and ejection velocities. The comparison of the flight times with the estimated evolutionary ages confirms that most stars are consistent with the runaway hypothesis, whereas the computed ejection velocities suggest a limit of 400 -- 500 $\mathrm{km}\, \mathrm{s}^{-1}$ for the ejection. This is consistent with the theoretical predictions of \citet{leonard2} and \citet{gvaramadze1}. The stars SB~357 and HS~1914+7139 were estimated to have flight times larger than their evolutionary ages by \citet{conlon3} and \citet{ramspeck}, in the case of SB~357, and by \citet{ramspeck} in the case of HS~1914+7139. The inclusion of proper motion data in the kinematical analysis permits us to confirm that SB~357 appears to be incompatible with ejection from the disc, whereas HS~1914+7139 now appears to have a flight time compatible with an ejection from the disc scenario. It is important to mention that the star HS~1914+7139 was also studied by \citet{heber1} who confirmed a flight time much higher than the estimated age, based on an analysis lacking proper motion data. Similarly, the star BD~--2 3766, analysed by \citet{conlon3}, who found a flight time higher than the estimated age of the star, appears to be compatible with ejection from the disc after the inclusion of proper motion data in the analysis. 

\subsection{The link with hypervelocity stars}
Hypervelocity stars are stars traveling on unbound orbits, generally believed to be ejected by the Supermassive Black Hole in the centre of the Milky Way \citep{brown3}, the so-called Hills mechanism \citep*{hills}. The apparent maximum ejection velocity in our sample is very close to the escape velocity in the solar neighbourhood ($555.3\ \mathrm{km}\, \mathrm{s}^{-1}$ for the adopted gravitational potential). This means that the group of stars with the highest ejection velocities (the high velocity group discussed on Section~\ref{s:ejection}, hereafter abbreviated to high velocity group) are near the regime of hypervelocity stars. Note that, depending on the direction of ejection, some of them could - in principle - exceed the escape velocity. It is however unclear if they constitute a low velocity tail of the hypervelocity population, or if they are simply extreme cases of runaway stars. It could also be the case that the distinction between hypervelocity and runaway stars is artificial and they all should be considered part of the same population, the former being just an extreme case of the latter.

There are at least two cases of hypervelocity stars which were shown to have been ejected far from the galactic centre. These are HD~271791 (\citealt{heber2} and \citealt{przybilla}) and HIP~60350 \citep{irrgang}, having ejection velocities of $\simeq 400\ \mathrm{km}\, \mathrm{s}^{-1}$ and $379\ \mathrm{km}\, \mathrm{s}^{-1}$, respectively. These velocities correspond to the observed limit in our study, as discussed in Section~\ref{s:ejection}, which makes the case for the existence of a real limit even stronger. Interestingly, from the stars in our sample with ejection velocities greater than $350\ \mathrm{km}\, \mathrm{s}^{-1}$, only two have points of origin in the disc compatible with, although not necessarily coinciding with, the centre of the galaxy, HIP~105912 and EC~19596--5356. A galactic centre origin can be ruled out for the other nine cases. Our results, combined with the results for HD~271791 and HIP~60350, imply that the high velocity group of runaway stars could not, in its entirety, have been ejected by the Hills mechanism.

Both the DEM and the BEM scenarios are able to produce stars with the required velocities of $\simeq 400\ \mathrm{km}\, \mathrm{s}^{-1}$. The mechanism proposed by \citet{gvaramadze1}, which is a variation of the classical DEM where the dynamical interaction is between a binary and a very massive star (mass greater than $50\ \mathrm{M}_\odot$) instead of another binary, can also produce both runaway and hypervelocity stars. In fact it is argued by \citet{gvaramadze2} that the star HD~271791 achieved its high velocity due to the DEM. However, the predicted velocity distribution is continuous in DEM models, with no existence of a natural velocity limit (with the exception of the $1400\ \mathrm{km}\, \mathrm{s}^{-1}$ limit established by \citealt{leonard5}). These predictions are at odds with the observed distribution, where both a limit velocity and a gap in the distribution are apparent. A variation of the classical BEM as described by \citet{przybilla} has been proposed to explain the ejection of both HD~271791 and HIP~60350. In this case a very close binary, composed of a massive star (mass $>40\ \mathrm{M}_{\odot}$) and a B type main sequence, undergoes a common envelope phase until the primary loses the envelope due to the deposition of the spiraling secondary's orbital energy, stopping the merger. The system will then be composed of a Wolf-Rayet star and a B type main sequence, with a very small separation between the two components, which can be as small as $13-15\ \mathrm{R}_{\odot}$. When the Wolf-Rayet ends as a supernova, the B type companion is released with a velocity close to its orbital velocity. This scenario establishes then a natural limit for the ejection velocity of around $400\ \mathrm{km}\, \mathrm{s}^{-1}$, close to the observed limit. Moreover, the small separation achieved during the common envelope phase, coupled with the fact that a Wolf-Rayet star is more compact than a main sequence star with the same mass, would explain the observed gap in the ejection velocity distribution. This follows because two different populations would correspond to different types of binaries: a binary with a very massive Wolf-Rayet star, evolving into a common envelope phase, as previously described, or the more traditional scenario (e.g. \citealt{zwart}) where the binary evolves either as a detached or semi-detached binary. Note that the shorter separation in the latter case implies higher ejection velocities.

On the other hand, it could be argued that the traditional mechanisms can only accelerate stars up to velocities of about $200\ \mathrm{km}\, \mathrm{s}^{-1}$ as has been traditionally believed (see for example \citealt{hoogerwerf} and \citealt{gvaramadze1}), what would mean that the stars with high ejection velocities in our sample were ejected by a different mechanism. This mechanism would be a third mechanism responsible for the ejection of hypervelocity stars far from the galactic centre, with the stars with high ejection velocities corresponding to the low velocity tail of the ejection velocity distribution produced by this mechanism. A candidate would be for example the mechanism proposed by \citet{gualandris1}, which consists in a dynamical encounter between a binary containing the ejected star and an intermediate-mass black hole in the centre of dense cluster. The fact that a gap in the ejection velocity distribution is observed suggests that they may correspond to two different populations, however only a better understanding of all selection effects would permit to draw a firm conclusion. This hypothesis suffers from the fact that direct evidence of the existence of intermediate-mass black holes is still lacking.

In summary, everything considered, the evidence suggests that there is no reason to separate runaway and hypervelocity stars in two separate populations. All mechanisms have to be considered together to understand the whole picture, including the BEM, the DEM, the Hills mechanism, and possibly the intermediate-mass black hole interaction mechanism. The most extreme ejection velocities are better explained by the Hills mechanism when they occur in the galactic centre, and by the BEM when they occur far from the galactic centre.

\section{Conclusions}
\label{s:conclusion}
With this work we wanted to achieve a better understanding of the mechanisms behind the runaway stars phenomenon. We aimed at producing an estimate of the ejection velocity distribution to serve as a constraint on the theoretical models. Stars far away from the galactic plane are ideal for this purpose because they have been ejected with higher velocities than most runaway stars still traveling in the disc. This is important since the models differ mostly in the high velocity of their ejection velocity distributions. We have greatly improved on previous analyses by: combining all the previously existing samples into a homogeneized one, including hitherto unavailable proper motion measurements in the kinematical analysis, doing a systematic treatment of the error propagation, and correcting for the effect of gravitational darkening. The main conclusions are:
\begin{enumerate}
\item a small sub-sample of three stars (SB~357, EC~20252--3137 and HIP~77131) cannot be directly explained by any of the ejection mechanisms, as they have flight times much larger than their evolutionary lifetimes. This is made clear by the example of the star SB~357 that was already suspected by \citet{ramspeck}, but whose case is made here more convincingly because we have included the proper motion of the star in the kinematical analysis. An independent confirmation of the atmospheric parameters is highly warranted. In contrast, some stars which appeared to be incompatible with ejection from the disc, are shown to be compatible once proper motions are included in the analysis. Note, for example, the cases of PHL~346 \citep{conlon4} and HS~1914+7139 \citep{ramspeck}.
\item we determined the ejection velocity distribution, which shows evidence for bimodality and a limiting velocity of 400 -- 500 $\mathrm{km}\, \mathrm{s}^{-1}$. The BEM, as presented by \citet{przybilla}, seems to be enough to produce the observed ejection velocity distribution. In this scenario, the limiting velocity seems to correspond to a real limit which arises naturally. The bimodality in the distribution could also be a natural consequence of the BEM. However, a contribution from the DEM was demonstrated by \citet{hoogerwerf}. Note that the observed group of high velocity stars makes the observed distribution deviate from the expected Maxwellian. More extensive modeling would be needed to assess if the mechanisms are able to produce the observed numbers.
\item the stars with the highest ejection velocities may be instead hypervelocity stars, at least in the general sense of corresponding to the lower velocity tail of a population of hypervelocity stars. Alternatively, hypervelocity stars can be seen as more extreme cases of runaway stars. There is at least some evidence for two populations given by the bimodality in the distribution. We have also determined that most of these high velocity stars are inconsistent with an origin close to the galactic centre. These results are consistent with the two hypervelocity stars which had their ejection velocities and places of origin in the galactic plane determined in other studies (\citealt{heber2} and \citealt{irrgang}). Since the BEM appears to be able to explain the features of the ejection velocity distribution, we conclude that it is probably the mechanism producing hypervelocity stars. It would be interesting to do a kinematical analysis of the remaining sample of hypervelocity stars and see if the ejection velocities and places of origin are still consistent with this scenario, however this is dependent on the delicate measurement of proper motions of very distant stars. The launch of the Gaia mission in 2012 will hopefully permit us to approach this problem.
\end{enumerate}
\section*{Acknowledgments}
M.S. gratefully acknowledges financial support by the University of Hertfordshire.
This research has made use of data obtained from the SuperCOSMOS Science Archive, prepared and hosted by the Wide Field Astronomy Unit, Institute for Astronomy, University of Edinburgh, which is funded by the UK Science and Technology Facilities Council.
This research has made use of NASA's Astrophysics Data System and of the SIMBAD database, operated at CDS, Strasbourg, France. 
We also thank Prof.~Elias~Brinks for helpful suggestions towards improving this paper.
\bibliographystyle{mn2e}
\bibliography{refs}

\begin{thebibliography}{}

\bibitem[\protect\citeauthoryear{{Abt}, {Levato} \& {Grosso}}{{Abt}
  et~al.}{2002}]{abt1}
{Abt} H.~A.,  {Levato} H.,    {Grosso} M.,  2002, \apj, 573, 359

\bibitem[\protect\citeauthoryear{{Allen} \& {Kinman}}{{Allen} \&
  {Kinman}}{2004}]{allen2}
{Allen} C.,  {Kinman} T.,  2004, in {Allen} C.,  {Scarfe} C.,  eds, RevMexAA
  (SC) Vol.~21 of RevMexAA (SC), {Young Stars far from the Galactic Plane:
  Runaways from Clusters}.
pp 121--127

\bibitem[\protect\citeauthoryear{{Allen} \& {Santillan}}{{Allen} \&
  {Santillan}}{1991}]{allen1}
{Allen} C.,  {Santillan} A.,  1991, RevMexAA, 22, 255

\bibitem[\protect\citeauthoryear{{Andrievsky}, {Sch{\"o}nberner} \&
  {Drilling}}{{Andrievsky} et~al.}{2000}]{andrievsky1}
{Andrievsky} S.~M.,  {Sch{\"o}nberner} D.,    {Drilling} J.~S.,  2000, \aap,
  356, 517

\bibitem[\protect\citeauthoryear{{Behr}}{{Behr}}{2003a}]{behr2}
{Behr} B.~B.,  2003a, \apjs, 149, 67

\bibitem[\protect\citeauthoryear{{Behr}}{{Behr}}{2003b}]{behr}
{Behr} B.~B.,  2003b, \apjs, 149, 101

\bibitem[\protect\citeauthoryear{{Behr}}{{Behr}}{2005}]{behr3}
{Behr} B.~B.,  2005, in {Barnes} III T.~G.,  {Bash} F.~N.,  eds, Cosmic
  Abundances as Records of Stellar Evolution and Nucleosynthesis Vol.~336 of
  ASP Conf. Ser., {Chemical Abundances Along the Horizontal Branch}.
pp 131--+

\bibitem[\protect\citeauthoryear{{Binney} \& {Tremaine}}{{Binney} \&
  {Tremaine}}{2008}]{binney}
{Binney} J.,  {Tremaine} S.,  2008, {Galactic Dynamics: Second Edition}.
Princeton University Press

\bibitem[\protect\citeauthoryear{{Blaauw}}{{Blaauw}}{1961}]{blaauw}
{Blaauw} A.,  1961, \bain, 15, 265

\bibitem[\protect\citeauthoryear{{Bressan}, {Fagotto}, {Bertelli} \&
  {Chiosi}}{{Bressan} et~al.}{1993}]{bressan1}
{Bressan} A.,  {Fagotto} F.,  {Bertelli} G.,    {Chiosi} C.,  1993, \aaps, 100,
  647

\bibitem[\protect\citeauthoryear{{Brown}, {Geller}, {Kenyon}, {Kurtz} \&
  {Bromley}}{{Brown} et~al.}{2007a}]{brown2}
{Brown} W.~R.,  {Geller} M.~J.,  {Kenyon} S.~J.,  {Kurtz} M.~J.,    {Bromley}
  B.~C.,  2007a, \apj, 660, 311

\bibitem[\protect\citeauthoryear{{Brown}, {Geller}, {Kenyon}, {Kurtz} \&
  {Bromley}}{{Brown} et~al.}{2007b}]{brown3}
{Brown} W.~R.,  {Geller} M.~J.,  {Kenyon} S.~J.,  {Kurtz} M.~J.,    {Bromley}
  B.~C.,  2007b, \apj, 671, 1708

\bibitem[\protect\citeauthoryear{{Cescutti}, {Matteucci}, {Fran{\c c}ois} \&
  {Chiappini}}{{Cescutti} et~al.}{2007}]{cescutti1}
{Cescutti} G.,  {Matteucci} F.,  {Fran{\c c}ois} P.,    {Chiappini} C.,  2007,
  \aap, 462, 943

\bibitem[\protect\citeauthoryear{{Conlon}}{{Conlon}}{1993}]{conlon4}
{Conlon} E.~S.,  1993, in {D.~D.~Sasselov} ed., Luminous High-Latitude Stars
  Vol.~45 of ASP Conf. Ser., {The Nature and Origin of Some High Galactic
  Latitude Hot Stars}.
pp 33--+

\bibitem[\protect\citeauthoryear{{Conlon}, {Brown}, {Dufton} \&
  {Keenan}}{{Conlon} et~al.}{1988}]{conlon1}
{Conlon} E.~S.,  {Brown} P.~J.~F.,  {Dufton} P.~L.,    {Keenan} F.~P.,  1988,
  \aap, 200, 168

\bibitem[\protect\citeauthoryear{{Conlon}, {Brown}, {Dufton} \&
  {Keenan}}{{Conlon} et~al.}{1989}]{conlon5}
{Conlon} E.~S.,  {Brown} P.~J.~F.,  {Dufton} P.~L.,    {Keenan} F.~P.,  1989,
  \aap, 224, 65

\bibitem[\protect\citeauthoryear{{Conlon}, {Dufton}, {Keenan} \&
  {Leonard}}{{Conlon} et~al.}{1990}]{conlon2}
{Conlon} E.~S.,  {Dufton} P.~L.,  {Keenan} F.~P.,    {Leonard} P.~J.~T.,  1990,
  \aap, 236, 357

\bibitem[\protect\citeauthoryear{{Conlon}, {Dufton}, {Keenan}, {McCausland} \&
  {Holmgren}}{{Conlon} et~al.}{1992}]{conlon3}
{Conlon} E.~S.,  {Dufton} P.~L.,  {Keenan} F.~P.,  {McCausland} R.~J.~H.,
  {Holmgren} D.,  1992, \apj, 400, 273

\bibitem[\protect\citeauthoryear{{Dehnen} \& {Binney}}{{Dehnen} \&
  {Binney}}{1998}]{dehnen}
{Dehnen} W.,  {Binney} J.~J.,  1998, \mnras, 298, 387

\bibitem[\protect\citeauthoryear{{Dorman}, {Rood} \& {O'Connell}}{{Dorman}
  et~al.}{1993}]{dorman}
{Dorman} B.,  {Rood} R.~T.,    {O'Connell} R.~W.,  1993, \apj, 419, 596

\bibitem[\protect\citeauthoryear{{Drimmel} \& {Spergel}}{{Drimmel} \&
  {Spergel}}{2001}]{drimmel1}
{Drimmel} R.,  {Spergel} D.~N.,  2001, \apj, 556, 181

\bibitem[\protect\citeauthoryear{{Ekstr{\"o}m}, {Meynet}, {Maeder} \&
  {Barblan}}{{Ekstr{\"o}m} et~al.}{2008}]{ekstrom1}
{Ekstr{\"o}m} S.,  {Meynet} G.,  {Maeder} A.,    {Barblan} F.,  2008, \aap,
  478, 467

\bibitem[\protect\citeauthoryear{{Evans}}{{Evans}}{1967}]{evans}
{Evans} D.~S.,  1967, in {A.~H.~Batten \& J.~F.~Heard} ed., Determination of
  Radial Velocities and their Applications Vol.~30 of IAU Symposium, {The
  Revision of the General Catalogue of Radial Velocities}.
pp 57--+

\bibitem[\protect\citeauthoryear{{Fagotto}, {Bressan}, {Bertelli} \&
  {Chiosi}}{{Fagotto} et~al.}{1994}]{fagotto1}
{Fagotto} F.,  {Bressan} A.,  {Bertelli} G.,    {Chiosi} C.,  1994, \aaps, 105,
  29

\bibitem[\protect\citeauthoryear{{Fr{\'e}mat}, {Zorec}, {Hubert} \&
  {Floquet}}{{Fr{\'e}mat} et~al.}{2005}]{fremat}
{Fr{\'e}mat} Y.,  {Zorec} J.,  {Hubert} A.-M.,    {Floquet} M.,  2005, \aap,
  440, 305

\bibitem[\protect\citeauthoryear{{Fuhrmann}}{{Fuhrmann}}{2004}]{fuhrmann1}
{Fuhrmann} K.,  2004, Astronomische Nachrichten, 325, 3

\bibitem[\protect\citeauthoryear{{Girard}, {Dinescu}, {van Altena}, {Platais},
  {Monet} \& {L{\'o}pez}}{{Girard} et~al.}{2004}]{girard}
{Girard} T.~M.,  {Dinescu} D.~I.,  {van Altena} W.~F.,  {Platais} I.,  {Monet}
  D.~G.,    {L{\'o}pez} C.~E.,  2004, \aj, 127, 3060

\bibitem[\protect\citeauthoryear{{Girardi}, {Bressan}, {Bertelli} \&
  {Chiosi}}{{Girardi} et~al.}{2000}]{girardi1}
{Girardi} L.,  {Bressan} A.,  {Bertelli} G.,    {Chiosi} C.,  2000, VizieR
  Online Data Catalog, 414, 10371

\bibitem[\protect\citeauthoryear{{Green}, {Schmidt} \& {Liebert}}{{Green}
  et~al.}{1986}]{green1}
{Green} R.~F.,  {Schmidt} M.,    {Liebert} J.,  1986, \apjs, 61, 305

\bibitem[\protect\citeauthoryear{{Greenstein} \& {Sargent}}{{Greenstein} \&
  {Sargent}}{1974}]{greenstein}
{Greenstein} J.~L.,  {Sargent} A.~I.,  1974, \apjs, 28, 157

\bibitem[\protect\citeauthoryear{{Gualandris} \& {Portegies
  Zwart}}{{Gualandris} \& {Portegies Zwart}}{2007}]{gualandris1}
{Gualandris} A.,  {Portegies Zwart} S.,  2007, \mnras, 376, L29

\bibitem[\protect\citeauthoryear{{Gvaramadze}}{{Gvaramadze}}{2009}]{gvaramadze%
2}
{Gvaramadze} V.~V.,  2009, \mnras, 395, L85

\bibitem[\protect\citeauthoryear{{Gvaramadze}, {Gualandris} \& {Portegies
  Zwart}}{{Gvaramadze} et~al.}{2009}]{gvaramadze1}
{Gvaramadze} V.~V.,  {Gualandris} A.,    {Portegies Zwart} S.,  2009, \mnras,
  396, 570

\bibitem[\protect\citeauthoryear{{Hambly}, {Rolleston}, {Keenan}, {Dufton} \&
  {Saffer}}{{Hambly} et~al.}{1997}]{hambly1}
{Hambly} N.~C.,  {Rolleston} W.~R.~J.,  {Keenan} F.~P.,  {Dufton} P.~L.,
  {Saffer} R.~A.,  1997, \apjs, 111, 419

\bibitem[\protect\citeauthoryear{{Hambly}, {Wood}, {Keenan} et~al.,}{{Hambly}
  et~al.}{1996}]{hambly3}
{Hambly} N.~C.,  {Wood} K.~D.,  {Keenan} F.~P.,    et~al., 1996, \aap, 306, 119

\bibitem[\protect\citeauthoryear{{Hanson}, {Klemola}, {Jones} \&
  {Monet}}{{Hanson} et~al.}{2003}]{hanson1}
{Hanson} R.~B.,  {Klemola} A.~R.,  {Jones} B.~F.,    {Monet} D.~G.,  2003,
  VizieR Online Data Catalog, 1283, 0

\bibitem[\protect\citeauthoryear{{Hauck} \& {Mermilliod}}{{Hauck} \&
  {Mermilliod}}{1998}]{hauck}
{Hauck} B.,  {Mermilliod} M.,  1998, \aaps, 129, 431

\bibitem[\protect\citeauthoryear{{Heber}, {Edelmann}, {Napiwotzki}, {Altmann}
  \& {Scholz}}{{Heber} et~al.}{2008}]{heber2}
{Heber} U.,  {Edelmann} H.,  {Napiwotzki} R.,  {Altmann} M.,    {Scholz} R.-D.,
   2008, \aap, 483, L21

\bibitem[\protect\citeauthoryear{{Heber}, {Moehler} \& {Groote}}{{Heber}
  et~al.}{1995}]{heber1}
{Heber} U.,  {Moehler} S.,    {Groote} D.,  1995, \aap, 303, L33+

\bibitem[\protect\citeauthoryear{{Hills}}{{Hills}}{1988}]{hills}
{Hills} J.~G.,  1988, \nat, 331, 687

\bibitem[\protect\citeauthoryear{{H{\o}g}, {Fabricius}, {Makarov}
  et~al.,}{{H{\o}g} et~al.}{2000}]{Hog}
{H{\o}g} E.,  {Fabricius} C.,  {Makarov} V.~V.,    et~al., 2000, \aap, 355, L27

\bibitem[\protect\citeauthoryear{{Holmberg} \& {Flynn}}{{Holmberg} \&
  {Flynn}}{2000}]{holmberg1}
{Holmberg} J.,  {Flynn} C.,  2000, \mnras, 313, 209

\bibitem[\protect\citeauthoryear{{Holmberg} \& {Flynn}}{{Holmberg} \&
  {Flynn}}{2004}]{holmberg2}
{Holmberg} J.,  {Flynn} C.,  2004, \mnras, 352, 440

\bibitem[\protect\citeauthoryear{{Hoogerwerf}, {de Bruijne} \& {de
  Zeeuw}}{{Hoogerwerf} et~al.}{2001}]{hoogerwerf}
{Hoogerwerf} R.,  {de Bruijne} J.~H.~J.,    {de Zeeuw} P.~T.,  2001, \aap, 365,
  49

\bibitem[\protect\citeauthoryear{{Irrgang}, {Przybilla}, {Heber}, {Fernanda
  Nieva} \& {Schuh}}{{Irrgang} et~al.}{2010}]{irrgang}
{Irrgang} A.,  {Przybilla} N.,  {Heber} U.,  {Fernanda Nieva} M.,    {Schuh}
  S.,  2010, \apj, 711, 138

\bibitem[\protect\citeauthoryear{{Keenan}, {Brown} \& {Lennon}}{{Keenan}
  et~al.}{1986}]{keenan3}
{Keenan} F.~P.,  {Brown} P.~J.~F.,    {Lennon} D.~J.,  1986, \aap, 155, 333

\bibitem[\protect\citeauthoryear{{Keenan} \& {Dufton}}{{Keenan} \&
  {Dufton}}{1983}]{keenan2}
{Keenan} F.~P.,  {Dufton} P.~L.,  1983, \mnras, 205, 435

\bibitem[\protect\citeauthoryear{{Keenan}, {Dufton} \& {McKeith}}{{Keenan}
  et~al.}{1982}]{keenan1}
{Keenan} F.~P.,  {Dufton} P.~L.,    {McKeith} C.~D.,  1982, \mnras, 200, 673

\bibitem[\protect\citeauthoryear{{Keenan}, {Lennon}, {Brown} \&
  {Dufton}}{{Keenan} et~al.}{1986}]{keenan4}
{Keenan} F.~P.,  {Lennon} D.~J.,  {Brown} P.~J.~F.,    {Dufton} P.~L.,  1986,
  \apj, 307, 694

\bibitem[\protect\citeauthoryear{{Kilian}}{{Kilian}}{1992}]{kilian1}
{Kilian} J.,  1992, \aap, 262, 171

\bibitem[\protect\citeauthoryear{{Kilian}}{{Kilian}}{1994}]{kilian2}
{Kilian} J.,  1994, \aap, 282, 867

\bibitem[\protect\citeauthoryear{{Kilkenny}}{{Kilkenny}}{1992}]{kilkenny}
{Kilkenny} D.,  1992, in {B.~Warner} ed., Variable Stars and Galaxies, in honor
  of M. W. Feast on his retirement Vol.~30 of Astronomical Society of the
  Pacific Conference Series, {High Latitude Blue Stars}.
pp 97--+

\bibitem[\protect\citeauthoryear{{Leonard}}{{Leonard}}{1991}]{leonard5}
{Leonard} P.~J.~T.,  1991, \aj, 101, 562

\bibitem[\protect\citeauthoryear{{Leonard}}{{Leonard}}{1993}]{leonard4}
{Leonard} P.~J.~T.,  1993, in {Sasselov} D.~D.,  ed., Luminous High-Latitude
  Stars Vol.~45 of ASP Conf. Ser., {Mechanisms for Ejecting Stars from the
  Galactic Plane}.
pp 360--+

\bibitem[\protect\citeauthoryear{{Leonard} \& {Dewey}}{{Leonard} \&
  {Dewey}}{1993}]{leonard3}
{Leonard} P.~J.~T.,  {Dewey} R.~J.,  1993, in {Sasselov} D.~D.,  ed., Luminous
  High-Latitude Stars Vol.~45 of ASP Conf. Ser., {Montecarlo Simulations of the
  Supernova Ejection Mechanism for the Runaway Stars}.
pp 239--+

\bibitem[\protect\citeauthoryear{{Leonard} \& {Duncan}}{{Leonard} \&
  {Duncan}}{1990}]{leonard2}
{Leonard} P.~J.~T.,  {Duncan} M.~J.,  1990, \aj, 99, 608

\bibitem[\protect\citeauthoryear{{Levenhagen} \& {Leister}}{{Levenhagen} \&
  {Leister}}{2006}]{levenhagen}
{Levenhagen} R.~S.,  {Leister} N.~V.,  2006, \mnras, 371, 252

\bibitem[\protect\citeauthoryear{{Lynn}, {Keenan}, {Dufton} et~al.,}{{Lynn}
  et~al.}{2004}]{lynn2}
{Lynn} B.~B.,  {Keenan} F.~P.,  {Dufton} P.~L.,    et~al., 2004, \mnras, 353,
  633

\bibitem[\protect\citeauthoryear{{Lynn}, {Keenan}, {Dufton}, {Saffer},
  {Rolleston} \& {Smoker}}{{Lynn} et~al.}{2004}]{lynn1}
{Lynn} B.~B.,  {Keenan} F.~P.,  {Dufton} P.~L.,  {Saffer} R.~A.,  {Rolleston}
  W.~R.~J.,    {Smoker} J.~V.,  2004, \mnras, 349, 821

\bibitem[\protect\citeauthoryear{{Maeder} \& {Meynet}}{{Maeder} \&
  {Meynet}}{2000}]{maeder1}
{Maeder} A.,  {Meynet} G.,  2000, \araa, 38, 143

\bibitem[\protect\citeauthoryear{{Magee}, {Dufton}, {Keenan}, {Rolleston},
  {Kilkenny}, {O'Donoghue}, {Koen} \& {Stobie}}{{Magee} et~al.}{2001}]{magee2}
{Magee} H.~R.~M.,  {Dufton} P.~L.,  {Keenan} F.~P.,  {Rolleston} W.~R.~J.,
  {Kilkenny} D.,  {O'Donoghue} D.,  {Koen} C.,    {Stobie} R.~S.,  2001,
  \mnras, 324, 747

\bibitem[\protect\citeauthoryear{{Martin}}{{Martin}}{2003}]{martin3}
{Martin} J.~C.,  2003, \pasp, 115, 49

\bibitem[\protect\citeauthoryear{{Martin}}{{Martin}}{2004}]{martin1}
{Martin} J.~C.,  2004, \aj, 128, 2474

\bibitem[\protect\citeauthoryear{{Martin}}{{Martin}}{2006}]{martin2}
{Martin} J.~C.,  2006, \aj, 131, 3047

\bibitem[\protect\citeauthoryear{{Mdzinarishvili} \&
  {Chargeishvili}}{{Mdzinarishvili} \& {Chargeishvili}}{2005}]{mdzi1}
{Mdzinarishvili} T.~G.,  {Chargeishvili} K.~B.,  2005, \aap, 431, L1

\bibitem[\protect\citeauthoryear{{Monet}, {Levine}, {Canzian} et~al.,}{{Monet}
  et~al.}{2003}]{monet}
{Monet} D.~G.,  {Levine} S.~E.,  {Canzian} B.,    et~al., 2003, \aj, 125, 984

\bibitem[\protect\citeauthoryear{{Moon} \& {Dworetsky}}{{Moon} \&
  {Dworetsky}}{1985}]{moon1}
{Moon} T.~T.,  {Dworetsky} M.~M.,  1985, \mnras, 217, 305

\bibitem[\protect\citeauthoryear{{Mooney}, {Rolleston}, {Keenan}, {Pinfield},
  {Pollacco}, {Dufton} \& {Katsiyannis}}{{Mooney} et~al.}{2000}]{mooney}
{Mooney} C.~J.,  {Rolleston} W.~R.~J.,  {Keenan} F.~P.,  {Pinfield} D.~J.,
  {Pollacco} D.~L.,  {Dufton} P.~L.,    {Katsiyannis} A.~C.,  2000, \aap, 357,
  553

\bibitem[\protect\citeauthoryear{{Napiwotzki}, {Sch{\"o}nberner} \&
  {Wenske}}{{Napiwotzki} et~al.}{1993}]{napiwotzki}
{Napiwotzki} R.,  {Sch{\"o}nberner} D.,    {Wenske} V.,  1993, \aap, 268, 653

\bibitem[\protect\citeauthoryear{{Odenkirchen} \& {Brosche}}{{Odenkirchen} \&
  {Brosche}}{1992}]{odenkirchen}
{Odenkirchen} M.,  {Brosche} P.,  1992, Astronomische Nachrichten, 313, 69

\bibitem[\protect\citeauthoryear{{Pauli}, {Napiwotzki}, {Heber}, {Altmann} \&
  {Odenkirchen}}{{Pauli} et~al.}{2006}]{pauli1}
{Pauli} E.-M.,  {Napiwotzki} R.,  {Heber} U.,  {Altmann} M.,    {Odenkirchen}
  M.,  2006, \aap, 447, 173

\bibitem[\protect\citeauthoryear{{Perets}}{{Perets}}{2009}]{perets}
{Perets} H.~B.,  2009, \apj, 698, 1330

\bibitem[\protect\citeauthoryear{{Pietrinferni}, {Cassisi}, {Salaris} \&
  {Castelli}}{{Pietrinferni} et~al.}{2004}]{pietrinferni1}
{Pietrinferni} A.,  {Cassisi} S.,  {Salaris} M.,    {Castelli} F.,  2004, \apj,
  612, 168

\bibitem[\protect\citeauthoryear{{Portegies Zwart}}{{Portegies
  Zwart}}{2000}]{zwart}
{Portegies Zwart} S.~F.,  2000, \apj, 544, 437

\bibitem[\protect\citeauthoryear{{Poveda}, {Ruiz} \& {Allen}}{{Poveda}
  et~al.}{1967}]{poveda}
{Poveda} A.,  {Ruiz} J.,    {Allen} C.,  1967, Boletin de los Observatorios
  Tonantzintla y Tacubaya, 4, 86

\bibitem[\protect\citeauthoryear{{Przybilla}, {Nieva}, {Heber} \&
  {Butler}}{{Przybilla} et~al.}{2008}]{przybilla}
{Przybilla} N.,  {Nieva} M.~F.,  {Heber} U.,    {Butler} K.,  2008, \apjl, 684,
  L103

\bibitem[\protect\citeauthoryear{{Ramspeck}, {Heber} \& {Moehler}}{{Ramspeck}
  et~al.}{2001}]{ramspeck}
{Ramspeck} M.,  {Heber} U.,    {Moehler} S.,  2001, \aap, 378, 907

\bibitem[\protect\citeauthoryear{{Robin}, {Reyl{\'e}}, {Derri{\`e}re} \&
  {Picaud}}{{Robin} et~al.}{2003}]{robin1}
{Robin} A.~C.,  {Reyl{\'e}} C.,  {Derri{\`e}re} S.,    {Picaud} S.,  2003,
  \aap, 409, 523

\bibitem[\protect\citeauthoryear{{Rolleston}, {Hambly}, {Dufton}
  et~al.,}{{Rolleston} et~al.}{1997}]{rolleston1}
{Rolleston} W.~R.~J.,  {Hambly} N.~C.,  {Dufton} P.~L.,    et~al., 1997,
  \mnras, 290, 422

\bibitem[\protect\citeauthoryear{{Rolleston}, {Hambly}, {Keenan}, {Dufton} \&
  {Saffer}}{{Rolleston} et~al.}{1999}]{rolleston2}
{Rolleston} W.~R.~J.,  {Hambly} N.~C.,  {Keenan} F.~P.,  {Dufton} P.~L.,
  {Saffer} R.~A.,  1999, \aap, 347, 69

\bibitem[\protect\citeauthoryear{{Saffer}, {Keenan}, {Hambly}, {Dufton} \&
  {Liebert}}{{Saffer} et~al.}{1997}]{saffer}
{Saffer} R.~A.,  {Keenan} F.~P.,  {Hambly} N.~C.,  {Dufton} P.~L.,    {Liebert}
  J.,  1997, \apj, 491, 172

\bibitem[\protect\citeauthoryear{{Sale}, {Drew}, {Knigge} et~al.,}{{Sale}
  et~al.}{2010}]{sale1}
{Sale} S.~E.,  {Drew} J.~E.,  {Knigge} C.,    et~al., 2010, \mnras, 402, 713

\bibitem[\protect\citeauthoryear{{Schaller}, {Schaerer}, {Meynet} \&
  {Maeder}}{{Schaller} et~al.}{1992}]{schaller}
{Schaller} G.,  {Schaerer} D.,  {Meynet} G.,    {Maeder} A.,  1992, \aaps, 96,
  269

\bibitem[\protect\citeauthoryear{{Schlegel}, {Finkbeiner} \&
  {Davis}}{{Schlegel} et~al.}{1998}]{schlegel1}
{Schlegel} D.~J.,  {Finkbeiner} D.~P.,    {Davis} M.,  1998, \apj, 500, 525

\bibitem[\protect\citeauthoryear{{Sch{\"o}nberner}}{{Sch{\"o}nberner}}{1979}]{%
schoenberner}
{Sch{\"o}nberner} D.,  1979, \aap, 79, 108

\bibitem[\protect\citeauthoryear{{Sch{\"o}nberner}, {Andrievsky} \&
  {Drilling}}{{Sch{\"o}nberner} et~al.}{2001}]{schonberner3}
{Sch{\"o}nberner} D.,  {Andrievsky} S.~M.,    {Drilling} J.~S.,  2001, \aap,
  366, 490

\bibitem[\protect\citeauthoryear{{Sch{\"o}nberner} \&
  {Napiwotzki}}{{Sch{\"o}nberner} \& {Napiwotzki}}{1994}]{schonberner2}
{Sch{\"o}nberner} D.,  {Napiwotzki} R.,  1994, \aap, 282, 106

\bibitem[\protect\citeauthoryear{{Stobie}, {Kilkenny}, {O'Donoghue}
  et~al.,}{{Stobie} et~al.}{1997}]{stobie1}
{Stobie} R.~S.,  {Kilkenny} D.,  {O'Donoghue} D.,    et~al., 1997, \mnras, 287,
  848

\bibitem[\protect\citeauthoryear{{Stone}}{{Stone}}{1991}]{stone1}
{Stone} R.~C.,  1991, \aj, 102, 333

\bibitem[\protect\citeauthoryear{{Tobin}}{{Tobin}}{1987}]{tobin1}
{Tobin} W.,  1987, in {Philip} A.~G.~D.,  {Hayes} D.~S.,   {Liebert} J.~W.,
  eds, IAU Colloq. 95: Second Conference on Faint Blue Stars {The faint,
  apparently-normal, high-latitude B stars - What might they be, and what
  observations are needed?}.
pp 149--158

\bibitem[\protect\citeauthoryear{{van Leeuwen}}{{van Leeuwen}}{2007}]{leeuwen1}
{van Leeuwen} F.,  2007, \aap, 474, 653

\bibitem[\protect\citeauthoryear{{Wenske} \& {Sch{\"o}nberner}}{{Wenske} \&
  {Sch{\"o}nberner}}{1993}]{wenske}
{Wenske} V.,  {Sch{\"o}nberner} D.,  1993, in {Weiss} W.~W.,  {Baglin} A.,
  eds, IAU Colloq. 137: Inside the Stars Vol.~40 of ASP Conf. Ser., {Influence
  of rotation on the stellar parameters}.
pp 162--+

\bibitem[\protect\citeauthoryear{{Xue}, {Rix}, {Zhao} et~al.,}{{Xue}
  et~al.}{2008}]{xue1}
{Xue} X.~X.,  {Rix} H.~W.,  {Zhao} G.,    et~al., 2008, \apj, 684, 1143

\bibitem[\protect\citeauthoryear{{Zacharias}, {Urban}, {Zacharias}, {Wycoff},
  {Hall}, {Monet} \& {Rafferty}}{{Zacharias} et~al.}{2004}]{zacharias}
{Zacharias} N.,  {Urban} S.~E.,  {Zacharias} M.~I.,  {Wycoff} G.~L.,  {Hall}
  D.~M.,  {Monet} D.~G.,    {Rafferty} T.~J.,  2004, \aj, 127, 3043

\end{thebibliography}

\appendix

\section{The galactic potential}
\label{ap:1}
The dynamical mass model created by \citet{allen1}, hereafter called the default model, has been used as a basis for several dynamical studies. It is constructed as a sum of three independent components corresponding roughly to the bulge, disc, and halo populations, with the dark mass contribution included in the Halo component. The model successfully reproduces the galaxy rotation curve and its main advantage is the fact that the functional form of the gravitational potential for the three components is quite simple, being also fully analytical, continuous, and with continuous derivatives everywhere. However the fact that it was created in 1991, in the pre-Hipparcos age, means it has some problems which we have addressed by updating it so it agrees with more recent results on the observable quantities. The problems we identified were: the disc component scale-length was too high, the local density was too high, as was the integrated surface density.

The disc scalelength used on the default model has the the value of $a_2=5.3178$. this value is a factor of 2 higher than recent estimates obtained from analysis of tracers of young populations. \citet{drimmel1} obtained a value of $a_2=2.25$ from infrared observations tracing the galactic dust in the disc, whereas \citet{sale1} obtained a value of $a_2=3.0$ by counting A-type stars in the outer disc. Changing the scale-length to the appropriate value in the default model produces ans excess of mass in the centre of the galaxy.

As a means of reducing the mass concentrated near the centre of the galaxy we introduced a second disc component, with negative mass, analogous to the ``hole'' component used by \citet{robin1} for their density model. In this case this component can also be thought of as a ``hole'' in the original disc, at least in the sense that the disc density is greatly reduced in the central regions. This new component has a functional form equal to the ``real'' disc component, hereafter first disc component, only with a negative sign, which ensures that we keep the simplicity and good mathematical properties of the default model. We designate this new model with four components by ``updated model'', and the new disc component by ``second disc component''.

To distinguish the parameters of the first and second disc components a superscript with the number 1 or 2, respectively, is used. The parameters used to constrain the model are summarized in Table~\ref{tbl:a3}. The procedure adopted to construct the updated model consisted of the following steps:

\begin{itemize}
	\item the first disc component scalelength, $a_2^1$ was set to $3\ \mathrm{kpc}$, for the reason explained before.
	\item the second disc component was established first, by choosing suitable parameters to satisfy two conditions: reduce the disc mass accumulated in the centre by as much as possible, while keeping the rotation curve as flat as possible in the region between $5 \textrm{ -- }20\ \mathrm{kpc}$. The vertical scale height was kept fixed and equal to the one in the first disc component. The parameter $M_2^2$ was chosen to be 10 per cent the parameter $M_2^1$.
	\item keeping the second disc and bulge components fixed, a first attempt at fitting a flat rotation curve over the interval $5 \textrm{ -- }20\ \mathrm{kpc}$, and slowly decreasing until $100\ \mathrm{kpc}$, by varying the first disc mass (parameter $M_2^1$), and the halo component mass and power law exponent (parameters $M_3$ and $\gamma$, respectively). The estimate obtained for the parameter $M_3$ resulting from this fit was taken as definitive.
	\item finally, estimates of the $\gamma$ and $M_2^1$ (and so also of $M_2^2$) parameters were obtained by simultaneously fitting a flat curve to the interval $5 \textrm{ -- }20\ \mathrm{kpc}$, imposing the observed local circular velocity, and the observed local density.
\end{itemize} 

\begin{table}
\caption{Physical quantities that were used to constraint the parameters of the models, compared with the measured values. The local density value is from \citet{holmberg1} and the circular velocity at the position of the Sun is from \citet{binney}. Values for two different distances $r_0$ are given to facilitate the comparison between the models,-- the distance to the Sun is assumed to be $8\, \mathrm{kpc}$ in the updated model, but $8.5\, \mathrm{kpc}$ in the default model.}
\begin{tabular}{lllllc}
\hline
 & \multicolumn{2}{l}{default} & \multicolumn{2}{l}{updated} & reference \\
 & \multicolumn{2}{l}{model} & \multicolumn{2}{l}{model} & values \\
\hline
$r_0\ (\mathrm{kpc})$ & 8.5 & 8.0 & 8.5 & 8.0 & 8.0 \\
$\rho_0\ (\mathrm{M}_\odot\, \mathrm{pc}^{-3})$ & 0.15 & 0.16 & 0.096 & 0.112 & $0.102\pm 0.010$ \\
$v_c^0\ (\mathrm{km}\, \mathrm{s}^{-1})$ & 220 & 220 & 218 & 217 & 220 \\
\hline
\end{tabular}
\label{tbl:a3}
\end{table}

\begin{table}
\caption{Updated model parameters and total masses derived for each component.}
\begin{tabular}{ccl}
\hline
Component & Mass ($\mathrm{M}_\odot$)& Parameters\\
\hline
  & & \\
Central mass & $1.41\times 10^{10}$ & $M_1=606.0$\\
 & & $b_1=0.3873$ \\
  & & \\
 &  & $M_2^1=3079.0$\\
First disc  & $7.14\times 10^{10}$ & $a_2^1=3.0$\\
 &  & $b_2^1=0.25$\\
  & & \\
 &  & $M_2^2=-307.9$\\
Second disc & $-7.14\times 10^{9}$ & $a_2^2=1.5$\\
 & & $b_2^2=0.25$ \\
  & & \\
Halo & $9.00\times 10^{11}$ & $M_3=4761.1$\\
 &  & $a_3=12.0$\\
   & & \\
\hline
\end{tabular}
\label{tbl:a1}
\end{table}

\begin{table}
\caption{Comparison of physical quantities computed from the model with measured values. The Oort constants and the escape velocity are from \citet{binney}, the local surface density is from \citet{holmberg2}, and the total mass from \citet{xue1}. Values for two different distances $r_0$ are given to facilitate the comparison between the models -- the distance to the Sun is assumed to be $8\, \mathrm{kpc}$ in the updated model, but $8.5\, \mathrm{kpc}$ in the default model.}
\begin{tabular}{lllllc}
\hline
 & \multicolumn{2}{l}{default} & \multicolumn{2}{l}{updated} & reference \\
 & \multicolumn{2}{l}{model} & \multicolumn{2}{l}{model} & values \\
\hline
$r_0\ (\mathrm{kpc})$ & 8.5 & 8.0 & 8.5 & 8.0 & 8.0 \\
$\Sigma_0\ (\mathrm{M}_\odot\, \mathrm{pc}^{-2})$ & 83 & 93 & 62 & 71 & $74\pm 6$ \\
A $(\mathrm{km}\, \mathrm{s}^{-1}\, \mathrm{kpc}^{-1})$ & 13.0 & 13.5 & 13.6 & 14.5 & $14.8\pm 0.8$ \\
B $(\mathrm{km}\, \mathrm{s}^{-1}\, \mathrm{kpc}^{-1})$ & -12.9 & -13.9 & -11.9 & -12.7 & $-12.4\pm 0.6$ \\
$v_e^0\ (\mathrm{km}\, \mathrm{s}^{-1})$ & 536 & 541 & 550 & 555 & $550\pm 50$ \\
$M_{tot}\ (\mathrm{M}_\odot)$ & \multicolumn{2}{l}{$9.00\times 10^{11}$} & \multicolumn{2}{l}{$1.00\times 10^{12}$} & $\simeq 1.00\times 10^{12}$\\
\hline
\end{tabular}
\label{tbl:a2}
\end{table}

\begin{figure}
\includegraphics[width=84 mm]{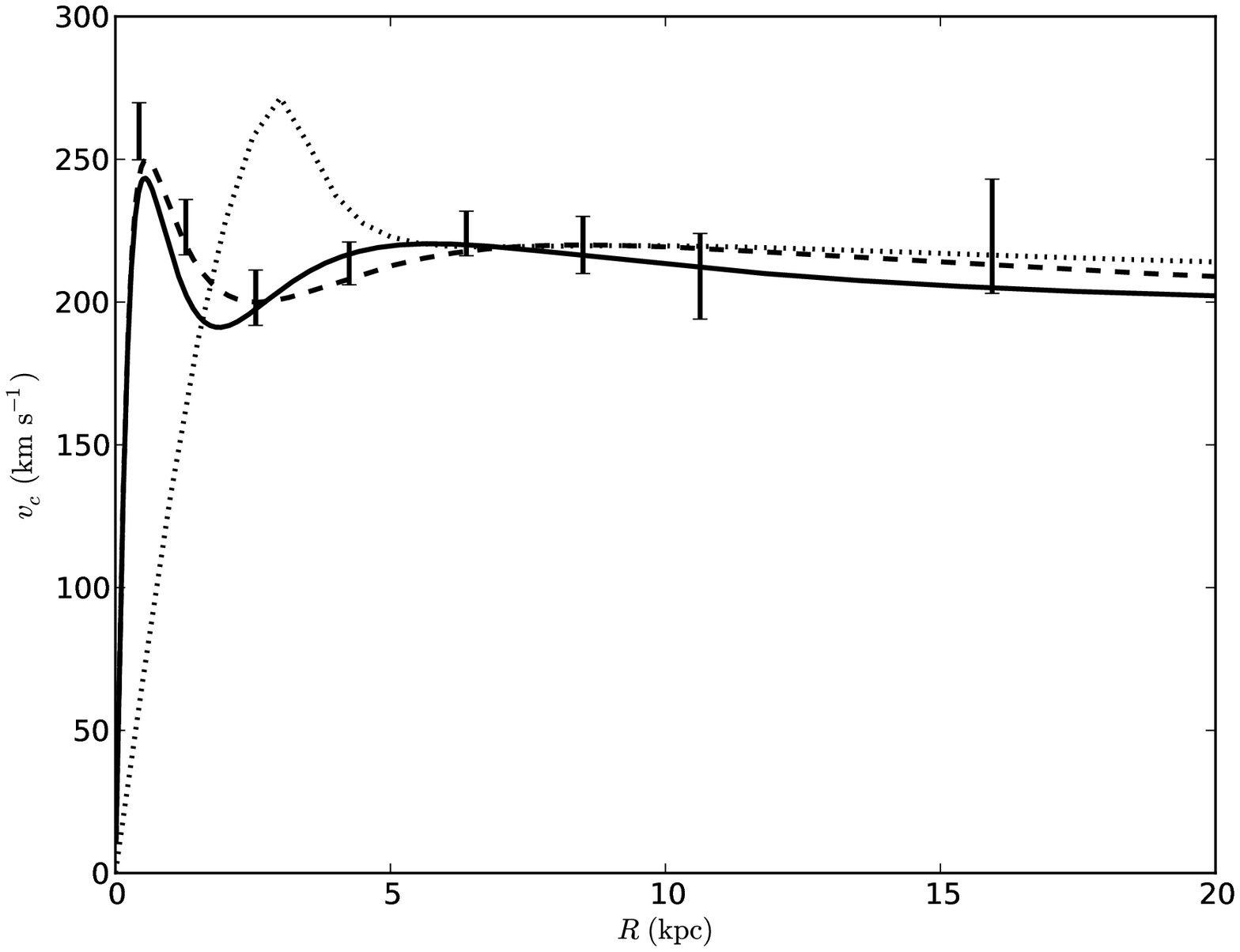}
\caption{Plot of the circular velocity as a function of radius over the interval $0 \textrm{ -- }20\ \mathrm{kpc}$. The dashed line corresponds to the default model, the dotted line correspond to the \citet{robin1} model, the solid line corresponds to the updated model, the error bars are data points from \citet{allen1}.}
\label{fig:a1}
\end{figure}

A summary of the parameters obtained through the described fitting procedure for the updated model can be seen in Table~\ref{tbl:a1}, together with the mass of each component. The quality of the model can be judged from Fig.~\ref{fig:a1} and from Table~\ref{tbl:a2}. In Fig.~\ref{fig:a1} we can see a comparison between the updated model and the default and \citet{robin1} models. The updated model behaves similarly to the others in the critical region between $5 \textrm{ -- }20\ \mathrm{kpc}$. Inside $5\ \mathrm{kpc}$ the behaviour is similar to the default model, which should not be surprising, since it is based on it. Note in particular that the agreement with the observed data points is quite good, although no effort was made to perform an explicit fit. 

Table~\ref{tbl:a2} verifies that the agreement of quantities computed from the model -- local surface density, Oort's A and B constants, escape velocity, and total mass of the galaxy -- is good, in particular when compared with the performance of the default model. Note that the default model was fine-tuned assuming a distance of the Sun from the galactic centre of $8.5\ \mathrm{kpc}$, whereas we assumed a distance of $8.0\ \mathrm{kpc}$. We conclude that the updated model is a better representation of the galactic gravitational potential than the default model.

\section{Data for main sequence stars}
\label{ap:2}
In this Appendix we present the relevant data for the stars classified as main sequence. In Tables~\ref{tbl:b1}~and~\ref{tbl:b2} the values used as input to compute the orbits are shown for the groups A and B respectively. The proper motions were obtained from a combination of the values from UCAC2 \citep{zacharias}, USNO-B \citep{monet}, Tycho-2 \citep{Hog} and Hipparcos \citep{leeuwen1}, SuperCOSMOS, SPM \citep{girard}, and NPM2 \citep{hanson1}, by a weighted average following the prescription in Section~2 of \citet{pauli1}. The gravities, temperatures and projected rotation velocities were taken from the literature or computed from Str\"omgren $uvby\beta$ photometry as indicated. Note that these are the measured gravities and temperatures not corrected for the gravitational darkening effect. The radial velocities are heliocentric corrected. The masses were derived from the evolutionary tracks of \citet{schaller}. See Section \ref{s:evol} for details of the procedure.

In Tables~\ref{tbl:b3}~and~\ref{tbl:b4} data concerning the present condition of the stars (present coordinates $(X,Y,Z)$, distances and evolutionary ages) and the condition at the moment of ejection from the disk (point of ejection in the plane $(X_0,Y_0)$, flight time and ejection velocities) are shown. All coordinates are given in a galactocentric Cartesian right-handed frame of reference where the $X$ axis passes through the position of the Sun, pointing in the opposite direction, and $Z$ points to the north galactic pole. Note that the distance from the galactic centre to the Sun was assumed to be $8\ \mathrm{kpc}$. Full three-dimensional velocities were computed from the knowledge of the distances, proper motions and radial velocities. The ejection velocities correspond to the orbital velocity for $Z=0$, corrected for the disk rotation velocity component. The evolutionary ages were obtained by interpolating the evolutionary tracks of \citet{schaller}.

The asymmetric error bounds correspond to confidence intervals enclosing the equivalent to $1-\sigma$ errors of the error distribution computed using the Monte Carlo procedure detailed in Section \ref{s:orb}.

\begin{table*}
\begin{center}

\caption{Input data for stars belonging to group A. The data shown are: the radial velocity, the proper motions in right ascension and declination directions, the surface gravity, temperature, mass and projected rotation velocity.}
\label{tbl:b1}

\scriptsize

\begin{tabular}{llllllll}
\hline
Name & r.v. & $\mathrm{p.m.}\alpha\cos\delta$ & $\mathrm{p.m.}\delta$ & $\log g$ & $T_{\mathrm{eff}}$ & $\mathrm{Mass}$ & $v\, \sin i$\\
 & $(\mathrm{km}\,\mathrm{s}^{-1})$ & $(\mathrm{mas}\, \mathrm{yr}^{-1})$ & $(\mathrm{mas}\, \mathrm{yr}^{-1})$ & $(\mathrm{cm}\, \mathrm{s}^{-2})$ & $\mathrm{(K)}$ & $(\mathrm{M}_\odot)$ & $(\mathrm{km}\, \mathrm{s}^{-1})$\\
\hline
EC 19337-6743&$-17\pm10$ (14) &$6.24\pm0.50$&$-5.64\pm0.61$&$3.90\pm0.20$ (14) &$10000\pm2000$ (14) &$2.8_{-0.7}^{+0.8}$&200 (14) \\
EC 05515-6231&$-12\pm10$ (14) &$-1.36\pm0.86$&$6.49\pm0.85$&$4.20\pm0.20$ (14) &$14000\pm2000$ (14) &$3.8_{-0.8}^{+0.9}$&65 (14) \\
EC 20089-5659&$-24\pm10$ (14) &$3.61\pm0.70$&$-7.97\pm0.72$&$3.90\pm0.20$ (14) &$14000\pm2000$ (14) &$4.4_{-0.9}^{+1.1}$&90 (14) \\
EC 03462-5813&$39\pm10$ (19) &$3.33\pm0.64$&$21.77\pm0.65$&$4.20\pm0.20$ (19) &$13500\pm675$ (19) &$3.6_{-0.4}^{+0.4}$&200 (19) \\
EC 05582-5816&$81\pm10$ (19) &$9.85\pm0.91$&$10.10\pm0.84$&$4.00\pm0.20$ (19) &$17000\pm850$ (19) &$5.6_{-0.6}^{+0.7}$&170 (19) \\
EC 10087-1411&$105\pm10$ (19) &$-9.34\pm0.76$&$-5.92\pm0.80$&$4.10\pm0.20$ (19) &$14500\pm725$ (19) &$4.2_{-0.5}^{+0.5}$&180 (19) \\
EC 05490-4510&$32\pm10$ (19) &$-1.75\pm1.34$&$2.83\pm1.37$&$4.20\pm0.20$ (19) &$17000\pm850$ (19) &$5.1_{-0.5}^{+0.6}$&30 (19) \\
EC 10549-2953&$-17\pm10$ (19) &$-8.81\pm1.09$&$1.03\pm1.07$&$4.10\pm0.20$ (19) &$14000\pm700$ (19) &$4.0_{-0.4}^{+0.5}$&200 (19) \\
EC 19071-7643&$-20\pm10$ (14) &$16.66\pm0.65$&$-20.15\pm0.64$&$4.20\pm0.20$ (14) &$21500\pm2000$ (14) &$7.5_{-1.2}^{+1.5}$&30 (14) \\
EC 19476-4109&$-19\pm10$ (14) &$-3.70\pm0.61$&$-4.51\pm0.62$&$4.00\pm0.20$ (14) &$17000\pm2000$ (14) &$5.6_{-1.1}^{+1.2}$&120 (14) \\
EC 20292-2414&$15\pm10$ (19) &$1.36\pm0.78$&$-11.32\pm0.85$&$4.10\pm0.20$ (19) &$25000\pm1250$ (19) &$10.3_{-1.3}^{+1.4}$&160 (19) \\
EC 13139-1851&$23\pm10$ (19) &$-9.15\pm1.39$&$-3.80\pm1.52$&$4.20\pm0.20$ (19) &$17000\pm850$ (19) &$5.1_{-0.6}^{+0.6}$&50 (19) \\
EC 20140-6935&$17\pm10$ (19) &$8.81\pm0.51$&$1.79\pm0.62$&$3.70\pm0.20$ (19) &$20500\pm1025$ (19) &$8.8_{-1.1}^{+1.5}$&65 (19) \\
EC 09414-1325&$71\pm10$ (19) &$-2.36\pm1.03$&$-1.47\pm1.10$&$4.10\pm0.20$ (19) &$14000\pm700$ (19) &$3.9_{-0.4}^{+0.5}$&260 (19) \\
EC 03240-6229&$-12\pm10$ (19) &$3.27\pm1.23$&$4.28\pm1.28$&$3.80\pm0.20$ (19) &$12000\pm600$ (19) &$3.6_{-0.4}^{+0.5}$&165 (19) \\
EC 20153-6731&$-40\pm10$ (19) &$5.07\pm1.07$&$-8.76\pm1.00$&$3.80\pm0.20$ (19) &$14500\pm725$ (19) &$4.8_{-0.6}^{+0.7}$&120 (19) \\
EC 19586-3823&$-102\pm10$ (14) &$-0.13\pm1.04$&$2.56\pm1.03$&$3.90\pm0.20$ (14) &$18500\pm2000$ (14) &$6.7_{-1.2}^{+1.4}$&150 (14) \\
EC 06387-8045&$60\pm10$ (19) &$-0.99\pm0.92$&$6.62\pm0.94$&$3.90\pm0.20$ (19) &$22500\pm1125$ (19) &$9.5_{-1.1}^{+1.4}$&190 (19) \\
PG 1533+467&$33\pm6$ (18) &$-11.45\pm0.53$&$11.41\pm0.54$&$4.09\pm0.10$ (18) &$18500\pm925$ (18) &$6.0_{-0.5}^{+0.5}$&215 (18) \\
EC 20252-3137&$26\pm10$ (14) &$-6.08\pm0.89$&$-3.59\pm0.91$&$4.00\pm0.20$ (14) &$23000\pm2000$ (14) &$9.2_{-1.5}^{+1.9}$&60 (14) \\
EC 05438-4741&$53\pm10$ (19) &$-0.35\pm1.07$&$-0.94\pm1.53$&$4.10\pm0.20$ (19) &$13500\pm675$ (19) &$3.8_{-0.4}^{+0.5}$&30 (19) \\
EC 20104-2944&$135\pm10$ (14) &$-0.45\pm1.70$&$-1.40\pm1.98$&$4.20\pm0.20$ (14) &$15000\pm2000$ (14) &$4.2_{-0.9}^{+1.0}$&50 (14) \\
EC 05515-6107&$89\pm10$ (19) &$-5.82\pm1.65$&$11.57\pm1.43$&$4.00\pm0.20$ (19) &$22000\pm1100$ (19) &$8.6_{-1.0}^{+1.1}$&290 (19) \\
PG 1205+228&$153\pm4$ (20) &$-15.44\pm0.55$&$0.05\pm0.48$&$4.10\pm0.20$ (20) &$16600\pm1000$ (20) &$5.2_{-0.6}^{+0.7}$&165 (20) \\
EC 23169-2235&$82\pm10$ (19) &$4.24\pm1.25$&$4.35\pm1.18$&$4.40\pm0.20$ (19) &$15000\pm750$ (19) &$3.9_{-0.4}^{+0.5}$&140 (19) \\
EC 09452-1403&$236\pm10$ (19) &$-3.55\pm1.54$&$-0.37\pm1.71$&$4.30\pm0.20$ (19) &$14000\pm700$ (19) &$3.6_{-0.4}^{+0.4}$&70 (19) \\
PG 2345+241&$80\pm3$ (20) &$-2.22\pm0.74$&$-3.09\pm0.67$&$4.20\pm0.20$ (20) &$18800\pm1000$ (20) &$6.0_{-0.7}^{+0.7}$&54 (20) \\
PHL 159&$88\pm3$ (18) &$-3.02\pm0.87$&$-9.38\pm0.69$&$3.59\pm0.10$ (18) &$18500\pm925$ (18) &$7.9_{-0.8}^{+0.8}$&21 (18) \\
EC 10500-1358&$99\pm10$ (19) &$-4.00\pm0.93$&$-2.17\pm1.01$&$3.80\pm0.20$ (19) &$15000\pm750$ (19) &$5.1_{-0.6}^{+0.7}$&100 (19) \\
PG 1511+367&$102\pm11$ (18) &$-3.64\pm0.88$&$-13.10\pm0.87$&$4.15\pm0.10$ (18) &$16100\pm805$ (18) &$4.7_{-0.4}^{+0.4}$&77 (18) \\
EC 20011-5005&$-171\pm10$ (14) &$2.06\pm1.61$&$-3.96\pm1.83$&$4.20\pm0.20$ (14) &$17000\pm2000$ (14) &$5.1_{-1.0}^{+1.1}$&30 (14) \\
BD -15 115&$93\pm4$ (18) &$7.04\pm1.00$&$0.46\pm0.76$&$3.81\pm0.10$ (18) &$20100\pm1005$ (18) &$8.0_{-0.8}^{+0.8}$&35 (18) \\
PG 2219+094&$-24\pm9$ (18) &$-3.53\pm1.17$&$-8.43\pm1.16$&$3.58\pm0.10$ (18) &$19500\pm975$ (18) &$8.1_{-0.8}^{+0.8}$&225 (20) \\
HS 1914+7139&$-39\pm10$ (6) &$-0.66\pm1.86$&$-0.14\pm2.60$&$3.90\pm0.10$ (18) &$17600\pm880$ (18) &$6.1_{-0.5}^{+0.5}$&250 (18) \\
PG 0855+294&$65\pm4$ (20) &$5.86\pm0.79$&$-3.98\pm0.82$&$3.80\pm0.20$ (20) &$18100\pm1000$ (20) &$6.8_{-0.8}^{+1.0}$&110 (20) \\
PG 0955+291&$76\pm14$ (20) &$-7.84\pm0.96$&$-4.27\pm1.14$&$3.80\pm0.20$ (20) &$13600\pm1000$ (20) &$4.3_{-0.5}^{+0.6}$&190 (20) \\
PG 2229+099&$-22\pm5$ (20) &$-1.22\pm1.85$&$-0.98\pm2.35$&$4.00\pm0.20$ (20) &$17600\pm1000$ (20) &$5.9_{-0.7}^{+0.8}$&16 (20) \\
PG 1610+239&$91\pm10$ (18) &$-0.20\pm2.80$&$-5.20\pm1.80$&$3.72\pm0.10$ (18) &$15500\pm775$ (18) &$5.6_{-0.5}^{+0.5}$&75 (18) \\
PG 2111+023&$-153\pm11$ (12) &$-6.98\pm1.76$&$-4.07\pm0.97$&$3.40\pm0.10$ (12) &$14000\pm700$ (12) &$5.2_{-0.4}^{+0.4}$&140 (12) \\
PG 0009+036&$142\pm18$ (20) &$-0.99\pm1.88$&$1.43\pm2.34$&$3.60\pm0.20$ (20) &$14800\pm1000$ (20) &$5.2_{-0.6}^{+0.7}$&350 (20) \\
PG 0122+214&$26\pm5$ (18) &$-1.10\pm0.58$&$-3.23\pm0.84$&$3.86\pm0.10$ (18) &$18300\pm915$ (18) &$6.5_{-0.6}^{+0.6}$&117 (18) \\
SB 357&$58\pm10$ (18) &$0.25\pm1.27$&$-2.81\pm1.57$&$3.90\pm0.10$ (18) &$19700\pm993$ (18) &$7.4_{-0.6}^{+0.7}$&180 (18) \\
EC 04420 -1908&$205\pm10$ (14) &$-1.27\pm1.81$&$-1.65\pm2.23$&$3.40\pm0.20$ (14) &$14000\pm2000$ (14) &$5.5_{-1.1}^{+1.3}$&180 (14) \\
PG 1332+137&$148\pm10$ (17) &$-5.99\pm0.77$&$-7.47\pm0.72$&$3.50\pm0.20$ (12) &$15000\pm750$ (12) &$5.9_{-0.7}^{+0.8}$&140 (12) \\
EC 19596-5356&$200\pm15$ (13) &$1.22\pm1.95$&$-0.52\pm2.47$&$3.75\pm0.20$ (13) &$16500\pm1500$ (13) &$5.9_{-0.9}^{+1.0}$&250 (13) \\
PHL 346&$63\pm4$ (18) &$4.12\pm1.23$&$-7.94\pm1.17$&$3.58\pm0.10$ (18) &$20700\pm1035$ (18) &$9.6_{-1.0}^{+1.2}$&45 (18) \\
PG 0914+001&$90\pm20$ (20) &$1.15\pm1.80$&$1.64\pm2.72$&$3.10\pm0.20$ (20) &$12300\pm1000$ (20) &$5.3_{-0.7}^{+0.8}$&325 (20) \\
PG 1209+263&$52\pm11$ (12) &$0.40\pm1.92$&$-0.74\pm2.38$&$3.00\pm0.20$ (12) &$12000\pm600$ (12) &$5.5_{-0.8}^{+0.8}$&70 (12) \\
\hline
\end{tabular}
\end{center}
References: (1) this study, from Str\"omgren $uvby\beta$ photometry; (2) \citet{behr2}; (3) \citet{conlon5}; (4) \citet{conlon3}; (5) \citet{evans}; (6) \citet{heber1}; (7) \citet{keenan1}; (8) \citet{keenan2}; (9) \citet{keenan3}; (10) \citet{kilkenny}; (11) \citet{levenhagen}; (12) \citet{lynn1}; (13) \citet{lynn2}; (14) \citet{magee2}; (15) \citet{martin3}; (16) \citet{martin1}; (17) \citet{martin2}; (18) \citet{ramspeck}; (19) \citet{rolleston1}; (20) \citet{rolleston2}.
\end{table*}

\begin{table*}
\begin{center}

\caption{Input data for stars belonging to group B. The data shown are: the radial velocity, the proper motions in right ascension and declination directions, the surface gravity, temperature, mass and projected rotation velocity.}
\label{tbl:b2}

\scriptsize

\begin{tabular}{llllllll}
\hline
Name & r.v. & $\mathrm{p.m.}\alpha\cos\delta$ & $\mathrm{p.m.}\delta$ & $\log g$ & $T_{\mathrm{eff}}$ & $\mathrm{Mass}$ & $v\, \sin i$\\
 & $(\mathrm{km}\,\mathrm{s}^{-1})$ & $(\mathrm{mas}\, \mathrm{yr}^{-1})$ & $(\mathrm{mas}\, \mathrm{yr}^{-1})$ & $(\mathrm{cm}\, \mathrm{s}^{-2})$ & $\mathrm{(K)}$ & $(\mathrm{M}_\odot)$ & $(\mathrm{km}\, \mathrm{s}^{-1})$\\
\hline
HIP 81153&$146\pm7$ (17) &$-8.64\pm0.65$&$-13.35\pm0.65$&$4.05\pm0.45$ (16) &$29891\pm684$ (16) &$15.4_{-3.1}^{+7.1}$&80 (16) \\
HIP 1241&$0\pm5$ (17) &$5.98\pm0.63$&$-0.43\pm0.49$&$3.98\pm0.15$ (1) &$11489\pm574$ (1) &$3.1_{-0.3}^{+0.3}$&130 (16) \\
HIP 28132&$54\pm3$ (17) &$-3.90\pm0.65$&$8.35\pm0.71$&$4.00\pm0.15$ (16) &$16948\pm140$ (16) &$5.5_{-0.3}^{+0.4}$&100 (16) \\
HIP 2702&$-26\pm10$ (14) &$-5.23\pm0.56$&$-5.77\pm0.57$&$4.20\pm0.20$ (14) &$15500\pm2000$ (14) &$4.4_{-0.9}^{+1.1}$&30 (14) \\
HIP 55051&$20\pm10$ (8) &$28.82\pm0.50$&$20.03\pm0.41$&$4.06\pm0.25$ (1) &$25518\pm1276$ (1) &$10.7_{-1.5}^{+1.8}$&150 (11) \\
HIP 114690&$83\pm7$ (7) &$4.98\pm0.71$&$-12.56\pm0.51$&$3.50\pm0.25$ (9) &$20008\pm1000$ (1) &$9.3_{-1.4}^{+1.9}$&238 (7) \\
HIP 11844&$-45\pm6$ (17) &$6.13\pm0.90$&$-2.76\pm0.95$&$4.14\pm0.08$ (16) &$11202\pm68$ (16) &$2.9_{-0.1}^{+0.1}$&130 (16) \\
HIP 16130&$28\pm6$ (17) &$8.65\pm0.63$&$2.59\pm0.70$&$4.11\pm0.15$ (1) &$14061\pm703$ (1) &$4.0_{-0.4}^{+0.4}$&60 (16) \\
HIP 58046&$20\pm2$ (17) &$8.12\pm0.83$&$6.73\pm0.82$&$4.25\pm0.25$ (16) &$13357\pm668$ (16) &$3.5_{-0.4}^{+0.5}$&230 (16) \\
BD +36 2242&$2\pm4$ (2) &$-5.92\pm0.52$&$1.68\pm0.50$&$3.87\pm0.15$ (1) &$11806\pm590$ (1) &$3.4_{-0.3}^{+0.4}$&77 (2) \\
HIP 98136&$29\pm6$ (17) &$-3.78\pm0.72$&$-10.84\pm0.56$&$3.42\pm0.20$ (1) &$21570\pm1079$ (1) &$11.6_{-1.7}^{+2.3}$&140 (16) \\
HIP 111396&$-33\pm6$ (17) &$-0.76\pm0.62$&$2.89\pm0.62$&$3.62\pm0.15$ (1) &$13662\pm683$ (1) &$4.9_{-0.5}^{+0.5}$&35 (16) \\
HIP 77131&$-9\pm5$ (5) &$-4.59\pm0.75$&$-1.76\pm0.71$&$3.86\pm0.25$ (16) &$32089\pm536$ (16) &$19.5_{-2.7}^{+5.5}$&250 (16) \\
HIP 113735&$-31\pm10$ (9) &$10.12\pm0.49$&$18.42\pm0.49$&$4.12\pm0.20$ (1) &$21645\pm1082$ (1) &$7.9_{-0.9}^{+1.0}$&?  \\
HIP 59955&$28\pm10$ (17) &$-10.82\pm0.72$&$-2.41\pm0.59$&$3.66\pm0.15$ (1) &$13860\pm693$ (1) &$4.7_{-0.4}^{+0.5}$&225 (16) \\
HIP 76161&$-51\pm7$ (15) &$0.61\pm0.76$&$-0.39\pm0.71$&$3.43\pm0.20$ (1) &$22331\pm1117$ (1) &$12.7_{-2.0}^{+3.0}$&?  \\
HIP 96130&$-51\pm6$ (17) &$1.32\pm0.72$&$-0.04\pm0.58$&$3.63\pm0.21$ (16) &$23141\pm239$ (16) &$11.8_{-1.6}^{+2.0}$&60 (16) \\
HIP 61800&$-48\pm13$ (17) &$0.07\pm0.45$&$-5.11\pm0.43$&$3.24\pm0.15$ (1) &$12247\pm612$ (1) &$4.8_{-0.5}^{+0.6}$&150 (16) \\
HIP 114569&$94\pm5$ (17) &$46.35\pm0.79$&$32.73\pm0.74$&$4.12\pm0.15$ (1) &$18518\pm926$ (1) &$6.1_{-0.6}^{+0.6}$&70 (16) \\
HIP 11809&$0\pm10$ (17) &$-18.90\pm0.88$&$-13.46\pm0.87$&$4.25\pm0.25$ (16) &$13265\pm663$ (16) &$3.5_{-0.4}^{+0.5}$&240 (16) \\
HIP 16758&$98\pm10$ (14) &$7.37\pm0.48$&$10.04\pm0.53$&$3.60\pm0.20$ (14) &$23453\pm1173$ (1) &$12.1_{-1.8}^{+2.5}$&90 (14) \\
HIP 108215&$-79\pm7$ (7) &$8.23\pm0.59$&$-8.49\pm0.38$&$3.06\pm0.15$ (1) &$13996\pm700$ (1) &$7.0_{-0.9}^{+1.2}$&169 (7) \\
HIP 115347&$23\pm7$ (7) &$-6.30\pm0.59$&$-7.01\pm0.46$&$3.49\pm0.20$ (1) &$21106\pm1055$ (1) &$10.6_{-1.6}^{+2.0}$&27 (7) \\
HIP 115729&$26\pm3$ (17) &$3.05\pm0.72$&$-1.21\pm0.61$&$3.46\pm0.15$ (1) &$17966\pm898$ (1) &$7.8_{-0.9}^{+1.1}$&25 (16) \\
HIP 3812&$19\pm10$ (14) &$0.80\pm0.70$&$-1.73\pm0.73$&$4.00\pm0.20$ (14) &$19000\pm2000$ (14) &$6.7_{-1.2}^{+1.3}$&219 (7) \\
PG 1530+212&$-7\pm25$ (2) &$4.04\pm0.52$&$6.60\pm0.53$&$4.00\pm0.25$ (2) &$15000\pm500$ (2) &$4.6_{-0.5}^{+0.7}$&104 (12) \\
HIP 79649&$19\pm6$ (17) &$-0.48\pm0.57$&$-2.97\pm0.62$&$3.69\pm0.20$ (1) &$21721\pm1086$ (1) &$9.8_{-1.3}^{+1.7}$&90 (16) \\
HIP 12320&$24\pm12$ (17) &$4.33\pm0.84$&$-1.25\pm0.92$&$4.00\pm0.25$ (16) &$13382\pm669$ (16) &$3.9_{-0.5}^{+0.6}$&225 (16) \\
HIP 109051&$72\pm8$ (17) &$1.43\pm0.72$&$-0.88\pm0.74$&$3.85\pm0.20$ (1) &$21021\pm1051$ (1) &$8.2_{-1.0}^{+1.3}$&105 (16) \\
HIP 111563&$42\pm7$ (7) &$5.13\pm0.47$&$-17.86\pm0.36$&$3.11\pm0.20$ (1) &$22545\pm1127$ (1) &$17.3_{-3.6}^{+6.0}$&105 (7) \\
HIP 13800&$-6\pm2$ (2) &$4.69\pm0.68$&$-3.63\pm0.80$&$3.67\pm0.15$ (1) &$16632\pm832$ (1) &$6.3_{-0.7}^{+0.7}$&12 (2) \\
HIP 48394&$14\pm12$ (17) &$3.34\pm0.93$&$-5.69\pm0.72$&$3.75\pm0.25$ (16) &$20021\pm1001$ (16) &$7.8_{-1.0}^{+1.4}$&225 (16) \\
HIP 107027&$117\pm8$ (17) &$-6.18\pm0.77$&$-11.99\pm0.64$&$3.65\pm0.21$ (16) &$21959\pm231$ (16) &$10.5_{-1.3}^{+1.8}$&150 (16) \\
HIP 1904&$-37\pm10$ (9) &$2.95\pm0.47$&$-6.27\pm0.49$&$3.03\pm0.15$ (1) &$12894\pm645$ (1) &$6.8_{-0.9}^{+1.2}$&?  \\
HIP 55461&$72\pm6$ (17) &$-0.97\pm0.74$&$-5.36\pm0.69$&$4.00\pm0.20$ (3) &$15600\pm1000$ (3) &$4.9_{-0.6}^{+0.7}$&140 (16) \\
HIP 112790&$15\pm4$ (17) &$-18.34\pm0.67$&$-13.53\pm0.69$&$3.53\pm0.15$ (1) &$15628\pm781$ (1) &$5.9_{-0.5}^{+0.4}$&70 (16) \\
HIP 59067&$35\pm5$ (17) &$-8.47\pm0.83$&$-5.87\pm0.70$&$3.89\pm0.15$ (16) &$14974\pm98$ (16) &$4.8_{-0.3}^{+0.4}$&70 (16) \\
BD +20 3004&$22\pm14$ (2) &$-10.91\pm0.51$&$5.28\pm0.55$&$3.40\pm0.15$ (1) &$14070\pm704$ (1) &$5.6_{-0.6}^{+0.8}$&105 (2) \\
HIP 105912&$-17\pm7$ (7) &$10.16\pm0.57$&$-11.61\pm0.47$&$3.29\pm0.20$ (1) &$22201\pm1110$ (1) &$13.6_{-2.4}^{+3.5}$&102 (7) \\
HIP 45904&$36\pm14$ (17) &$3.27\pm0.79$&$-9.63\pm0.60$&$3.48\pm0.20$ (1) &$21016\pm1051$ (1) &$10.7_{-1.5}^{+2.0}$&277 (12) \\
HIP 37903&$84\pm10$ (17) &$2.32\pm0.83$&$-5.77\pm0.76$&$3.00\pm0.25$ (16) &$18000\pm900$ (16) &$10.8_{-2.2}^{+3.3}$&150 (16) \\
HIP 70275&$241\pm6$ (2) &$4.24\pm0.69$&$-10.74\pm0.64$&$3.64\pm0.20$ (1) &$22790\pm1140$ (1) &$11.3_{-1.6}^{+2.1}$&68 (2) \\
BD -2 3766&$41\pm10$ (4) &$1.77\pm0.59$&$17.26\pm0.62$&$3.70\pm0.20$ (4) &$22000\pm1000$ (4) &$10.1_{-1.3}^{+1.7}$&200 (16) \\
PB 5418&$152\pm10$ (4) &$8.25\pm1.48$&$-4.74\pm1.63$&$4.00\pm0.20$ (4) &$19310\pm966$ (1) &$6.8_{-0.8}^{+0.9}$&150 (10) \\
HIP 56322&$254\pm9$ (17) &$3.54\pm0.73$&$11.52\pm0.69$&$3.57\pm0.25$ (1) &$25501\pm1275$ (1) &$14.2_{-2.5}^{+4.2}$&160 (16) \\
Ton S 308&$89\pm10$ (4) &$-0.38\pm1.57$&$-0.96\pm1.76$&$4.00\pm0.20$ (4) &$17821\pm891$ (1) &$6.0_{-0.7}^{+0.8}$&120 (4) \\
PHL 2018&$108\pm10$ (4) &$3.87\pm1.19$&$-11.27\pm1.17$&$3.70\pm0.20$ (4) &$19095\pm955$ (1) &$7.7_{-0.9}^{+1.1}$&150 (4) \\
HIP 52906&$84\pm11$ (17) &$-7.39\pm0.61$&$1.34\pm0.60$&$3.39\pm0.15$ (1) &$18898\pm945$ (1) &$9.0_{-1.0}^{+1.2}$&160 (16) \\
\hline
\end{tabular}
\end{center}
References: (1) this study, from Str\"omgren $uvby\beta$ photometry; (2) \citet{behr2}; (3) \citet{conlon5}; (4) \citet{conlon3}; (5) \citet{evans}; (6) \citet{heber1}; (7) \citet{keenan1}; (8) \citet{keenan2}; (9) \citet{keenan3}; (10) \citet{kilkenny}; (11) \citet{levenhagen}; (12) \citet{lynn1}; (13) \citet{lynn2}; (14) \citet{magee2}; (15) \citet{martin3}; (16) \citet{martin1}; (17) \citet{martin2}; (18) \citet{ramspeck}; (19) \citet{rolleston1}; (20) \citet{rolleston2}.
\end{table*}

\begin{table*}
\begin{center}

\caption{Computed values for stars belonging to group A. The values shown are: the present coordinates of the star $(X,Y,Z)$ (in galactocentric Cartesian coordinates), the distance, the evolutionary age, the flight time, the ejection velocity and the coordinates $(X_0,Y_0)$ of the point of ejection in the galactic plane.}
\label{tbl:b3}

\scriptsize

\begin{tabular}{llllllllll}
\hline
Name & $X$ & $Y$ & $Z$ & $d$ & $\mathrm{Age}$ & $t_{\mathrm{flight}}$ & $v_{\mathrm{ejec}}$  & $X_0$ & $Y_0$\\
 & $\mathrm{(kpc)}$ & $\mathrm{(kpc)}$ & $\mathrm{(kpc)}$ & $\mathrm{(kpc)}$ & $\mathrm{(Myr)}$ & $\mathrm{(Myr)}$ & $(\mathrm{km}\,\mathrm{s}^{-1})$  & $\mathrm{(kpc)}$ & $\mathrm{(kpc)}$\\
\hline
EC 19337-6743&$-7.60_{-0.10}^{+0.20}$&$-0.30_{-0.10}^{+0.10}$&$-0.30_{-0.10}^{+0.10}$&$0.56_{-0.22}^{+0.30}$&$223.2_{-132.0}^{+314.4}$&$21.0_{-3.8}^{+5.8}$&$28_{-8}^{+11}$&$-5.57_{-0.62}^{+0.83}$&$-4.72_{-1.05}^{+0.74}$\\
EC 05515-6231&$-8.00_{-0.00}^{+0.00}$&$-0.60_{-0.30}^{+0.20}$&$-0.40_{-0.10}^{+0.10}$&$0.71_{-0.22}^{+0.29}$&$36.7_{-36.7}^{+68.1}$&$31.0_{-4.7}^{+5.0}$&$35_{-9}^{+10}$&$-4.78_{-0.80}^{+0.94}$&$-7.03_{-0.95}^{+0.94}$\\
EC 20089-5659&$-7.40_{-0.20}^{+0.30}$&$-0.20_{-0.10}^{+0.00}$&$-0.40_{-0.20}^{+0.10}$&$0.81_{-0.25}^{+0.35}$&$83.6_{-37.1}^{+65.7}$&$25.9_{-3.0}^{+3.6}$&$53_{-14}^{+21}$&$-4.21_{-0.74}^{+0.77}$&$-4.90_{-0.67}^{+0.58}$\\
EC 03462-5813&$-8.00_{-0.00}^{+0.00}$&$-0.60_{-0.20}^{+0.20}$&$-0.60_{-0.20}^{+0.10}$&$0.83_{-0.21}^{+0.27}$&$1.8_{-1.8}^{+75.3}$&$12.2_{-2.1}^{+2.2}$&$79_{-16}^{+20}$&$-6.59_{-0.50}^{+0.67}$&$-3.20_{-0.66}^{+0.63}$\\
EC 05582-5816&$-8.10_{-0.00}^{+0.10}$&$-1.10_{-0.40}^{+0.20}$&$-0.60_{-0.30}^{+0.10}$&$1.31_{-0.33}^{+0.43}$&$29.0_{-28.4}^{+13.4}$&$35.8_{-7.4}^{+6.4}$&$162_{-24}^{+24}$&$-2.02_{-2.25}^{+2.27}$&$-3.80_{-0.39}^{+0.52}$\\
EC 10087-1411&$-8.20_{-0.10}^{+0.00}$&$-0.90_{-0.30}^{+0.20}$&$0.60_{-0.10}^{+0.20}$&$1.10_{-0.27}^{+0.36}$&$32.7_{-32.7}^{+37.8}$&$18.7_{-6.9}^{+9.7}$&$141_{-22}^{+35}$&$-6.48_{-0.91}^{+1.85}$&$-2.84_{-0.64}^{+0.79}$\\
EC 05490-4510&$-8.50_{-0.10}^{+0.10}$&$-1.40_{-0.50}^{+0.30}$&$-0.80_{-0.30}^{+0.20}$&$1.74_{-0.43}^{+0.56}$&$18.3_{-18.3}^{+24.7}$&$23.7_{-4.7}^{+7.1}$&$51_{-8}^{+10}$&$-6.47_{-0.64}^{+1.09}$&$-6.03_{-1.33}^{+1.09}$\\
EC 10549-2953&$-7.90_{-0.00}^{+0.10}$&$-1.50_{-0.50}^{+0.30}$&$0.80_{-0.20}^{+0.20}$&$1.71_{-0.42}^{+0.55}$&$36.4_{-36.4}^{+42.7}$&$42.9_{-7.2}^{+8.9}$&$56_{-11}^{+14}$&$-0.81_{-1.95}^{+2.48}$&$-7.86_{-0.64}^{+0.63}$\\
EC 19071-7643&$-6.70_{-0.30}^{+0.50}$&$-1.20_{-0.40}^{+0.30}$&$-0.90_{-0.40}^{+0.20}$&$2.03_{-0.56}^{+0.75}$&$6.2_{-6.2}^{+12.7}$&$6.6_{-0.3}^{+0.5}$&$226_{-63}^{+82}$&$-5.21_{-0.68}^{+0.91}$&$-2.18_{-0.18}^{+0.15}$\\
EC 19476-4109&$-6.30_{-0.50}^{+0.70}$&$0.00_{-0.10}^{+0.00}$&$-0.90_{-0.40}^{+0.20}$&$1.93_{-0.55}^{+0.77}$&$37.1_{-22.1}^{+27.0}$&$34.3_{-2.1}^{+2.0}$&$111_{-31}^{+44}$&$-1.98_{-0.82}^{+0.75}$&$-4.61_{-0.64}^{+0.83}$\\
EC 20292-2414&$-6.40_{-0.40}^{+0.50}$&$0.60_{-0.20}^{+0.20}$&$-1.00_{-0.40}^{+0.20}$&$1.97_{-0.49}^{+0.67}$&$7.4_{-7.4}^{+4.6}$&$16.2_{-1.5}^{+1.6}$&$131_{-36}^{+49}$&$-5.92_{-0.49}^{+0.59}$&$-1.55_{-0.51}^{+0.67}$\\
EC 13139-1851&$-7.20_{-0.20}^{+0.30}$&$-0.90_{-0.30}^{+0.20}$&$1.10_{-0.30}^{+0.40}$&$1.65_{-0.41}^{+0.54}$&$18.7_{-18.7}^{+24.5}$&$26.1_{-4.5}^{+4.5}$&$122_{-32}^{+45}$&$-4.14_{-1.28}^{+1.51}$&$-4.03_{-0.44}^{+0.44}$\\
EC 20140-6935&$-6.50_{-0.40}^{+0.50}$&$-1.00_{-0.40}^{+0.20}$&$-1.20_{-0.40}^{+0.30}$&$2.14_{-0.55}^{+0.79}$&$22.4_{-4.3}^{+4.9}$&$12.0_{-0.8}^{+0.8}$&$114_{-27}^{+40}$&$-5.76_{-0.56}^{+0.76}$&$-3.90_{-0.52}^{+0.45}$\\
EC 09414-1325&$-8.90_{-0.30}^{+0.20}$&$-2.30_{-0.70}^{+0.60}$&$1.30_{-0.30}^{+0.50}$&$2.78_{-0.70}^{+0.91}$&$1.2_{-1.2}^{+61.6}$&$35.7_{-10.2}^{+14.4}$&$105_{-18}^{+27}$&$-5.01_{-1.47}^{+2.54}$&$-6.14_{-0.94}^{+1.02}$\\
EC 03240-6229&$-7.80_{-0.00}^{+0.10}$&$-1.40_{-0.50}^{+0.30}$&$-1.50_{-0.60}^{+0.40}$&$2.12_{-0.54}^{+0.73}$&$140.0_{-31.7}^{+33.7}$&$45.5_{-6.4}^{+7.2}$&$76_{-13}^{+16}$&$-0.98_{-1.51}^{+1.73}$&$-8.39_{-0.81}^{+0.72}$\\
EC 20153-6731&$-6.10_{-0.50}^{+0.70}$&$-1.20_{-0.40}^{+0.30}$&$-1.50_{-0.50}^{+0.40}$&$2.71_{-0.68}^{+0.96}$&$77.7_{-15.4}^{+17.9}$&$23.1_{-2.4}^{+2.8}$&$154_{-38}^{+50}$&$-1.52_{-0.92}^{+1.01}$&$-3.72_{-0.59}^{+0.66}$\\
EC 19586-3823&$-5.30_{-0.70}^{+1.10}$&$0.10_{-0.00}^{+0.10}$&$-1.60_{-0.60}^{+0.50}$&$3.20_{-0.91}^{+1.25}$&$30.3_{-11.9}^{+16.8}$&$39.1_{-5.0}^{+5.5}$&$153_{-23}^{+34}$&$1.93_{-0.75}^{+0.83}$&$-6.74_{-1.00}^{+0.73}$\\
EC 06387-8045&$-6.70_{-0.30}^{+0.40}$&$-3.10_{-1.10}^{+0.80}$&$-1.70_{-0.60}^{+0.40}$&$3.75_{-0.94}^{+1.29}$&$12.4_{-9.0}^{+3.9}$&$24.4_{-4.0}^{+5.0}$&$134_{-18}^{+17}$&$-3.10_{-1.47}^{+1.96}$&$-5.25_{-0.65}^{+0.56}$\\
PG 1533+467&$-7.70_{-0.00}^{+0.10}$&$1.30_{-0.20}^{+0.20}$&$1.80_{-0.30}^{+0.30}$&$2.23_{-0.33}^{+0.38}$&$0.3_{-0.3}^{+9.6}$&$16.1_{-0.9}^{+1.0}$&$213_{-30}^{+37}$&$-4.32_{-0.65}^{+0.77}$&$-2.30_{-0.17}^{+0.19}$\\
EC 20252-3137&$-5.40_{-0.70}^{+1.00}$&$0.50_{-0.10}^{+0.20}$&$-1.80_{-0.60}^{+0.50}$&$3.21_{-0.86}^{+1.24}$&$12.5_{-8.4}^{+6.5}$&$45.7_{-5.7}^{+7.5}$&$155_{-43}^{+59}$&$-2.77_{-0.84}^{+0.65}$&$-4.95_{-0.80}^{+0.98}$\\
EC 05438-4741&$-8.90_{-0.20}^{+0.30}$&$-3.10_{-1.00}^{+0.80}$&$-1.90_{-0.60}^{+0.50}$&$3.67_{-0.91}^{+1.26}$&$82.1_{-80.4}^{+32.3}$&$35.5_{-8.4}^{+13.3}$&$100_{-18}^{+24}$&$-6.82_{-1.19}^{+1.79}$&$-8.07_{-1.62}^{+1.39}$\\
EC 20104-2944&$-4.70_{-1.00}^{+1.30}$&$0.70_{-0.20}^{+0.30}$&$-1.90_{-0.80}^{+0.50}$&$3.90_{-1.14}^{+1.59}$&$26.7_{-26.7}^{+51.2}$&$17.9_{-3.8}^{+5.4}$&$172_{-23}^{+49}$&$-5.87_{-0.99}^{+1.34}$&$-3.15_{-1.29}^{+0.94}$\\
EC 05515-6107&$-8.00_{-0.00}^{+0.00}$&$-3.40_{-1.20}^{+0.80}$&$-2.00_{-0.70}^{+0.50}$&$4.01_{-1.00}^{+1.33}$&$4.2_{-4.2}^{+9.2}$&$13.9_{-2.3}^{+3.2}$&$188_{-39}^{+65}$&$-4.51_{-1.17}^{+1.84}$&$-6.02_{-1.59}^{+1.24}$\\
PG 1205+228&$-8.20_{-0.10}^{+0.00}$&$-0.30_{-0.20}^{+0.00}$&$2.20_{-0.60}^{+0.70}$&$2.22_{-0.56}^{+0.76}$&$18.2_{-18.2}^{+24.5}$&$14.4_{-3.7}^{+4.5}$&$254_{-40}^{+60}$&$-5.42_{-1.18}^{+2.01}$&$-2.06_{-0.17}^{+0.28}$\\
EC 23169-2235&$-7.30_{-0.20}^{+0.20}$&$0.60_{-0.20}^{+0.20}$&$-2.30_{-0.70}^{+0.60}$&$2.47_{-0.61}^{+0.82}$&$0.0_{-0.0}^{+33.3}$&$20.4_{-3.5}^{+3.9}$&$151_{-24}^{+34}$&$-5.39_{-0.95}^{+1.27}$&$-4.80_{-1.02}^{+0.86}$\\
EC 09452-1403&$-9.50_{-0.40}^{+0.40}$&$-4.00_{-1.30}^{+1.00}$&$2.40_{-0.60}^{+0.80}$&$4.89_{-1.20}^{+1.61}$&$1.4_{-1.4}^{+72.1}$&$26.1_{-8.7}^{+13.5}$&$272_{-31}^{+62}$&$-4.85_{-1.90}^{+3.82}$&$-3.27_{-1.22}^{+1.07}$\\
PG 2345+241&$-9.00_{-0.40}^{+0.20}$&$3.80_{-0.90}^{+1.30}$&$-2.90_{-1.00}^{+0.70}$&$4.93_{-1.25}^{+1.65}$&$11.2_{-11.2}^{+18.5}$&$26.7_{-4.4}^{+5.4}$&$149_{-17}^{+20}$&$-9.27_{-1.30}^{+0.79}$&$-3.80_{-1.14}^{+0.93}$\\
PHL 159&$-6.00_{-0.30}^{+0.40}$&$3.30_{-0.50}^{+0.60}$&$-3.00_{-0.50}^{+0.50}$&$4.87_{-0.74}^{+0.88}$&$31.5_{-5.3}^{+6.4}$&$21.4_{-2.4}^{+3.0}$&$188_{-20}^{+25}$&$-9.32_{-0.92}^{+0.63}$&$-0.07_{-0.82}^{+0.99}$\\
EC 10500-1358&$-8.40_{-0.20}^{+0.10}$&$-4.20_{-1.50}^{+1.10}$&$3.50_{-0.90}^{+1.20}$&$5.46_{-1.38}^{+1.93}$&$68.7_{-13.8}^{+15.6}$&$55.1_{-13.3}^{+17.6}$&$224_{-41}^{+53}$&$0.04_{-2.80}^{+2.77}$&$-4.16_{-1.21}^{+2.23}$\\
PG 1511+367&$-6.90_{-0.10}^{+0.20}$&$1.90_{-0.30}^{+0.30}$&$3.60_{-0.50}^{+0.70}$&$4.27_{-0.63}^{+0.72}$&$23.5_{-22.7}^{+20.8}$&$20.5_{-2.2}^{+2.6}$&$256_{-28}^{+32}$&$-10.22_{-0.98}^{+0.71}$&$0.37_{-0.77}^{+0.99}$\\
EC 20011-5005&$-1.40_{-1.90}^{+2.50}$&$-1.30_{-0.50}^{+0.40}$&$-4.20_{-1.60}^{+1.20}$&$7.87_{-2.24}^{+3.12}$&$16.2_{-16.2}^{+31.1}$&$34.7_{-8.3}^{+17.2}$&$274_{-33}^{+37}$&$4.81_{-1.59}^{+3.69}$&$-3.18_{-2.24}^{+1.99}$\\
BD -15 115&$-8.30_{-0.10}^{+0.00}$&$1.00_{-0.20}^{+0.20}$&$-4.60_{-0.90}^{+0.70}$&$4.73_{-0.72}^{+0.87}$&$25.4_{-3.9}^{+5.1}$&$30.4_{-2.6}^{+2.7}$&$249_{-35}^{+48}$&$-1.78_{-1.24}^{+1.42}$&$-3.77_{-0.63}^{+0.68}$\\
PG 2219+094&$-6.30_{-0.30}^{+0.30}$&$5.60_{-0.90}^{+1.00}$&$-4.60_{-0.90}^{+0.70}$&$7.42_{-1.11}^{+1.37}$&$25.3_{-4.0}^{+4.7}$&$39.3_{-7.7}^{+11.9}$&$200_{-40}^{+58}$&$-14.07_{-3.97}^{+2.57}$&$2.28_{-2.69}^{+2.62}$\\
HS 1914+7139&$-10.50_{-0.40}^{+0.40}$&$10.60_{-1.50}^{+1.80}$&$4.80_{-0.70}^{+0.80}$&$11.87_{-1.71}^{+1.95}$&$22.7_{-11.7}^{+9.2}$&$58.3_{-26.6}^{+54.5}$&$212_{-84}^{+175}$&$-6.37_{-7.78}^{+8.60}$&$-2.60_{-12.79}^{+8.02}$\\
PG 0855+294&$-13.80_{-2.10}^{+1.50}$&$-1.70_{-0.60}^{+0.40}$&$4.90_{-1.20}^{+1.80}$&$7.81_{-1.99}^{+2.80}$&$34.6_{-7.5}^{+8.7}$&$24.1_{-3.1}^{+3.8}$&$278_{-62}^{+92}$&$-16.03_{-3.98}^{+2.68}$&$-2.99_{-0.68}^{+1.07}$\\
PG 0955+291&$-11.60_{-1.10}^{+0.90}$&$-1.30_{-0.40}^{+0.30}$&$4.90_{-1.30}^{+1.40}$&$6.14_{-1.51}^{+1.87}$&$85.5_{-25.7}^{+28.4}$&$78.2_{-20.3}^{+48.0}$&$277_{-39}^{+41}$&$8.27_{-4.92}^{+11.26}$&$-1.67_{-1.76}^{+2.58}$\\
PG 2229+099&$-6.50_{-0.40}^{+0.50}$&$6.00_{-1.50}^{+2.00}$&$-5.10_{-1.70}^{+1.30}$&$7.98_{-1.92}^{+2.70}$&$34.5_{-20.5}^{+11.6}$&$63.1_{-19.6}^{+43.8}$&$178_{-38}^{+72}$&$-3.35_{-5.05}^{+4.23}$&$-7.05_{-9.05}^{+5.02}$\\
PG 1610+239&$-4.00_{-0.60}^{+0.70}$&$3.40_{-0.50}^{+0.60}$&$5.20_{-0.70}^{+0.90}$&$7.45_{-1.15}^{+1.34}$&$62.1_{-10.4}^{+13.0}$&$42.0_{-13.1}^{+29.6}$&$212_{-23}^{+38}$&$-8.65_{-5.20}^{+2.50}$&$-3.56_{-6.95}^{+3.71}$\\
PG 2111+023&$-2.40_{-0.80}^{+0.80}$&$7.50_{-1.10}^{+1.20}$&$-5.30_{-0.90}^{+0.70}$&$10.86_{-1.50}^{+1.76}$&$81.0_{-13.1}^{+16.3}$&$186.6_{-61.0}^{+44.2}$&$241_{-35}^{+29}$&$-6.58_{-5.11}^{+3.33}$&$-5.09_{-2.30}^{+3.78}$\\
PG 0009+036&$-8.90_{-0.30}^{+0.20}$&$3.50_{-0.90}^{+1.10}$&$-5.60_{-1.90}^{+1.40}$&$6.67_{-1.65}^{+2.16}$&$34.2_{-33.3}^{+17.7}$&$40.5_{-10.7}^{+20.3}$&$256_{-40}^{+60}$&$-5.91_{-2.66}^{+3.72}$&$-9.95_{-8.64}^{+4.16}$\\
PG 0122+214&$-12.50_{-0.80}^{+0.70}$&$4.70_{-0.70}^{+0.90}$&$-5.60_{-1.00}^{+0.80}$&$8.62_{-1.29}^{+1.55}$&$34.5_{-6.3}^{+7.6}$&$39.9_{-5.4}^{+7.0}$&$156_{-21}^{+30}$&$-12.93_{-1.97}^{+1.77}$&$-3.39_{-2.14}^{+2.02}$\\
SB 357&$-7.50_{-0.10}^{+0.10}$&$-0.80_{-0.20}^{+0.10}$&$-5.90_{-1.10}^{+0.80}$&$6.02_{-0.89}^{+1.03}$&$6.1_{-5.9}^{+8.1}$&$53.3_{-6.5}^{+8.0}$&$198_{-26}^{+46}$&$-5.42_{-1.96}^{+1.72}$&$-6.64_{-2.06}^{+2.01}$\\
EC 04420-1908&$-14.70_{-2.70}^{+2.00}$&$-5.10_{-2.10}^{+1.50}$&$-6.20_{-2.50}^{+1.90}$&$10.44_{-3.16}^{+4.35}$&$68.6_{-26.0}^{+47.6}$&$29.3_{-8.6}^{+17.6}$&$257_{-54}^{+94}$&$-11.43_{-4.97}^{+4.13}$&$-7.40_{-4.29}^{+2.74}$\\
PG 1332+137&$-6.20_{-0.40}^{+0.60}$&$-0.60_{-0.20}^{+0.20}$&$6.30_{-1.70}^{+2.00}$&$6.54_{-1.70}^{+2.13}$&$58.2_{-11.8}^{+13.4}$&$32.4_{-6.9}^{+9.8}$&$413_{-77}^{+38}$&$-5.35_{-0.98}^{+1.22}$&$1.99_{-2.22}^{+4.08}$\\
EC 19596-5356&$3.20_{-2.90}^{+4.00}$&$-3.10_{-1.10}^{+0.80}$&$-7.30_{-2.60}^{+1.90}$&$13.81_{-3.63}^{+4.80}$&$33.5_{-19.5}^{+16.6}$&$29.2_{-9.1}^{+20.1}$&$475_{-83}^{+74}$&$-1.67_{-4.01}^{+4.41}$&$-6.08_{-6.77}^{+4.10}$\\
PHL 346&$-4.60_{-0.50}^{+0.70}$&$3.00_{-0.50}^{+0.60}$&$-7.30_{-1.40}^{+1.20}$&$8.55_{-1.33}^{+1.61}$&$21.4_{-3.5}^{+4.5}$&$28.5_{-3.1}^{+3.7}$&$418_{-47}^{+49}$&$-3.89_{-1.22}^{+1.28}$&$4.28_{-2.11}^{+2.93}$\\
PG 0914+001&$-18.90_{-3.50}^{+2.90}$&$-13.80_{-4.50}^{+3.70}$&$10.90_{-2.90}^{+3.60}$&$20.62_{-5.28}^{+6.74}$&$73.8_{-17.7}^{+23.2}$&$45.0_{-18.3}^{+61.1}$&$369_{-157}^{+240}$&$-14.82_{-9.71}^{+11.88}$&$-24.72_{-12.55}^{+9.27}$\\
PG 1209+263&$-11.80_{-0.90}^{+0.90}$&$-2.90_{-0.80}^{+0.60}$&$30.70_{-7.10}^{+7.70}$&$30.93_{-7.28}^{+7.94}$&$76.1_{-20.1}^{+32.6}$&$192.6_{-43.7}^{+37.2}$&$390_{-100}^{+293}$&$-26.70_{-66.82}^{+35.18}$&$-32.44_{-68.29}^{+39.87}$\\
\hline
\end{tabular}
\end{center}
\end{table*}

\begin{table*}
\begin{center}

\caption{Computed values for stars belonging to group B. The values shown are: the present coordinates of the star $(X,Y,Z)$ (in galactocentric Cartesian coordinates), the distance, the evolutionary age, the flight time, the ejection velocity and the coordinates $(X_0,Y_0)$ of the point of ejection in the galactic plane.}
\label{tbl:b4}

\scriptsize

\begin{tabular}{llllllllll}
\hline
Name & $X$ & $Y$ & $Z$ & $d$ & $\mathrm{Age}$ & $t_{\mathrm{flight}}$ & $v_{\mathrm{ejec}}$  & $X_0$ & $Y_0$\\
 & $\mathrm{(kpc)}$ & $\mathrm{(kpc)}$ & $\mathrm{(kpc)}$ & $\mathrm{(kpc)}$ & $\mathrm{(Myr)}$ & $\mathrm{(Myr)}$ & $(\mathrm{km}\,\mathrm{s}^{-1})$  & $\mathrm{(kpc)}$ & $\mathrm{(kpc)}$\\
\hline
HIP 81153&$-7.00_{-0.40}^{+1.20}$&$0.20_{-0.10}^{+0.20}$&$0.50_{-0.20}^{+0.60}$&$1.18_{-0.54}^{+1.20}$&$4.2_{-4.2}^{+2.1}$&$6.8_{-3.0}^{+5.3}$&$169_{-11}^{+32}$&$-7.88_{-0.09}^{+0.39}$&$-0.79_{-0.15}^{+0.38}$\\
HIP 1241&$-8.10_{-0.00}^{+0.10}$&$0.30_{-0.10}^{+0.00}$&$-0.60_{-0.10}^{+0.10}$&$0.64_{-0.13}^{+0.16}$&$174.8_{-50.8}^{+44.1}$&$29.1_{-2.9}^{+2.8}$&$43_{-8}^{+9}$&$-5.27_{-0.53}^{+0.56}$&$-5.47_{-0.34}^{+0.38}$\\
HIP 28132&$-9.00_{-0.20}^{+0.20}$&$-1.10_{-0.30}^{+0.20}$&$-0.60_{-0.10}^{+0.10}$&$1.59_{-0.29}^{+0.37}$&$43.5_{-15.5}^{+4.3}$&$19.9_{-3.0}^{+3.7}$&$58_{-7}^{+12}$&$-6.42_{-0.50}^{+0.72}$&$-5.55_{-1.08}^{+0.86}$\\
HIP 2702&$-7.70_{-0.10}^{+0.10}$&$-0.40_{-0.20}^{+0.10}$&$-0.70_{-0.30}^{+0.20}$&$0.87_{-0.25}^{+0.35}$&$24.9_{-24.9}^{+44.1}$&$62.6_{-9.1}^{+10.3}$&$59_{-11}^{+16}$&$-0.64_{-1.45}^{+1.42}$&$-9.43_{-0.43}^{+0.54}$\\
HIP 55051&$-8.10_{-0.00}^{+0.00}$&$-0.60_{-0.20}^{+0.20}$&$0.70_{-0.20}^{+0.30}$&$0.92_{-0.28}^{+0.41}$&$6.6_{-6.6}^{+4.6}$&$6.1_{-0.7}^{+0.6}$&$188_{-52}^{+77}$&$-8.43_{-0.26}^{+0.16}$&$-2.54_{-0.71}^{+0.55}$\\
HIP 114690&$-7.90_{-0.10}^{+0.00}$&$0.60_{-0.20}^{+0.20}$&$-0.70_{-0.30}^{+0.20}$&$0.89_{-0.27}^{+0.40}$&$19.8_{-4.3}^{+4.4}$&$6.9_{-1.5}^{+1.5}$&$104_{-12}^{+18}$&$-7.96_{-0.03}^{+0.05}$&$-1.02_{-0.09}^{+0.11}$\\
HIP 11844&$-8.30_{-0.10}^{+0.00}$&$-0.10_{-0.00}^{+0.00}$&$-0.80_{-0.10}^{+0.10}$&$0.87_{-0.09}^{+0.10}$&$175.0_{-47.5}^{+27.2}$&$70.9_{-4.5}^{+4.6}$&$77_{-6}^{+7}$&$1.23_{-0.73}^{+0.69}$&$-8.06_{-0.28}^{+0.31}$\\
HIP 16130&$-8.50_{-0.20}^{+0.10}$&$-0.30_{-0.10}^{+0.00}$&$-0.90_{-0.20}^{+0.20}$&$1.10_{-0.22}^{+0.27}$&$70.3_{-53.3}^{+26.2}$&$38.2_{-6.0}^{+6.1}$&$71_{-11}^{+12}$&$-3.08_{-1.31}^{+1.49}$&$-6.30_{-0.37}^{+0.52}$\\
HIP 58046&$-8.10_{-0.10}^{+0.00}$&$-0.20_{-0.10}^{+0.00}$&$0.90_{-0.30}^{+0.30}$&$0.92_{-0.27}^{+0.37}$&$1.8_{-1.8}^{+87.0}$&$16.7_{-3.1}^{+3.5}$&$86_{-19}^{+26}$&$-7.75_{-0.09}^{+0.10}$&$-4.51_{-1.23}^{+0.98}$\\
BD +36 2242&$-8.20_{-0.10}^{+0.00}$&$0.10_{-0.10}^{+0.00}$&$1.10_{-0.20}^{+0.30}$&$1.16_{-0.23}^{+0.29}$&$166.6_{-32.3}^{+35.8}$&$32.3_{-3.5}^{+3.9}$&$63_{-10}^{+12}$&$-4.47_{-0.77}^{+0.92}$&$-6.20_{-0.43}^{+0.48}$\\
HIP 98136&$-5.50_{-0.60}^{+1.00}$&$1.10_{-0.30}^{+0.40}$&$-1.10_{-0.40}^{+0.30}$&$2.98_{-0.80}^{+1.11}$&$15.5_{-3.3}^{+3.8}$&$21.0_{-2.1}^{+2.4}$&$163_{-45}^{+50}$&$-5.64_{-0.52}^{+0.54}$&$-0.90_{-0.94}^{+1.27}$\\
HIP 111396&$-7.50_{-0.10}^{+0.10}$&$0.60_{-0.10}^{+0.20}$&$-1.10_{-0.30}^{+0.20}$&$1.38_{-0.28}^{+0.33}$&$91.4_{-17.0}^{+20.4}$&$49.2_{-2.7}^{+2.9}$&$79_{-10}^{+11}$&$-0.59_{-0.64}^{+0.64}$&$-7.63_{-0.37}^{+0.29}$\\
HIP 77131&$-5.50_{-0.80}^{+1.20}$&$-0.50_{-0.30}^{+0.10}$&$1.20_{-0.40}^{+0.60}$&$2.79_{-0.85}^{+1.39}$&$4.5_{-2.9}^{+0.6}$&$17.9_{-1.5}^{+1.7}$&$124_{-41}^{+72}$&$-3.70_{-1.04}^{+1.46}$&$-3.18_{-0.38}^{+0.55}$\\
HIP 113735&$-7.20_{-0.20}^{+0.20}$&$-0.50_{-0.20}^{+0.10}$&$-1.20_{-0.50}^{+0.30}$&$1.56_{-0.38}^{+0.51}$&$12.5_{-12.3}^{+7.3}$&$15.3_{-1.3}^{+2.0}$&$168_{-39}^{+54}$&$-5.05_{-0.47}^{+0.54}$&$-5.63_{-0.57}^{+0.52}$\\
HIP 59955&$-7.90_{-0.10}^{+0.00}$&$-0.50_{-0.10}^{+0.10}$&$1.40_{-0.30}^{+0.40}$&$1.50_{-0.30}^{+0.38}$&$76.1_{-15.0}^{+15.9}$&$27.3_{-4.3}^{+4.3}$&$120_{-22}^{+28}$&$-4.31_{-1.11}^{+1.37}$&$-4.29_{-0.37}^{+0.38}$\\
HIP 76161&$-5.00_{-0.80}^{+1.20}$&$-0.90_{-0.40}^{+0.20}$&$1.40_{-0.30}^{+0.60}$&$3.44_{-0.93}^{+1.38}$&$14.0_{-3.5}^{+3.9}$&$30.1_{-3.1}^{+3.7}$&$109_{-26}^{+41}$&$-1.11_{-0.84}^{+0.84}$&$-6.12_{-0.57}^{+0.53}$\\
HIP 96130&$-4.20_{-1.00}^{+1.50}$&$0.90_{-0.20}^{+0.30}$&$-1.40_{-0.60}^{+0.30}$&$4.17_{-1.11}^{+1.58}$&$14.4_{-2.1}^{+1.4}$&$18.3_{-2.5}^{+2.7}$&$166_{-48}^{+92}$&$-1.44_{-0.81}^{+0.96}$&$-2.88_{-0.66}^{+0.68}$\\
HIP 61800&$-8.20_{-0.00}^{+0.00}$&$0.20_{-0.10}^{+0.00}$&$1.50_{-0.30}^{+0.40}$&$1.51_{-0.29}^{+0.42}$&$100.9_{-24.9}^{+26.6}$&$58.2_{-7.6}^{+7.9}$&$94_{-15}^{+18}$&$-2.17_{-1.34}^{+1.37}$&$-7.34_{-0.28}^{+0.36}$\\
HIP 114569&$-7.50_{-0.10}^{+0.10}$&$0.40_{-0.10}^{+0.10}$&$-1.50_{-0.30}^{+0.30}$&$1.60_{-0.31}^{+0.40}$&$23.3_{-20.0}^{+10.8}$&$7.6_{-0.7}^{+0.7}$&$408_{-71}^{+89}$&$-4.46_{-0.96}^{+1.24}$&$-2.33_{-0.34}^{+0.31}$\\
HIP 11809&$-9.20_{-0.50}^{+0.40}$&$0.30_{-0.10}^{+0.10}$&$-1.60_{-0.60}^{+0.50}$&$2.01_{-0.60}^{+0.84}$&$2.0_{-2.0}^{+89.8}$&$11.6_{-0.8}^{+0.8}$&$230_{-66}^{+94}$&$-11.10_{-1.43}^{+1.02}$&$-2.61_{-0.25}^{+0.22}$\\
HIP 16758&$-8.10_{-0.10}^{+0.00}$&$-1.40_{-0.50}^{+0.40}$&$-1.60_{-0.70}^{+0.40}$&$2.13_{-0.57}^{+0.82}$&$13.5_{-2.7}^{+3.1}$&$21.0_{-4.9}^{+5.9}$&$153_{-20}^{+19}$&$-4.23_{-1.60}^{+2.59}$&$-4.16_{-0.71}^{+0.81}$\\
HIP 108215&$-6.80_{-0.30}^{+0.30}$&$0.40_{-0.10}^{+0.10}$&$-1.60_{-0.40}^{+0.40}$&$2.34_{-0.51}^{+0.68}$&$42.6_{-11.7}^{+16.3}$&$27.1_{-2.2}^{+2.6}$&$256_{-52}^{+78}$&$-1.11_{-0.35}^{+0.36}$&$-1.98_{-0.74}^{+0.84}$\\
HIP 115347&$-7.70_{-0.10}^{+0.10}$&$0.80_{-0.20}^{+0.20}$&$-1.60_{-0.60}^{+0.40}$&$1.79_{-0.47}^{+0.66}$&$18.4_{-4.0}^{+4.7}$&$33.8_{-5.5}^{+6.6}$&$84_{-15}^{+19}$&$-7.66_{-0.64}^{+0.40}$&$-5.71_{-0.62}^{+0.59}$\\
HIP 115729&$-7.70_{-0.10}^{+0.10}$&$0.70_{-0.10}^{+0.20}$&$-1.60_{-0.50}^{+0.30}$&$1.82_{-0.36}^{+0.47}$&$32.9_{-6.6}^{+7.4}$&$26.6_{-2.4}^{+2.5}$&$88_{-14}^{+18}$&$-5.44_{-0.50}^{+0.60}$&$-4.75_{-0.31}^{+0.29}$\\
HIP 3812&$-7.50_{-0.10}^{+0.20}$&$-0.80_{-0.30}^{+0.20}$&$-1.70_{-0.60}^{+0.50}$&$1.91_{-0.52}^{+0.69}$&$0.2_{-0.2}^{+15.5}$&$37.2_{-6.4}^{+6.8}$&$85_{-17}^{+22}$&$-4.39_{-1.02}^{+1.09}$&$-6.77_{-0.67}^{+0.75}$\\
PG 1530+212&$-6.90_{-0.30}^{+0.50}$&$0.70_{-0.20}^{+0.30}$&$1.70_{-0.50}^{+0.70}$&$2.15_{-0.63}^{+0.89}$&$63.3_{-52.8}^{+14.1}$&$37.9_{-8.0}^{+8.3}$&$139_{-37}^{+49}$&$-1.97_{-2.14}^{+2.26}$&$-8.71_{-1.92}^{+1.48}$\\
HIP 79649&$-7.90_{-0.00}^{+0.10}$&$1.90_{-0.50}^{+0.70}$&$1.90_{-0.50}^{+0.60}$&$2.70_{-0.70}^{+0.98}$&$18.3_{-3.6}^{+4.0}$&$28.7_{-4.2}^{+5.0}$&$89_{-15}^{+18}$&$-6.89_{-0.31}^{+0.37}$&$-4.47_{-0.64}^{+0.52}$\\
HIP 12320&$-9.70_{-0.80}^{+0.50}$&$0.50_{-0.20}^{+0.10}$&$-2.00_{-0.90}^{+0.60}$&$2.72_{-0.79}^{+1.15}$&$72.8_{-72.5}^{+29.4}$&$47.6_{-9.4}^{+10.4}$&$105_{-25}^{+38}$&$-2.76_{-1.77}^{+2.13}$&$-6.27_{-0.72}^{+0.80}$\\
HIP 109051&$-6.80_{-0.30}^{+0.50}$&$1.20_{-0.30}^{+0.40}$&$-2.00_{-0.70}^{+0.50}$&$2.63_{-0.67}^{+0.92}$&$19.9_{-6.1}^{+4.8}$&$21.1_{-3.2}^{+3.3}$&$123_{-17}^{+27}$&$-6.03_{-0.66}^{+0.85}$&$-3.87_{-0.49}^{+0.44}$\\
HIP 111563&$-7.10_{-0.20}^{+0.50}$&$0.90_{-0.30}^{+0.50}$&$-2.00_{-1.00}^{+0.60}$&$2.40_{-0.70}^{+1.10}$&$9.4_{-2.8}^{+3.5}$&$15.8_{-1.7}^{+1.8}$&$223_{-62}^{+86}$&$-7.22_{-0.19}^{+0.21}$&$0.10_{-0.96}^{+1.83}$\\
HIP 13800&$-9.80_{-0.50}^{+0.30}$&$0.00_{-0.00}^{+0.10}$&$-2.20_{-0.60}^{+0.40}$&$2.89_{-0.60}^{+0.75}$&$47.7_{-8.3}^{+10.6}$&$59.5_{-4.9}^{+5.3}$&$145_{-31}^{+41}$&$-2.02_{-1.07}^{+1.13}$&$-5.88_{-0.72}^{+1.02}$\\
HIP 48394&$-9.90_{-0.90}^{+0.50}$&$0.80_{-0.30}^{+0.30}$&$2.20_{-0.60}^{+1.10}$&$3.07_{-0.91}^{+1.39}$&$21.4_{-9.9}^{+5.2}$&$25.4_{-4.1}^{+4.7}$&$122_{-29}^{+41}$&$-9.27_{-1.32}^{+0.77}$&$-3.05_{-0.58}^{+0.79}$\\
HIP 107027&$-6.20_{-0.50}^{+0.70}$&$1.30_{-0.40}^{+0.40}$&$-2.20_{-0.90}^{+0.60}$&$3.15_{-0.83}^{+1.20}$&$16.2_{-2.3}^{+1.2}$&$23.7_{-5.3}^{+7.8}$&$181_{-22}^{+34}$&$-9.43_{-1.64}^{+0.67}$&$-1.63_{-0.49}^{+1.19}$\\
HIP 1904&$-7.60_{-0.10}^{+0.10}$&$-0.20_{-0.10}^{+0.00}$&$-2.30_{-0.60}^{+0.50}$&$2.32_{-0.51}^{+0.69}$&$46.2_{-13.2}^{+18.8}$&$57.6_{-3.9}^{+3.9}$&$164_{-31}^{+45}$&$-0.12_{-0.70}^{+0.65}$&$-5.10_{-0.71}^{+0.93}$\\
HIP 55461&$-8.50_{-0.20}^{+0.10}$&$-1.10_{-0.30}^{+0.30}$&$2.40_{-0.60}^{+0.80}$&$2.64_{-0.67}^{+0.90}$&$54.2_{-27.7}^{+20.1}$&$29.9_{-7.1}^{+9.1}$&$145_{-24}^{+38}$&$-6.79_{-0.50}^{+0.80}$&$-4.39_{-0.42}^{+0.50}$\\
HIP 112790&$-7.10_{-0.10}^{+0.10}$&$1.20_{-0.20}^{+0.10}$&$-2.40_{-0.30}^{+0.40}$&$2.84_{-0.44}^{+0.33}$&$60.3_{-9.3}^{+12.0}$&$170.1_{-48.3}^{+49.6}$&$169_{-18}^{+11}$&$-19.89_{-6.20}^{+5.53}$&$-14.23_{-2.14}^{+2.16}$\\
HIP 59067&$-8.10_{-0.00}^{+0.10}$&$-0.90_{-0.20}^{+0.20}$&$2.60_{-0.50}^{+0.60}$&$2.75_{-0.52}^{+0.65}$&$72.7_{-8.6}^{+3.2}$&$39.2_{-3.8}^{+3.5}$&$234_{-46}^{+59}$&$-1.73_{-1.25}^{+1.28}$&$-3.24_{-0.67}^{+1.00}$\\
BD +20 3004&$-6.70_{-0.20}^{+0.40}$&$0.50_{-0.10}^{+0.20}$&$2.90_{-0.60}^{+0.80}$&$3.23_{-0.63}^{+0.90}$&$71.0_{-17.7}^{+19.2}$&$18.8_{-1.4}^{+1.2}$&$247_{-46}^{+67}$&$-2.70_{-1.13}^{+1.38}$&$-2.65_{-0.29}^{+0.37}$\\
HIP 105912&$-5.20_{-0.70}^{+1.20}$&$0.70_{-0.20}^{+0.30}$&$-3.00_{-1.20}^{+0.80}$&$4.17_{-1.14}^{+1.70}$&$12.8_{-3.3}^{+4.1}$&$15.3_{-0.6}^{+0.7}$&$457_{-133}^{+130}$&$-2.48_{-1.15}^{+1.59}$&$0.71_{-1.07}^{+1.52}$\\
HIP 45904&$-11.10_{-1.20}^{+0.80}$&$0.60_{-0.10}^{+0.30}$&$3.10_{-0.80}^{+1.10}$&$4.42_{-1.15}^{+1.66}$&$15.4_{-2.9}^{+3.1}$&$24.8_{-3.7}^{+4.3}$&$240_{-54}^{+71}$&$-10.20_{-1.55}^{+0.98}$&$-0.25_{-1.19}^{+2.16}$\\
HIP 37903&$-13.10_{-2.60}^{+1.70}$&$2.30_{-0.80}^{+1.20}$&$3.20_{-1.10}^{+1.70}$&$6.42_{-2.11}^{+3.36}$&$19.1_{-6.7}^{+8.9}$&$26.4_{-5.8}^{+8.1}$&$279_{-80}^{+123}$&$-8.83_{-1.64}^{+1.21}$&$-0.04_{-1.60}^{+3.54}$\\
HIP 70275&$-5.40_{-0.60}^{+0.90}$&$-1.10_{-0.30}^{+0.30}$&$3.20_{-0.80}^{+1.10}$&$4.26_{-1.08}^{+1.48}$&$15.1_{-2.8}^{+3.5}$&$38.3_{-15.8}^{+38.4}$&$271_{-8}^{+7}$&$-13.65_{-8.48}^{+2.95}$&$-2.59_{-0.56}^{+0.85}$\\
BD -2 3766&$-5.90_{-0.50}^{+0.80}$&$-1.00_{-0.40}^{+0.20}$&$3.50_{-0.90}^{+1.30}$&$4.22_{-1.10}^{+1.50}$&$15.2_{-3.0}^{+3.0}$&$14.8_{-1.3}^{+1.2}$&$425_{-109}^{+151}$&$-3.82_{-1.13}^{+1.54}$&$-8.13_{-2.45}^{+1.74}$\\
PB 5418&$-7.90_{-0.10}^{+0.00}$&$3.80_{-1.00}^{+1.20}$&$-4.80_{-1.60}^{+1.20}$&$6.09_{-1.49}^{+2.03}$&$24.6_{-15.0}^{+7.5}$&$16.5_{-2.3}^{+2.8}$&$415_{-100}^{+141}$&$-5.18_{-1.06}^{+1.58}$&$1.13_{-1.38}^{+2.16}$\\
HIP 56322&$-8.60_{-0.30}^{+0.20}$&$-2.90_{-1.40}^{+0.90}$&$5.30_{-1.70}^{+2.70}$&$6.09_{-1.92}^{+3.17}$&$9.2_{-2.2}^{+2.1}$&$13.1_{-2.8}^{+3.4}$&$471_{-99}^{+189}$&$-7.21_{-0.38}^{+0.76}$&$-8.13_{-5.46}^{+2.96}$\\
Ton S 308&$-10.60_{-0.90}^{+0.70}$&$-2.00_{-0.70}^{+0.50}$&$-5.90_{-2.00}^{+1.50}$&$6.76_{-1.69}^{+2.24}$&$33.0_{-18.9}^{+10.3}$&$47.9_{-10.6}^{+14.6}$&$169_{-24}^{+43}$&$-7.36_{-3.14}^{+3.37}$&$-8.84_{-3.35}^{+2.24}$\\
PHL 2018&$-5.10_{-0.70}^{+1.10}$&$1.80_{-0.40}^{+0.60}$&$-6.00_{-2.10}^{+1.50}$&$6.93_{-1.77}^{+2.39}$&$26.6_{-5.3}^{+5.7}$&$24.4_{-3.7}^{+4.8}$&$399_{-66}^{+68}$&$-5.90_{-0.92}^{+1.11}$&$4.21_{-2.64}^{+4.72}$\\
HIP 52906&$-11.60_{-0.80}^{+0.70}$&$-0.10_{-0.10}^{+0.00}$&$6.80_{-1.40}^{+1.60}$&$7.72_{-1.53}^{+1.85}$&$25.3_{-4.4}^{+5.4}$&$90.6_{-23.8}^{+46.6}$&$277_{-31}^{+25}$&$16.67_{-8.09}^{+16.54}$&$-10.62_{-4.90}^{+2.44}$\\
\hline
\end{tabular}
\end{center}
\end{table*}

\end{document}